\documentclass[11pt,a4paper]{article}
\usepackage{amsmath,amssymb,graphicx,cite,color,url,hyperref,setspace,bm}
\usepackage{bold-extra}
\newcommand*{\defeq}{\stackrel{\text{def}}{=}}
\begin{document}

\begin{flushright}
TIFR/TH/23-3
\end{flushright}

\begin{center}

\makebox{{\bfseries{\scshape{\LARGE
The Higgs Boson and its Physics --- an Overview}}}}

\setstretch{1.15}
{\sl Sreerup Raychaudhuri}\\ [2mm]
\setstretch{1.05}

{\small Department of Theoretical Physics, Tata Institute of Fundamental 
Research,\\ Homi Bhabha Road, Mumbai 400 005, India.} \\ E-mail: {\sf 
sreerup@theory.tifr.res.in}\\ [4mm] \setstretch{1.15} \today

\bigskip\medskip

\centerline{\Large\bf Abstract}
\end{center}
\vspace*{-0.27in}
\begin{quotation}
\small
\setstretch{1.05}
\noindent The Higgs boson plays a central role in the Standard Model, as 
well as in theories which go beyond it. This article is therefore 
divided into two parts. The first takes s historical approach and shows 
how the mass problem entered weak interaction theory from the beginning 
and how it was solved by invoking the Higgs boson. This is followed by a 
construction of the Glashow-Salam-Weinberg model, again stressing the 
role played by the Higgs doublet. This part culminates in the Higgs 
boson discovery of 2012. The second part first discusses the 
shortcomings of the Standard Model and then touches upon the major 
theories which try to improve upon it, mostly with profound consequences 
on the Higgs sector. This is followed up by short descriptions of a 
number of popular extensions of the Higgs sector, and culminates in a 
brief introduction to effective field theory approaches to studying the 
Higgs sector.
\end{quotation}
\normalsize
\setstretch{1.1}

\centerline{\sl Invited article to appear in a Special Issue of the {\em Indian Journal of Physics} }

\begin{center}
\Large{\textbf{{\textsc{Part 1}} : Roads to the Higgs boson}}
\end{center}

The 2012 discovery of the Higgs boson created a worldwide sensation and 
was followed a year later by the award of a much-belated Nobel Prize to 
two of the seven scientists who had originally worked out the basic 
theory of this elusive particle in 1964. While it was widely reported 
that the search for the Higgs boson was the culmination of an effort 
which had lasted the 48 years from 1964 to 2012, the real journey had 
started much earlier. In fact, it started in 1934 when Enrico Fermi made 
a facile assumption in an effort to explain beta decay. The Higgs boson 
discovery was, therefore, the culmination of a 78 year-long series of 
developments.

In order to appreciate the proper significance of the Higgs boson as the 
keystone, as it were, of the electroweak theory, it is necessary to 
retrace this long path and see how the current formalism was arrived at. 
The first part of this article, therefore, does just this. It starts 
with the Fermi theory of weak interactions and explains successive 
developments in a part-historical part-pedagogic manner and ends with 
the Higgs boson discovery.

However, the Higgs boson, now that it is known to exist, throws up 
almost as many questions as it answers. Is it an elementary particle or 
a composite? Are there more scalars in the theory? Can we understand why 
at high energies, it behaves as if tachyonic? All of this has led to 
intense speculation about the nature and properties of this particle, 
and suggestions that it is just a part of some additional structures in 
the electroweak sector -- the tip of the iceberg, as it were. The second 
part of this article will discuss these ideas in some detail.

This article will deal mostly with theoretical and phenomenological 
aspects of the Higgs story. The purely experimental aspects are 
undoubtedly fascinating, but they call for a different article and a 
different author with the appropriate expertise.

The titles of the different sections invoke different evolutionary 
stages in the historical development of Western art. This is not 
whimsical but a serious attempt to draw parallels with parallel fields 
of human creativity, of which science forms just one, even though it is 
expressed in a different form.

\section{\large\bf Beginnings: Gauge Theory and Electroweak Unification} 

The concept of a {\it symmetry}, though understood in geometry from 
ancient times, and intuitively used by Newton and his followers, finds 
its precise expression in the variational formulation of classical 
mechanics. A symmetry is really a {\it no-go theorem}, for it means that 
we cannot physically distinguish between two different configurations of 
the dynamical variables describing a system, i.e. the action remains 
invariant. It was Emmy N\"other who famously discovered in 1918 that 
there is an intimate connection between symmetries of the action and 
conserved quantities \cite{Nother}, but the fact that there is an 
intimate connection between symmetries and the fundamental forces of 
Nature was really an epiphany central to Einstein's 1915 theory of 
general relativity \cite{Einstein-GR}. The {\it equivalence principle}, 
the foundation on which general relativity is based, is simply the 
statement that general coordinate invariance -- a symmetry -- becomes 
manifest in experiments as a gravitational force \cite{Hilbert}. General 
coordinate invariance is, of course, just a no-go theorem for attaining 
an absolute frame of reference $\grave{\rm a}$ la Newton. Soon, Hermann 
Weyl \cite{Weyl} discovered another symmetry of the free Einstein 
equations, viz., under dilatations, to which he gave the name 'gauge' 
symmetry.

The advent of quantum mechanics \cite{quantum} in 1925-26 led to a 
similar idea being applied by V.~Fock and, independently, by F.~London 
\cite{London} in the context of the Schr\"odinger equation.  The force, 
in this case, turned out to be another well-known one, viz., 
electromagnetism. The name given by Weyl to a dilatation, i.e., gauge 
transformation, was now re-purposed to describe a phase transformation. 
In modern parlance, it can be a combination of the two. However, since 
the core idea invokes special relativity, it is better expressed in a 
fully relativistic formalism.
 
We start, therefore, with the Lorentz-invariant action for a free 
complex scalar field
\begin{equation}
S \defeq \int d^4x \left[ \partial^\mu \varphi^*(x) \ \partial_\mu 
\varphi(x) + M^2 \varphi^*(x) \varphi(x) \right]
\label{eqn:freescalar}
\end{equation}
and note that it is invariant under a phase transformation
\begin{equation}
\varphi(x) \to \varphi'(x) = e^{-ie\alpha} \; \varphi(x)
\label{eqn:U1gauge}
\end{equation}
if $\alpha$ is independent of spacetime coordinates $x$, but not 
otherwise. Here $e$ is a book-keeping parameter, which can be different 
for different fields. However, a spacetime-independent $\alpha$ would 
mean the same phase transformation at all points in spacetime --- which 
would require a signal to propagate instantaneously. Relativity forbids 
this and therefore, demands that $\alpha$ should be spacetime-dependent, 
i.e. $\alpha = \alpha(x)$. On the other hand, the equation of motion, 
which arises from this action is
\begin{equation}
\left( \Box + M^2 \right) \; \varphi(x) = 0
\label{eqn:KleinGordon}
\end{equation}  
indicating that $\alpha$ is the phase of the scalar wavefunction, when 
we make the transition from a field theory to a single-particle 
equation. In quantum mechanics, the probability interpretation tells us 
that the phase of this wavefunction does not affect any physics, and 
this means that there must exist a symmetry under phase transformations 
--- another no-go theorem that the phase cannot be measured. Translated 
to the field theory we have a paradox, viz., if $ \alpha$ is local, then 
this phase invariance is lost. This may be compared to the loss of local 
Lorentz invariance of the spacetime derivative $\partial_\mu$ in a 
curved space, and it has the same solution as in general relativity, 
viz., to add a field of force $A_\mu(x)$ which will nullify the effect 
and restore the symmetry. Thus, we rewrite the action as
\begin{equation}
S \defeq \int d^4x \left[ D^\mu \varphi^*(x) \ D_\mu \varphi(x) + M^2 
\varphi^*(x) \varphi(x) \right]
\label{eqn:scalarcovariant}
\end{equation}
where $D_\mu = \partial_\mu + iA_\mu(x)$ is called a (gauge) covariant 
derivative and we require that the transformations
\begin{equation}
\varphi(x) \to \varphi'(x) = e^{-i\alpha} \; \varphi(x)
\qquad\qquad
A_\mu(x) \to A'_\mu(x) = A_\mu(x) + \partial_\mu \alpha(x)
\label{eqn:localU1gt}
\end{equation}   
occur simultaneously. Taken together, we refer to this as a local {\it 
gauge transformstion}. The new field $A_\mu(x)$ -- which is known as the 
{\it gauge field} -- must have its own dynamics, which involves the 
construct $ \partial_\mu A_\nu(x)$. The minimal Lorentz and gauge 
invariant construct with this object is $F^{\mu\nu}F_{\mu\nu}$, where 
$F_{\mu\nu}(x) = \partial_\mu A_\nu(x) - \partial_\nu A_ \mu(x) $. 
Adding this to the action, we obtain
\begin{equation}
S \defeq \int d^4x \left[ D^\mu \varphi^*(x) \ D_\mu \varphi(x) + M^2 \varphi^*(x) 
\varphi(x) -\frac{1}{4}\;F^{\mu\nu}(x) \, F_{\mu\nu}(x)   \right]
\label{eqn:sEDaction}
\end{equation} 
from which the equations of motion become
\begin{eqnarray}
\left( D_\mu + iA_\mu \right)^2 \varphi(x) & = & 0 \nonumber \\
\partial_\mu F^{\mu\nu} & = & J^\nu(x)
\label{eqn:Maxwellequations} 
\end{eqnarray} 
where
\begin{equation}
J_\mu(x) \defeq i\left[\varphi^*(x)\; \partial_\mu \;\varphi(x) - 
\varphi(x)\; \partial_\mu \;\varphi^*(x) \right]
\label{eqn:scalarcurrent}
\end{equation}
is a conserved N\"other current corresponding to the phase symmetry. 
These correspond so perfectly to the equations of a charged scalar 
$\varphi(x)$ in an electromagnetic field $A_\mu(x)$ that it is quite 
natural to identify the gauge field $A_\mu$ with the force field of 
Maxwellian electrodynamics. This is the essence of Fock and London's 
understanding of the electromagnetic field as a consequence of the phase 
symmetry of quantum mechanics.

We note that the addition of a mass term 
\begin{equation}
S_m \defeq \int d^4x \  M_A^2 \, A^\mu(x) A_\mu(x)
\label{eqn:U1massterm}
\end{equation}
would result in a mass $M_A$ for the quanta of the electromagnetic 
field, i.e. the photon. However, such a term is clearly not gauge 
invariant and hence the photon must perforce remain massless --- as 
indeed, it is.

It is somewhat mind-boggling to think that all electromagnetic phenomena 
-- from lightning bolts in the sky to the tiny currents in our brains -- 
arise from an inability to measure the phase of so tenuous a thing as a 
wavefunction \footnote{It certainly failed to convince Einstein himself, 
whose original idea it stemmed from. This led to Einstein attacking the 
very basis of phase invariance, viz., the probability interpretation of 
quantum mechanics, coining the dictum 'God does not play dice'. But we 
now have plenty of evidence that, in fact, He does.}. However, this is 
no more surprising than that a universal force like gravity, felt so 
acutely on the Earth's surface, should stem from our inability to detect 
the state of acceleration of an observer. A more pedantic statement 
would be that our mathematical understanding of forces in terms of 
fields of force lend themselves to an understanding in terms of symmetry 
\cite{Tong}.

The discovery of the Dirac equation in 1928 \cite{Dirac} immediately led 
to a reformulation of the above theory for a charged fermion field, and 
this led to a description of {\it quantum elctrodynamics}, or QED, which 
has the classical action
\begin{equation}
S_{\rm QED} \defeq \int d^4x \left[ i\overline{\psi}(x) \gamma^\mu D_\mu 
\psi(x) - m \overline{\psi}(x) \psi(x) -\frac{1}{4}\;F^{\mu\nu}(x) \, 
F_{\mu\nu}(x)\right]
\label{eqn:QED}
\end{equation}
where $D_\mu = \partial_\mu + ieQ_e A_\mu$. This has local gauge 
invariance under
\begin{equation}
\psi(x) \to \psi'(x) = e^{-ieQ_e\alpha} \; \psi (x)
\qquad\qquad
A_\mu(x) \to A'_\mu(x) = A_\mu(x) + \partial_\mu \alpha(x)
\label{eqn:localU1gtfermion}
\end{equation}     
where $e$ is the proton charge and $Q_e = -1$. This also leads to the 
conservation of the N\"other current
\begin{equation}
J_\mu (x) \defeq e\overline{\psi}(x) \gamma_\mu \psi(x) \qquad\qquad 
\partial_\mu J^\mu (x) = 0
\label{eqn:electroncurrent}
\end{equation}
which is nothing but the electromagnetic current familiar to us in 
everyday life. It is this formulation which initiated all the triumphs 
of the 1930s and 1940s in quantum electrodynamics.

Almost simultaneously with the invention of quantum field theory 
\cite{Pauli}, however came the realisation that there are two more 
fundamental fields of force. Chadwick's discovery of the neutron in 1932 
\cite{Chadwick} inspired Heisenberg to postulate the existence of the 
{\it strong} (nuclear) force \cite{Heisenberg}, which was soon 
formulated as a field theory by Yukawa \cite{Yukawa}. But also, 
simultaneously, came the first understanding of weak interactions. As 
far back as 1899, Rutherford had identified alpha and beta particles as 
separate components of radioactive emanations, with greatly different 
penetrating power \cite{Rutherford}. It was, however, Becquerel who 
identified the beta rays as high-energy electrons \cite{Becquerel}. The 
fact that they have energies around an MeV was a clear proof that they 
did not come from among the atomic orbitals which have binding energies 
of a few tens of keV at most. However, the question of how high energy 
electrons can remain bound inside the positively-charged nucleus 
remained a mystery till the advent of quantum mechanics. It was 
Ambartzumian and Iwanenko who proved \cite{Ivanenko}, using the 
Uncertainty Principle, that electrons bound inside nuclei must have 
energies close to a GeV{\footnote{This is a textbook exercise today.}, 
whereas the beta particles have energies an order of magnitude less. It 
was clear, therefore, that the beta rays must be produced {\it at the 
instant} of interaction. The extremely long lives of nuclides undergoing 
beta decay made it clear that this interaction was not the strong 
interaction.

The situation was clarified by the neutron discovery. The mass 
difference between a neutron and a proton is just about 1.2~MeV, and 
therefore, the beta decay process could be interpreted as a neutron 
decaying into a proton and an electron. Pauli (1930) then postulated an 
additional massless, neutral particle -- the neutrino -- to ensure 
energy-momentum conservation \cite{neutrino}. This novel object would be 
a participant only in the {\it weak} (nuclear) interaction, which must 
be the cause of beta decay processes of the form
\begin{equation}
{}_1n^0 \to {}_1p^+ + e^- + \bar{\nu}_e
\label{eqn:betadecay}
\end{equation} 
where we have written $\bar{\nu}_e$ instead of just $\nu$ with the 
benefit of hindsight. The ingredients were all in place now for Fermi to 
put together his heuristic theory of beta decay \cite{Fermi}, with the 
famous current-current form of the interaction\footnote{It is worth 
mentioning that Fermi had submitted his paper to {\it Nature}, which had 
rejected it '{\it because it contained speculations too remote from 
reality to be of interest to the reader}' \cite{PaisInward}.}, i.e.
\begin{equation}
S_\beta = \int d^4x \ G_W \; J^\mu_{(h)}(x) \, J_\mu^{(\ell)}(x)
\label{eqn:currentcurrent}
\end{equation}
\vspace*{-0.4in} 
\begin{minipage}{4.0in}
where the {\it hadronic} current is
\begin{equation}
J^\mu_{(h)}(x) \defeq \overline{\psi}_p(x) \gamma^\mu \psi_n(x)
\label{eqn:hadroncurent} 
\end{equation}
and the {\it leptonic} current is
\begin{equation}
J_\mu^{(\ell)}(x) \defeq \overline{\psi}_e(x) \gamma_\mu \psi_\nu(x)
\label{eqn:leptoncurrent} 
\end{equation}
while $G_W$ is a (dimensionful) coupling constant. 
\end{minipage}
\hskip 16pt
\begin{minipage}{2.4in}
\begin{center} 
\includegraphics[height = 0.22\textheight, 
width=0.95\textwidth]{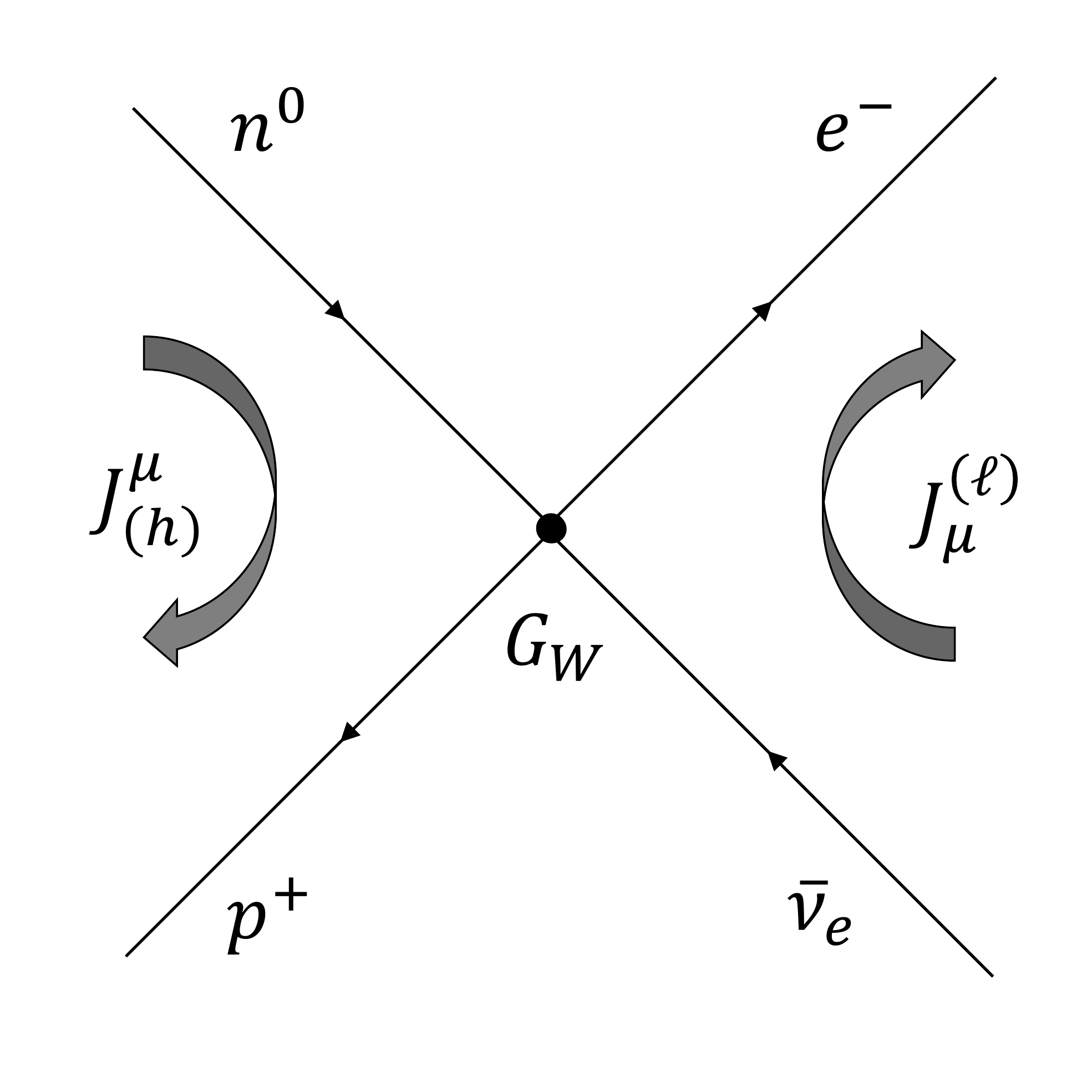}
\end{center}
\end{minipage}

\bigskip

\begin{minipage}{3.0in}
\begin{center} 
\includegraphics[height = 0.22\textheight, 
width=1.15\textwidth]{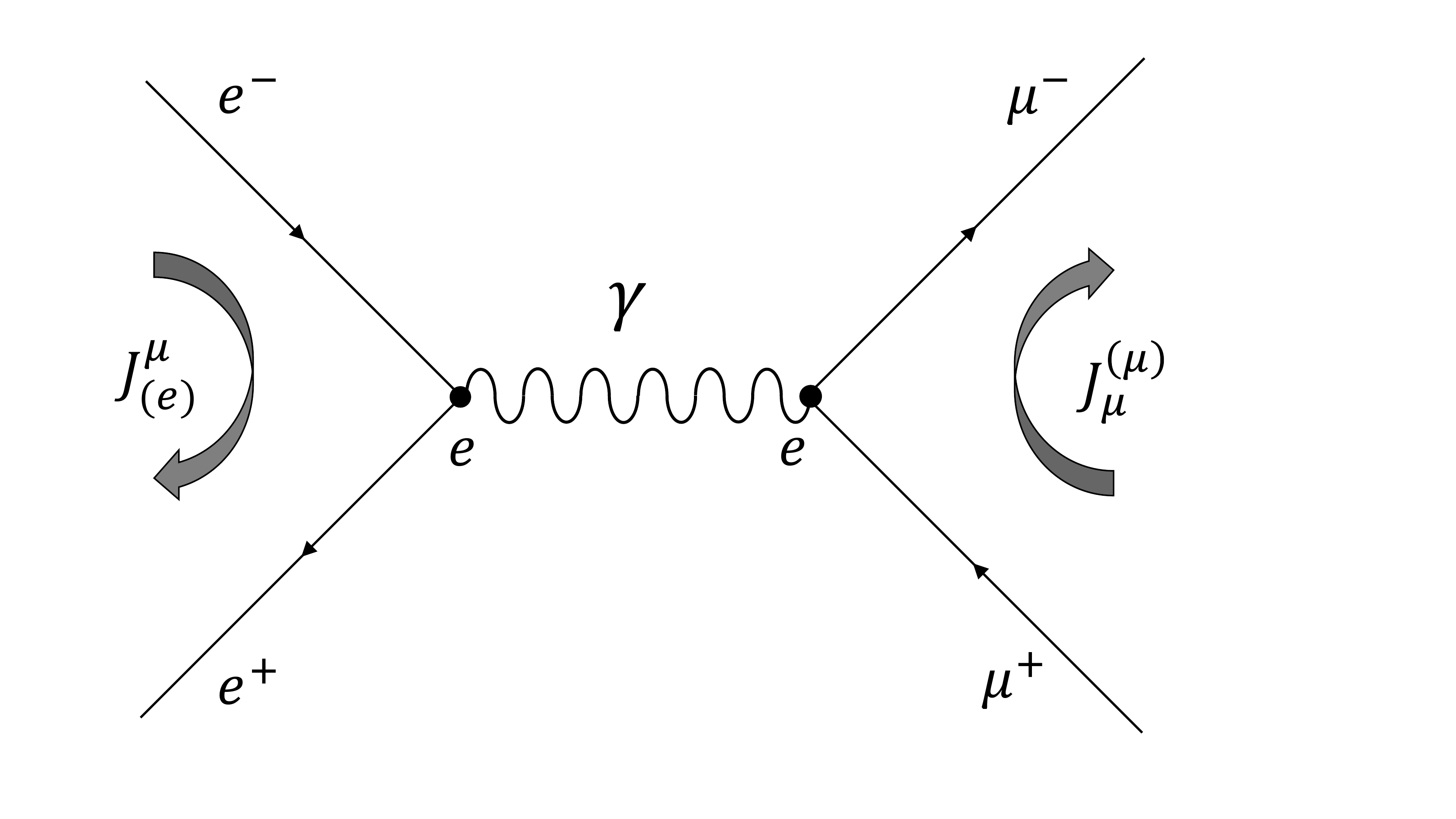}
\end{center}
\end{minipage}
\hskip 6pt
\begin{minipage}{3.4in}

This is a classic four-fermion interaction, and was clearly inspired 
\cite{Lee} by the form of the effective interaction for an 
electromagnetic scattering of the type, say, $e^+ e^- \to \mu^+\mu^-$, 
which takes the form
\begin{eqnarray}
{\cal L}_{\rm eff} & = & e^2 \, J^\mu_{\rm (e)} \frac{g_{\mu\nu}}{E_{\rm cm}^2} J^\nu_{(\mu)} \\
& = & \frac{e^2}{E_{\rm cm}^2} \; \overline{\psi}_e(x) \gamma^\mu \psi_e(x) \ 
\overline{\psi}_\mu(x) \gamma^\mu \psi_\mu(x)  \nonumber
\label{eqn:ee2mumuQED} 
\end{eqnarray}
\noindent In QED the effective coupling $e^2/E_{\rm cm}^2$ is 
energy-dependent. Whatever was the reasoning which led
\end{minipage}
\vspace*{-0.2in}

Fermi to set this as to a constant for weak interactions eventually 
proved to be a very deep insight --- and, as a matter of fact, it lies 
at the very core of the theme of this article. The choice of vector 
currents, instead of any other Lorentz covariant structures, is more 
understandable as inspired by the example of QED, but it proved to be an 
equally brilliant choice\footnote{In fact, when eventually published, 
Fermi's work was considered less remarkable for its form of the 
interaction than for making a connection between the interaction action 
and the decay rate, which he later formulated as the Golden Rule (see 
E.~Fermi, {\it Nuclear Physics} U. of Chicago Press (1950)). The 
groundwork for this, however, had already been done by Dirac 
\cite{golden}}.

\newpage

\begin{figure}[h!]
\begin{center}
\includegraphics[width=0.5\textwidth]{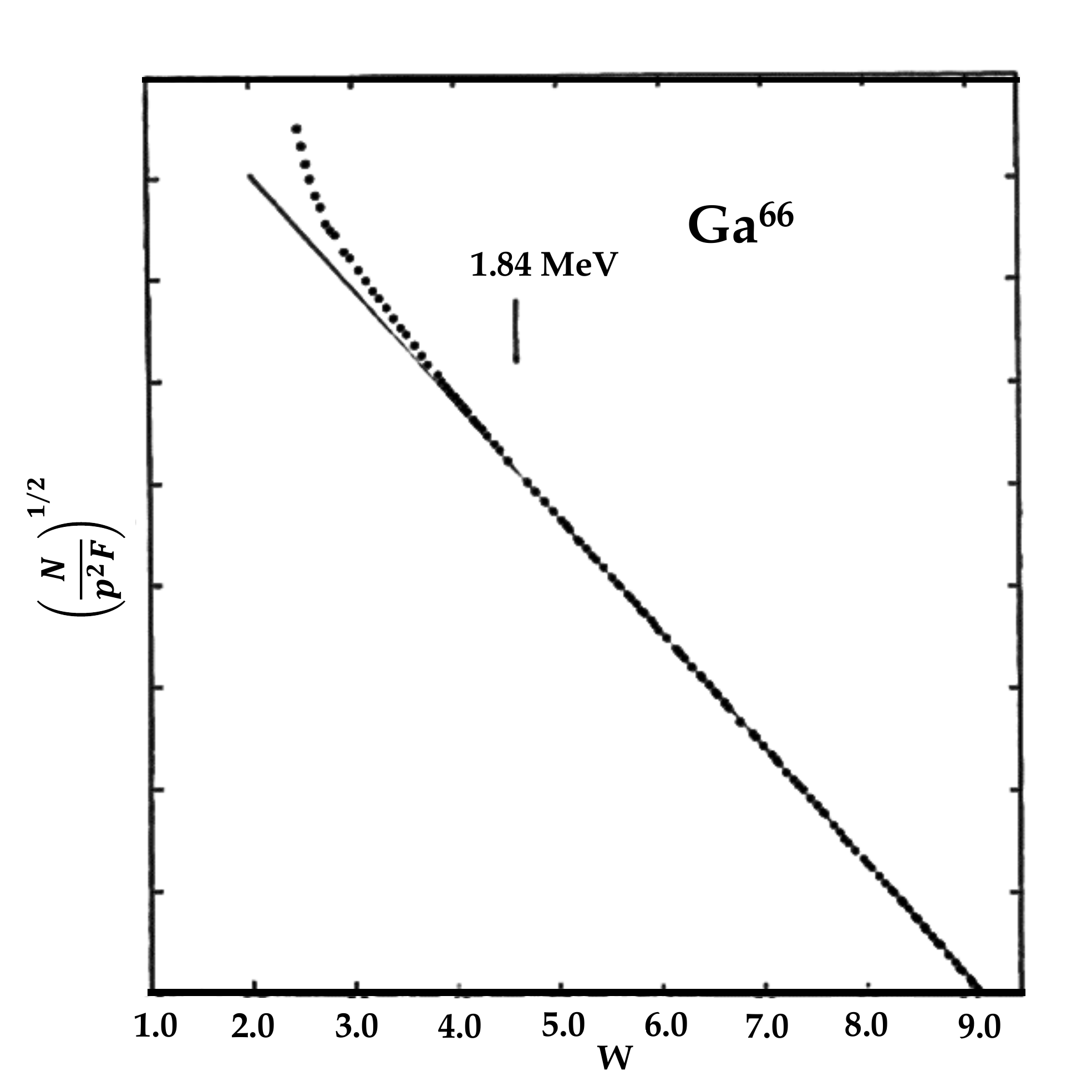}  
\caption{\small Fermie-Kurie plot of $\beta$ decay of Ga-66. Reproduced from 
Ref.~\cite{Camp} (1963).}
\label{fig:Kurieplot}
\end{center}
\end{figure}

With this new interaction, Fermi was able to predict the energy 
distribution of the beta particles.  His original graphs were 
qualitative, but soon they were found to stand up extremely well to 
experimental measurements \cite{Kurie}. A later comparison \cite{Camp} 
is shown in Figure~1 which shows a Fermi-Kurie plot in radioactive 
Gallium-66, where the horizontal axis is $W$, the energy of the beta 
particle, and the vertical axis is the quantity $(N/p^2F)^{1/2}$, where 
$N$ is the number of beta particles with momentum $p$ and $F$ is the 
so-called Fermi factor.

The (solid) straight line is the prediction of the Fermi theory and it 
may be seen that the dots, which are the experimental data, correspond 
very closely with the theoretical line. Deviations at low energies may 
be attributed to scattering of the beta particles in the solid material. 
Apart from this, the excellent straight line formed by the data points 
is a strong vindication of Fermi's theory of beta decay. Moreover, the 
fact that the line remains straight as it intersects the horizontal axis 
is a proof that the neutrino mass is zero --- at least so far as this 
experiment goes. In fact the best upper bounds on neutrino mass do come 
from beta-decay data\footnote{Though we now know that neutrinos have 
tiny but nonzero masses, no beta decay experiment has ever reached the 
sensitivity to see a deviation in the Fermi-Kurie line close to its 
intersection with the horizontal axis. The latest results are reported 
by \cite{nu-mass}.}.

It later turned out that the Fermi theory with vector-vector 
interactions cannot explain the energy plots in many cases of beta 
decay, prompting Gamow and Teller to add an axial vector-axial vector 
current to the Fermi interaction \cite{Teller}. Ever since, beta decay 
processes are classified in textbooks as as {\it Fermi transitions} and 
{\it Gamow-Teller transitions}. It was also not long before the weak 
interaction was found to be {\it universal}, i.e. a similar four-fermion 
current- current interaction could correctly predict the muon lifetime 
with the same value of $G_W$ as required for beta decay \cite{Klein}. 
The Fermi theory remains, to this day, the best explanation of the low 
energy behaviour of weakly-decaying particles.

Trouble arises, however, when we go to higher energies. The reason is 
not difficult to understand. If we go back to the QED process $e^+ e^- 
\to \mu^+\mu^-$, we can calculate the total cross-section, neglecting 
fermion masses, as \cite{Halzen}
\begin{equation}
\sigma_{\rm QED} = \frac{1}{12\pi } \left(\frac{e^2}{E_{\rm cm}}\right)^2 
\label{eqn:CSee2mumu}
\end{equation}
which arises (see above) when we have an effective coupling constant 
$e^2/E_{\rm cm}^2$. If we replace this, for a weak interaction between 
massless fermions, by $G_W$, then we should get
\begin{equation}
\sigma_{\rm weak} = \frac{e^4}{12\pi} \frac{1}{E^2_{\rm cm}} \times 
\left( \frac{G_W}{e^2/E_{\rm cm}^2} \right)^2 = \frac{G_W^2}{12\pi} 
E_{\rm cm}^2
\label{eqn:weakee2mumu}
\end{equation}
Unlike the QED case, this grows without limit as $E_{\rm cm}$ grows, and 
will sooner or later render the $S$-matrix non-unitary\footnote{In 
technical language, it will violate perturbative unitarity.}. Since this 
is not acceptable in any quantum theory, we must conclude that the Fermi 
theory breaks down at higher energies. On the other hand, making $G_W$ 
energy-dependent $\grave{\rm a}$ la QED would lead to large deviations 
in the Fermi-Kurie plot.

The above paradox is nicely solved by the so-called {\it intermediate 
vector boson} (IVB) hypothesis \cite{IVB}. For, if the massless photon 
$\gamma$ is replaced by a massive vector boson $W$ in the $e^+ e^- \to 
\mu^+\mu^-$ process, we will get a diagram of the form
\vspace*{-0.2in}
\begin{center} 
\begin{figure}[h!] 
\centerline{\includegraphics[width=0.65\textwidth]{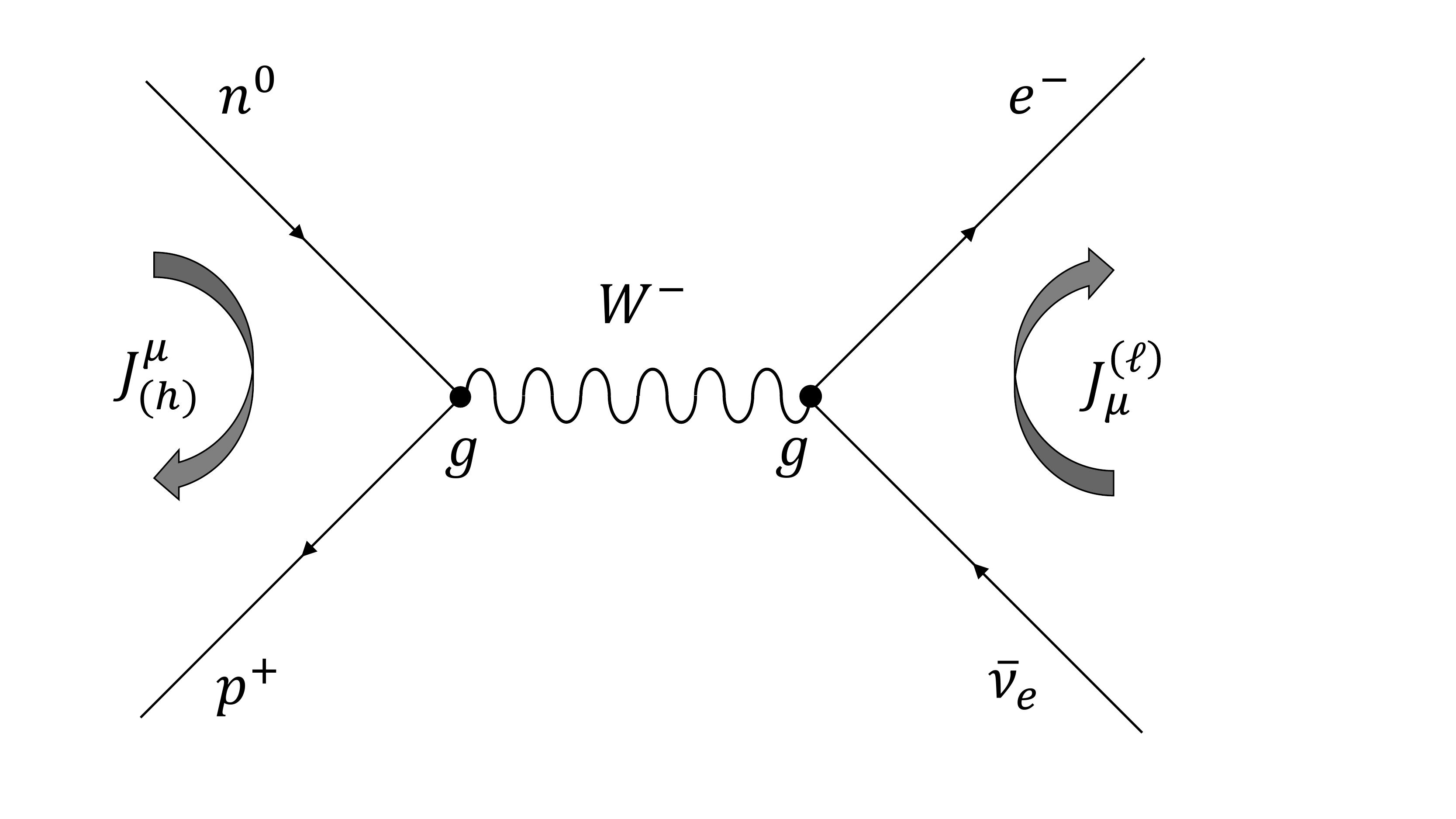} } 
\end{figure} 
\end{center} 
\vspace*{-0.8in}
where the overall factor in the cross-section will be replaced by
\begin{equation}
\left(\frac{e^2}{E_{\rm cm}^2}\right)^2  \to 
\left(\frac{g^2}{E_{\rm cm}^2 - M_W^2}\right)^2 
\label{eqn:replacement}
\end{equation} 
where $M_W$ is the mass of the $W$ boson and $g$ replaces $e$ as a weak 
coupling constant. Now, obviously, in the low- and high- energy limits, 
we will have
\begin{equation}
\left(\frac{g^2}{E_{\rm cm}^2 - M_W^2}\right)^2 \to \left\{ 
\begin{array}{l}
 \left(g^2/M_W^2\right)^2 \ \ \ {\rm when} \ E_{\rm cm} \ll M_W \\
 \left(g^2/E_{\rm cm}^2\right)^2 \ \ \ {\rm when} \ E_{\rm cm} \gg M_W
\end{array} \right.
\label{eqn:IVBlimits}
\end{equation} 
The low energy limit is a constant $G_W^2 = \left(g^2/M_W^2\right)^2$ 
--- exactly as guessed by Fermi --- and in the high energy limit we get 
the suppression factor $E_{\rm cm}^{-4}$, which is sufficient, as in 
QED, to ensure perturbative unitarity.

In the above, however, we have cheated, for the amplitude with the full 
expression for the massive vector boson propagator is
\begin{equation}
{\cal L}_{\rm eff} = e^2 \, J^\mu_{\rm (h)} \frac{g_{\mu\nu} - p_\mu p_\nu/M_W^2}{E_{\rm cm}^2 - M_W^2} J^\nu_{(\ell)}
\label{eqn:cheating}
\end{equation} 
where $p$ is the four-momentum flowing through the $W$ propagator. While 
the first (photon-like) term in the propagator does indeed lead to a 
good high-energy behaviour, the second $p$-dependent term again provides 
a factor of $E_{\rm cm}^2$ in the amplitude. Thus, with this term, the 
bad behaviour, viz. unitarity violation, comes back. For this reason, 
though the IVB hypothesis was first made in 1949 \footnote{The IVB 
hypothesis was originally proposed, almost casually, in 
Ref.~\cite{IVB}.}, it began to be taken seriously only after the 
establishment of the V-A form of the Fermi current \cite{VminusA}. This 
was because a solution for the problem had been discovered by Schwinger 
\cite{Schwinger} in 1957. His idea was to point out that the weak 
amplitude can be written
\begin{equation}
{\cal L}_{\rm eff} = e^2 \, J^\mu_{\rm (h)} \frac{g_{\mu\nu}}{E_{\rm cm}^2 - M_W^2} J^\nu_{(\ell)} 
- \frac{e^2}{M_W^2} \,                                                                                                                                       
\frac{p_\mu J^\mu_{\rm (h)} \ p_\nu J^\nu_{(\ell)}}{E_{\rm cm}^2 - M_W^2}
\label{eqn:Schwinger} 
\end{equation}   
and this will become exactly like a massive photon if
\begin{equation}
p_\mu J^\mu_{\rm (h)} =  p_\nu J^\nu_{(\mu)} = 0
\label{eqn:gaugeweak}    
\end{equation}
which, in position space, becomes
\begin{equation}
\partial_\mu J^\mu_{\rm (h)} =  \partial_\nu J^\nu_{(\ell)} = 0
\label{eqn:weakNoether}   
\end{equation}
Now these are conserved currents, and therefore, by the inverse of 
N\"other's theorem.  there must be an underlying symmetry. Taken 
together with the existence of a new vector boson, it was not difficult 
to guess that this would surely be a gauge symmetry, with the gauge 
boson as the IVB. In fact, it would have to be not one gauge boson, but 
{\it two}, for the $W^-$ in the figure must have its antiparticle, the 
$W^+$.

Schwinger actually went further --- he assumed that the photon may be a 
neutral companion to these two charged gauge bosons and hence what we 
would then have is a {\it unified} theory of electromagnetism and weak 
interactions, i.e. an {\it electroweak} theory. Such a triplet would 
arise as a vector in an isospin-like gauge theory\footnote{In group 
theoretic language: as the adjoint representation in an $SU(2)$ gauge 
theory}, just as Yang, Mills and Shaw had shown \cite{YangMills} in 
1954. It was an exciting idea, and so Schwinger made his student Glashow 
work out the phenomenological details of this first idea of electroweak 
interactions for his 1959 Ph.D. thesis \cite{GlashowA}. A similar idea 
was also proposed by Salam and Ward around the same time \cite{SalamA}. 
In fact, Salam and Ward stuck to this problem, writing a series of 
papers, of increasing sophistication, trying to set up a gauge theory of 
electroweak interactions. This eventually led to Salam being included in 
the Nobel Prize for developing the Standard Model.

The realisation soon dawned on Glashow that if we take the iso-triplet 
$(\!\!\begin{array}{ccc} W^+ & W^0 & W^- \end{array}\!\!)$ then the 
group symmetry will compel the $W^0$ to couple to a pair of neutrinos. 
In fact, the couplings of the $W$ triplet to electrons and (electron) 
neutrinos would have the form
\begin{eqnarray}
{\cal L}_W & = & \frac{g_2}{\sqrt{2}}\; \overline{e}(x) \gamma^\mu 
\nu_e(x) W_\mu^-(x) + \frac{g_2}{\sqrt{2}}\; \overline{\nu}_e(x) 
\gamma^\mu e(x) W_\mu^+(x) \nonumber \\ & & \!\!\!\! - \frac{g_2}{2}\; 
\overline{e}(x) \gamma^\mu e(x) W_\mu^0(x) + \frac{g_2}{2}\; 
\overline{\nu}_e(x) \gamma^\mu \nu_e(x) W_\mu^0(x)
\label{eqn:WcouplingsSU2} 
\end{eqnarray}
If the last term is to be absent, we must put $g_2 = 0$, which will 
remove {\it all} the interactions giving rise to beta and muon decay. 
Therefore, one cannot associate the $W^0_\mu$ field with the photon as 
he, Schwinger, as well as Salam and Ward, had done.

Sticking manfully to his task, however, Glashow in 1961 decided to 
extend his model by adding an extra $U(1)$ symmetry as a direct product 
to the $SU$ symmetry \cite{GlashowB}. Thus, there were now {\it two} 
neutral gauge bosons $W^0$ and $B^0$.
\begin{eqnarray}
{\cal L}_{W,B}^0 & = & \frac{g_2}{2}\; \overline{\nu}_e(x) \gamma^\mu \nu_e(x) W_\mu^0(x)
- \frac{g_2}{2}\; \overline{e}(x) \gamma^\mu e(x) W_\mu^0(x) \nonumber \\
& & \!\!\!\! - \frac{g_1}{2}\; \overline{\nu}_e(x) \gamma^\mu \nu_e(x) B_\mu^0(x)
- \frac{g_1}{2}\; \overline{e}(x) \gamma^\mu e(x) B_\mu^0(x)
\label{eqn:NCinSU2xU1} 
\end{eqnarray}
Though the $B_\mu$ corresponds to a $U(1)$ symmetry in 
Eqn.~(\ref{eqn:NCinSU2xU1}) it cannot be the photon, for it will only 
decouple from neutrinos if $g_1 = 0$, which is trivial. Glashow then 
made the heuristic assumption that the $W^0$ and $B^0$ are {\it mixed} 
states (for some unknown reason), i.e.
\vspace*{-0.1in} 
\begin{eqnarray}
W_\mu^0 & \defeq & Z_\mu(x) \cos\theta - A_\mu(x) \sin\theta \nonumber \\
B_\mu^0 & \defeq & Z_\mu(x) \sin\theta + A_\mu(x) \cos\theta
\label{eqn:WBmixing}  
\end{eqnarray}
\vspace*{-0.1in} 
leading to
\begin{eqnarray}
{\cal L}_{W,B}^0 & = & \frac{1}{2}\left(g_2 \sin\theta - g_1 \cos\theta 
\right) \ \overline{\nu}_e(x) \gamma^\mu \nu_e(x) \ A_\mu(x) \nonumber 
\\
& & \!\!\!\!\! -\, \frac{1}{2}\left(g_2 \sin\theta + g_1 \cos\theta 
\right) \ \overline{e}(x) \gamma^\mu e(x) \ A_\mu(x) \dots
\label{eqn:ZAinteractions}
\end{eqnarray} 
By {\it tuning} the mixing angle $\theta$ such that
\begin{equation}
\tan\theta \defeq \frac{g_1}{g_2}
\label{eqn:angletuning}
\end{equation}
Glashow was able to ensure that the $A_\mu(x)$'s coupling to a neutrino 
pair would cancel --- and thus the $A_\mu(x)$ could be identified with 
the photon, with $e = g_2 \sin\theta$ to get the right QED coupling of 
electrons to photons\footnote{It is often asked why Glashow chose the 
direct product group $SU(2)\times U(1)$ rather than the simpler $U(2)$. 
The reason is as follows. In the above, that would mean putting $g_1 = 
g_2$, i.e. $\theta = \pi/4$. We would then have $e = g_2/\sqrt{2}$ and 
hence $G_W = 2e^2/M_W^2$, leading to $M_W \approx 23$~GeV. In Glashow's 
time, this would have seemed absurdly high. Today, on the other hand -- 
with hindsight -- we know that this is actually about 3.5 times too 
small.}. It was the introduction of an additional $U(1)$ symmetry -- and 
hence an additional IVB -- which did the trick, by cancelling the $W^0$ 
coupling to neutrinos against the extra $U(1)$ contribution to the same. 
This produced a phenomenologically viable model of leptons\footnote{It 
proved ultimately good enough to win Glashow a Nobel Prize (1979).}, 
albeit with a fine-tuned parameter $\theta$, for which fine-tuning there 
was no natural explanation. We may note, however, that if we replace the 
lepton doublet $\overline{L} = (\!\!\begin{array}{cc} \overline{\nu}_e & 
\overline{e} \end{array}\!\!)$ by a nucleon doublet $ \overline{N} = 
(\!\!\begin{array}{cc} \overline{p} & \overline{n} \end{array}\!\!)$ we 
will obtain a similar-looking interaction
\begin{eqnarray}
{\cal L}_{W,B}^0 & = & \frac{1}{2}\left(g_2 \cos\theta - g_1 \sin\theta 
\right) \ \overline{p}(x) \gamma^\mu p(x) \ A_\mu(x) \nonumber \\
& & \!\!\!\! - \frac{1}{2}\left(g_2 \sin\theta + g_1 \cos\theta \right) \ 
\overline{n}(x) \gamma^\mu n(x) \ A_\mu(x) + \dots
\label{eqn:ZAinteractionswrong}
\end{eqnarray} 
To make the photon-neutron coupling vanish, however, we must have 
\begin{equation}
\tan\theta = - \frac{g_1}{g_2}
\label{eqn:angletuningwrong}
\end{equation}
which would then give the right sign for the photon-proton coupling. But 
we cannot have both the fine-tuning relations (\ref{eqn:angletuning}) 
and (\ref{eqn:angletuningwrong}) simultaneously, unless $g_1 = 0$, which 
is trivial. It may have been for this reason that Glashow's 1961 paper 
restricts itself to electroweak interactions of leptons\footnote{The 
failure of the $SI(2)\times U(1)$ theory for hadrons has its explanation 
in the simple fact that hadrons are not elementary articles. When the 
same idea is applied to an $SU(2)$ doublet of quarks, it works 
perfectly. However, when Glashow wrote his paper in 1961, composite 
nucleons would have seemed too far-fetched an idea. This came in only 
with the quark model in 1964.}.

Moreover, in Glashow's work, no explanation was attempted for the fact 
that the photon is massless and the $W^\pm$ are massive (and so, 
presumably is the $W^0$, or its mixed version). This, in fact, was the 
elephant in the room, '{\it a stumbling block we must overlook}', as 
Glashow, somewhat apologetically, put it\footnote{Italics by the present 
author.}.  However, that it did bother him is clear from the fact that 
he modestly put the words 'partial symmetry' into the title of his 
paper, that is to say, a symmetry which would not apply to the mass 
terms for the $W$-bosons\footnote{In justification, Glashow quoted two 
well-known examples -- the scale and chiral invariance of massless 
theories. These are discussed in the next section.}, but to the rest of 
the Lagrangian.
     
Glashow's {\it tour de force}, like many another famous work, 'fell 
stillborn from the press'. By the time Weinberg's classic paper on 
electroweak interactions appeared in 1967, Glashow had acquired only 4 
(non-self) citations. The tension between gauge theory and IVB masses 
proved too much for most scientists to swallow, and as a result, gauge 
theory (except QED) was cast into the shade. The 1960s were a period 
when current algebras and $S$-matrix theory dominated the world of 
quantum field theory, and, on the phenomenological side, the composite 
nature of hadrons was postulated and proved. Only a few intrepid souls 
were still bothered about electroweak unification.

Today Glashow's 1961 paper has well over 9,000 citations and is regarded 
as the seminal paper in the development of the Standard Model. What made 
the difference? It was the solution of the mass problem, the removal of 
the 'stumbling block' --- to which we must now turn.


\section{Dark Ages: The Mass Problem} 

How does one postulate both a symmetry and also introduce terms that 
break it? This is not difficult to envisage. To take a common example, 
the Earth may be approximated, for many purposes, as a perfect sphere, 
i.e. a solid having rotational symmetry (red curves in Figure~2). 
However, we are aware of irregularities (sketched highly exaggerated as 
the black curve in Figure~2), which matter only when the Earth's 
gravitational field has to be considered to a high level of accuracy. In 
technical language, the Earth's gravitational potential has a dominant 
monopole term, which represents the spherical symmetry. Then there are 
small dipole terms, quadrupole terms, etc., i.e. terms which represent 
the breaking of rotational symmetry.

\begin{figure}[h!]
\begin{center}
\includegraphics[width=0.7\textwidth]{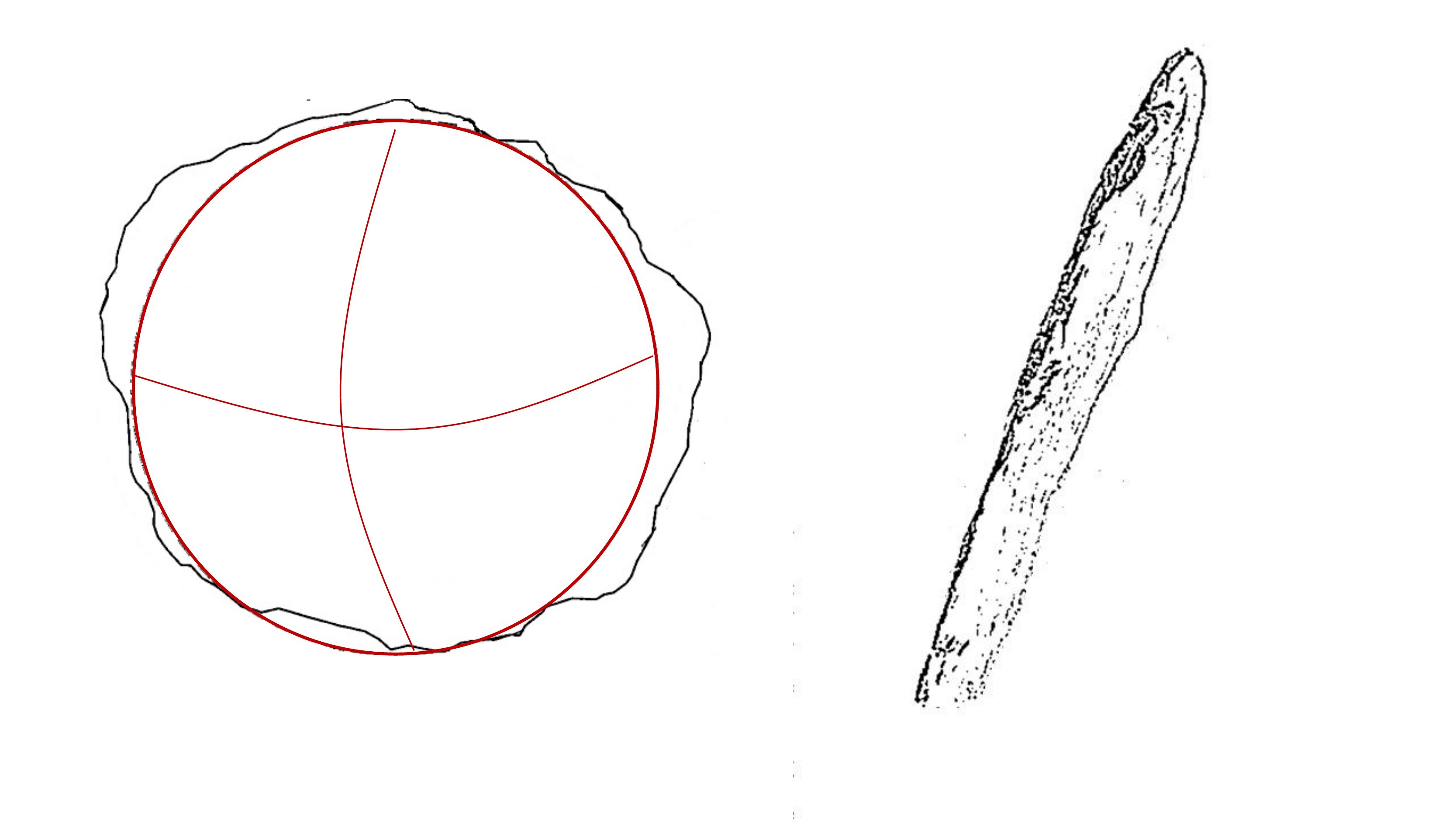}  
\caption{\small Sketch of the Earth (left) and Oumuamua (right).}
\label{fig:Symmetry}
\end{center}
\end{figure}
\vspace*{-0.2in} 
The above is an example of {\it explicit} symmetry-breaking, where the 
higher multipole terms which do not respect the spherical symmetry are 
very small. However, this cannot work in the case when the deviation 
from the symmetry is very large. For example, no one would dream of 
approximating the cigar-shaped interstellar object Oumuamua (see 
Figure~2) as a sphere with small corrections. In particle physics, large 
mass terms, such as are demanded for the $W^\pm$ bosons, would 
correspond to these large deviations from gauge theory. Where could they 
be coming from?

A similar question had plagued the theory of strong interactions as 
well. In the 1950s, the dominant theory was still the 1935 theory of 
Yukawa \cite{Yukawa}, with some group theoretic modifications over the 
years. This theory of pions and nucleons has a nucleon $N$ of mass 
around 938~MeV and a pion $\vec{\pi}$ of mass around 140~MeV, 
transforming under $SU(2)$ of isospin as a doublet and a triplet 
respectively. However, the free theory has a large global symmetry 
$SU(2)_L \times SU(2)_R$ if it is written in terms of {\it massless} 
left- and right-handed nucleons, viz.
\begin{equation}
{\cal L}_{\pi N\bar{N}} \defeq i\overline{N}_L \not{\!\partial} N_L + 
i\overline{N}_R \not{\!\partial} N_R + \partial^\mu\vec{\pi}\cdot 
\partial_\mu \vec{\pi}
\label{eqn:Yukawatheory}  
\end{equation} 
where $\overline{N} \defeq (\!\!\begin{array}{cc} \overline{p} & 
\overline{n} \end{array}\!\!)$ is the nucleon doublet and $\vec{\pi} 
\defeq (\!\!\begin{array}{ccc} \pi^+ & \pi^0 & \pi^- \end{array}\!\!)$ 
is the pion triplet. Obviously, this is invariant under global 
transformations
\begin{equation}
N_L \to N_L' = \mathbb{U}_L N_L \qquad\qquad N_R \to N_R' = \mathbb{U}_R 
N_R
\label{eqn:chiralsymmetry}
\end{equation} 
where $\mathbb{U}_L$ and $\mathbb{U}_R$ are $SU(2)$ matrices. However, a 
mass term
\begin{equation}
{\cal L}_m = m_N \left(\overline{N}_L N_R + \overline{N}_R  N_L \right)
\label{eqn:nucleonmass}
\end{equation}
immediately breaks this symmetry, unless, indeed $\mathbb{U}_L = 
\mathbb{U}_R$, i.e. the global $SU(2)_L \times SU(2)_R$ breaks down to 
the diagonal $SU(2)$ of isospin. At low energies, the nucleon mass $m_N 
\approx 938$~MeV is a large value, which cannot be attributed to 
explicit symmetry-breaking.

The solution had, in fact, already been discovered in condensed matter 
physics, where the theory of superconductivity incorporates a phenomenon 
now called {\it spontaneous symmetry-breaking}. It had appeared, in a 
non-relativistic form, in the 1950 theory of Landau and Ginzburg 
\cite{Landau}, where the photon effectively acquires a mass inside a 
superconducting medium below the critical temperature and cannot 
propagate --- making the macroscopic medium at once a perfect conductor 
and a perfect diamagnet. This phase transition occurs when a dynamical 
variable of the material, called an {\it order parameter} by Landau, is 
close to a ground state, which does not obey the symmetry of the 
Lagrangian. The idea was given a firm basis in the BCS theory of 
superconductivity \cite{BCS}, where the hitherto-esoteric order 
parameter assumes a concrete form as the density of a Bose-Einstein 
condensate of paired electrons (Cooper pairs), forming quasi-bosons. As 
one would expect, the BCS ground state breaks the gauge symmetry of 
electromagnetism, which would otherwise preclude a photon mass.

Could there be a similar phenomenon happening in the pion-nucleon 
regime? This was the question asked by Nambu and Jona-Lasinio in 1960 
\cite{Nambu}, and they managed to construct an elaborate theory based 
closely on the BCS theory, with a nucleonic condensate $\langle 
N\bar{N}\rangle$ which would break the chiral $SU(2)_L \times SU(2)_R$ 
down to $SU(2)$ of isospin. Once the symmetry is broken spontaneously, 
there is no restriction on the mass of the nucleon, for it is related to 
the energy gap between the ground state and the first excited state of 
the condensate. Some mechanism was postulated to create the $\langle 
N\bar{N}\rangle$ bound state --- this would be the analogue of the 
Fr\"ohlich effective interaction which creates the Cooper pairs. 
Inspired by the success of Nambu and Jona-Lasinio, Gell-Mann and L\'evy 
\cite{Levy} then created a simpler model where the condensate $\langle 
N\bar{N} \rangle$ is replaced by a scalar field $\sigma$. This came to 
be known as the (linear) sigma model.  In this model, one creates an 
$SU(2)$ doublet
\begin{equation}
\Phi(x) \defeq \mathbb{I} \; \sigma(x) + i \; \vec{\mathbb{T}} \cdot \vec{\pi}(x)
\label{eqn:sigmacombo} 
\end{equation}    
and adds to the isospin Hamiltonian a potential energy term of the form
\begin{equation}
V(\Phi) \defeq - \frac{\mu^2}{2} \Phi^\dagger \Phi + \frac{\lambda}{4!} (\Phi^\dagger \Phi)^4
\label{eqn:Mexicanhat} 
\end{equation}
which is closely analogous to the Ginzburg-Landau Hamiltonian if $\mu$ 
and $\lambda$ are real and positive. This potential breaks the $SU(2)$ 
symmetry spontaneously and permits the nucleons to have large 
masses\footnote{Some of the details are indicated, in the electroweak 
context, in the next section.}. At the same time, the pions remain 
necessarily massless, following a 1962 theorem proved by Goldstone, 
Salam and Weinberg\cite{Gold62}. As a matter of fact, Goldstone had 
discovered \cite{Gold61} in 1961 that theories of the sigma model type 
always contain massless bosons, but a year later, he teamed up with 
Salam and Weinberg to prove the {\it Goldstone theorem}, which states 
\cite{Gold62} that whenever a continuous global (or local) symmetry is 
spontaneously broken, there will be massless bosons. Today they are 
called {\it Goldstone bosons} and are ubiquitous in quantum field 
theories. The pions in the sigma model are Goldstone bosons, and 
necessarly massless. However, it is possible to introduce pion mass 
terms as {\it explicit} symmetry-breaking terms, which necessarily makes 
the pion masses small. Here, it seems, lay the explanation of why the 
pions are so much lighter than the nucleons.

Nucleons and pions are not, however, the only hadrons. Throughout the 
1950s and early 1960s, new hadrons with exotic properties were being 
discovered \cite{Close}, and these were then postulated by 
Gell-Mann\cite{quarks} and Zweig\cite{zweig} in 1964 to be composites of 
a set of three basic quarks. The analysis of deep-inelastic scattering 
data in 1969 established the existence of these quarks\cite{partons}, 
and a gauge interaction between them was established by the early 1970s 
\cite{QCD}. The sigma model, therefore, never really achieved the status 
of a fundamental theory; however, it proved to be the first and paradigm 
case for a whole class of {\it effective} field theories. But it was in 
the theory of electroweak interactions that the ideas introduced in the 
sigma model secured their biggest triumph.

\section{Early Renaissance: The Higgs Mechanism} 

It may have seemed natural, in the early 1960s, to create a model of 
electroweak interactions on the analogy of the sigma model. This would 
certainly have been an excellent means of generating masses for the IVB 
particles, and would nicely round off Glashow's unified electroweak 
theory.  However, the concept foundered on the rock of the Goldstone 
theorem. A sigma model analogue would indeed make the vector bosons 
massive (except for the photon), but in their place we would have three 
massless scalars, with interactions of the same strength. Such bosons 
would have surely shown up long before in cosmic rays, or nuclear 
reactions, for the weak interactions are ubiquitous.  There were not 
even light bosons analogous to the pions, to which one could attach an 
explicit symmetry-breaking. For this reason, no one interested in 
particle physics hastened to create a sigma model for electroweak 
interactions.

Once again, the clue came from condensed matter physics. In 1963, 
Anderson published a study \cite{Anderson} of superconductivity, in 
which he suggested that the Goldstone degree of freedom arising from 
spontaneous symmetry-breakdown in a metallic superconductor actually 
reappears as a longitudinal mode of the photon\footnote{which 
corresponds to longitudinal oscillations in the free electron gas, 
quantised as 'plasmons'.}, rather than as an independent massless boson. 
The two massless bosons 'cancel', in Anderson's language, leaving a 
massive (vector) boson. This is, in effect, what is known today as the 
{\it Higgs mechanism}. There was really no reason for Anderson not to go 
on to rewrite his non-relativistic model in a relativistic avatar, since 
he did mention Yang-Mills theory in passing. However, he drew back from 
this step, contenting himself with the statement that 'it is not at all 
clear' that a similar mechanism would work in Sakurai's theory 
\cite{Sakurai} of the strong interactions.

In this 1960 paper, Sakurai had tried to write down a Yang-Mills theory 
for the $SU(2)_I \times U(1)_Y$ flavour symmetry of hadrons 
\cite{Sakurai}. This had not been taken seriously for the same reason as 
Glashow's electroweak paper, because the gauge bosons were necessarily 
massless, leading to a long-range interaction; and even if the gauge 
symmetry could be spontaneously broken, there would be massless 
strongly-interacting Goldstone scalars which are known not to exist. 
Interestingly, the group symmetry is the same as that discovered by 
Glashow, and the mathematical structure of Sakurai's theory is almost 
identical to that of Weinberg's electroweak model. Of course, Sakurai's 
model did not survive the discovery that flavour symmetries are global 
in nature --- and inexact to boot. But it is ironic that what was 
dismissed as 'not at all clear' in Anderson's 1963 paper, was precisely 
what became crystal clear in Weinberg's 1967 paper\footnote{Weinberg 
later wrote, in a characteristically insightful article 
\cite{WeinAnders}: "In fact, Anderson was right: the reason for the 
exception noted by Higgs {\it et al.} is that it is not possible to 
quantize a theory with a local symmetry in a way that preserves both 
manifest Lorentz invariance and the usual rules of quantum mechanics, 
including the requirement that probabilities be positive. In fact, there 
are two ways to quantize theories with local symmetries: one way that 
preserves positive probabilities but loses manifest Lorentz invariance, 
and another that preserves manifest Lorentz invariance but seems to lose 
positive probabilities, so in fact these theories actually do respect 
both Lorentz invariance and positive probabilities; they just don't 
respect our theorem." Here, by 'our' theorem, he meant the Goldstone 
theorem\cite{Gold62}.}.

It did not take long for the relativistic generalisation to come. Three 
groups arrived at practically the same formulation independently during 
1964. In those pre-arXiv days, neither of them knew of the others' work 
till it was actually published. The first off the mark were Englert and 
Brout \cite{Englert} at Brussels, followed by Higgs \cite{Higgs} at 
Edinburgh, and a little later, by Guralnik, Hagen and Kibble 
\cite{Guralnik} at London. Thus, the 
Anderson-Brout-Englert-Guralnik-Hagen-Higgs-Kibble mechanism -- or Higgs 
mechanism for short if we do not wish to mention all the 'magnificent 
seven' -- came into being. Its application to the Abelian gauge theory 
of Section~1 is outlined below in terms of three 'miracles'.

We start with the scalar gauge theory of Eqn.~(\ref{eqn:sEDaction}) 
keeping the scalar massless and add to it two self-interaction terms 
similar to those of the sigma model, i.e.  Eqn.~(\ref{eqn:Mexicanhat}), 
thereby getting
\begin{eqnarray}
S & \defeq & \int d^4x \left[ (D^\mu \varphi)^*(x) \ D_\mu \varphi(x) 
-\frac{1}{4}\;F^{\mu\nu}(x) \, F_{\mu\nu}(x) - V(\varphi) \right] 
\nonumber \\ V(\varphi) & = & -\frac{\mu^2}{2} \; \varphi^*(x) 
\varphi(x) + \frac{\lambda}{4!} \left\{\varphi^*(x) \varphi(x)\right\}^2
\label{eqn:ssbU1scalar}
\end{eqnarray}  
where, as in the sigma model, $\mu$ and $\lambda$ are real and positive 
constants. It should be noted that the $\mu$ term has the wrong sign for 
a mass term\footnote{If we try to identify it as a mass term, then 
$\varphi$ would be a {\it tachyon} and the theory would therefore be 
unphysical.} and should be regarded as an interaction term.

If we plot the potential energy term $V(\varphi)$ as a function of 
$\varphi$, we will obtain the so-called 'Mexican hat' potential shown in 
Figure~3. Since the potential depends only on $|\varphi|^2$, it will 
have rotational symmetry in the complex $\varphi$-plane. Close to the 
origin, i.e. at small values of $|\varphi|^2$, the quadratic term will 
dominate, producing approximately an inverted paraboloid of revolution, 
but at higher values of $|\varphi|$ the quartic term will begin to 
dominate, causing a turnaround and making the potential positive.  These 
contrary terms will therefore create a minimum for some value $|\varphi| 
= v$, which is easily calculated as
\begin{equation}
\frac{v}{\sqrt{2}} = \sqrt{\frac{6\mu^2}{\lambda}}
\label{eqn:vev}
\end{equation} 

\begin{figure}[h!]
\begin{center}
\includegraphics[width=0.55\textwidth]{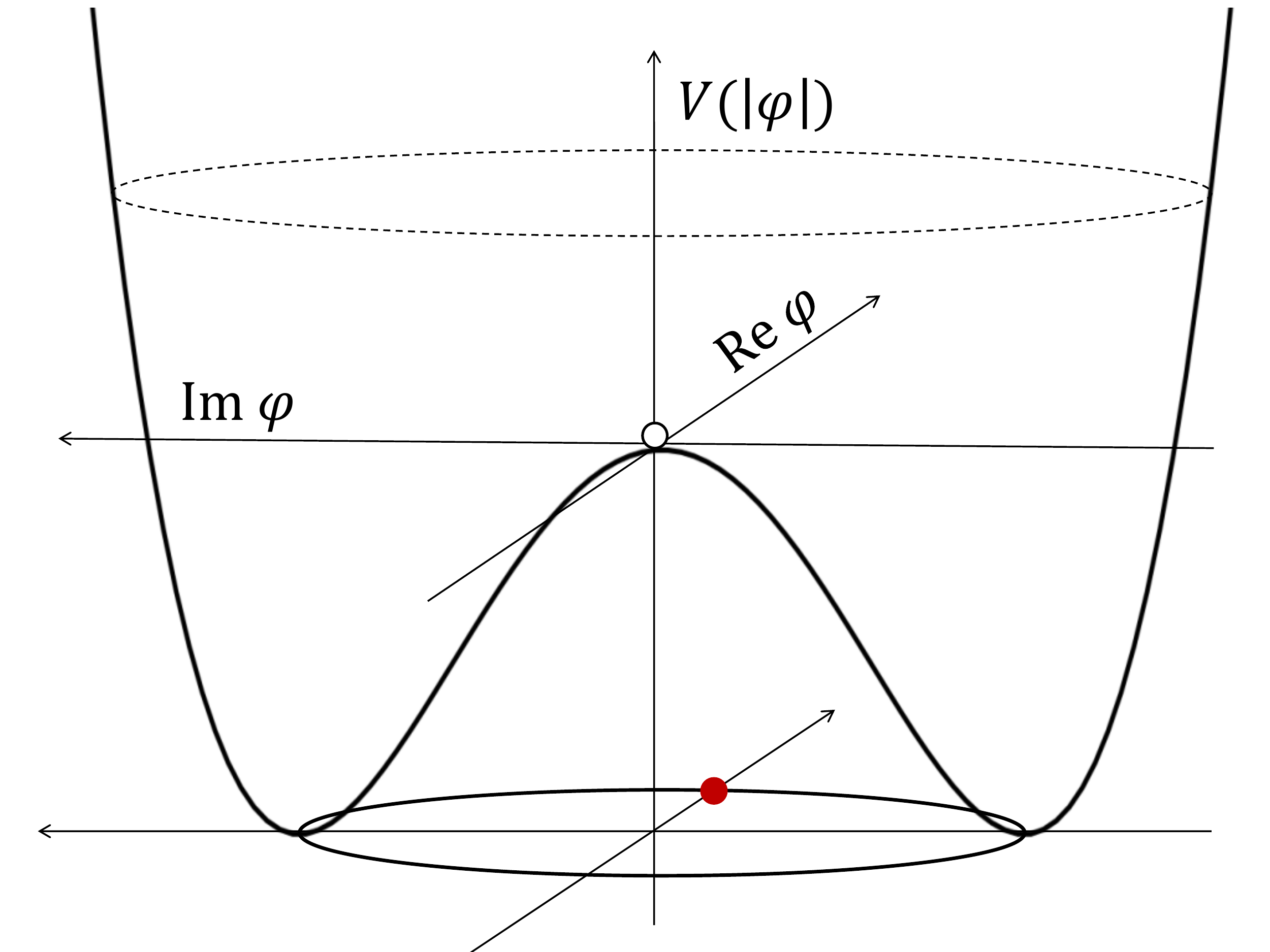}  
\caption{\small Spontaneous symmetry-breaking potential in a $U(1)$ gauge theory..}
\label{fig:Mexicanhat}
\end{center}
\end{figure}
\vspace*{-0.2in}

However, there is no restriction on the {\it phase} of $\varphi$, which 
indicates that in the complex $\varphi$-plane there is a ring of 
possible minima of the potential energy with radius $|\varphi| = v$. In 
the quantum theory, these will correspond to degenerate vacua 
$\langle\varphi \rangle = v/\sqrt{2}$. Of these, only one can be the 
physical vacuum, and this will be a random choice denoted in Figure~3 by 
a red dot. The unstable point $\varphi = 0$ is denoted by a white dot. 
We now notice that the phase invariance --- or $U(1)$ gauge invariance 
--- is spontaneously broken by the choice of a single vacuum state along 
the ring.

Recognising that at energies (temperatures) far above $V(v/\sqrt{2})$ 
the $U(1)$ symmetry will be there, but it will only be at low energies 
(temperatures) close to $V(v/\sqrt{2})$, that the symmetry is broken, we 
realise that this is really a {\it phase transition} from a symmetric 
phase to a broken-symmetric phase. As the system in the symmetric phase 
cools down, the false vacuum at $\varphi = 0$ will become manifestly 
unstable, and thus any quantum fluctuation will cause the system to 
'roll down' to ons of the degenerate true vacua.

To write down a quantisable field theory, we must now expand $\varphi$ 
around the vacuum state. Before we do so, however, it is convenient to 
choose the axes in the complex $\varphi$-plane such that the vacuum 
point lies along the real axis --- this can be done without loss of 
generality and at the same time enables us to eliminate an extraneous 
phase parameter, which would otherwise have to be carried through the 
calculation. We now write
\begin{equation}
\varphi \defeq \frac{v}{\sqrt{2}} + \omega(x)
\label{eqn:shiftscalar}
\end{equation}            
where $\omega(x)$ is the scalar degree of freedom which is amenable for 
a quantum field theory.  In terms of this, the action becomes
\begin{equation}
S = \int d^4x \left[ (\partial^\mu + i e A^\mu)(\frac{v}{\sqrt{2}} + 
\omega^*) \ (\partial_\mu - ieA_ \mu) (\frac{v}{\sqrt{2}} + \omega) - 
\frac{1}{4}\;F^{\mu\nu} \, F_{\mu\nu} - V(\omega) \right]
\label{eqn:ssbU1shifted}
\end{equation}
with a potential energy term
\begin{equation}
V(\omega) = \mu^2 \left({\rm Re}\;\omega\right)^2 + \mu^2 \; \omega^* 
\omega \; {\rm Re}\;\omega + \frac{\lambda}{4!}\left( \omega^* \omega 
\right)^2
\label{eqn:potU1shifted}
\end{equation}
where we need to substitute $\mu^2 = \lambda v^2/12$ and a constant term 
has been dropped. The kinetic term in Eqn.~(\ref{eqn:ssbU1shifted}), 
when expanded, contains a term
\begin{equation}
{\cal L}_m = + \frac{1}{2} \; e^2 v^2 \; A^\mu A_\mu 
\label{eqn:photonmass}
\end{equation}
The above is a mass term for the photon, and therefore, we identify 
$M_\gamma = ev$. This is the {\it first miracle} of the sigma model, or 
more generally of the spontaneously-broken theory, that the choice of 
one out of multiple possible vacua generates a gauge boson mass. If we 
consider the quadratic term in Eqn.~(\ref{eqn:potU1shifted}), we will 
note that it is a mass term with the correct sign, i.e., the real scalar 
field ${\rm Re}\;\omega$ has a mass $\sqrt{2}\mu$. This is the {\it 
second miracle} of Higgs \cite{Higgs}, that the tachyonic scalar theory, 
once the proper vacuum is chosen, becomes a real scalar theory with a 
real mass. This scalar is known as the {\it Higgs boson}\footnote{This 
name was first used, with justification, by B.W.~Lee \cite{Sample}, 
since it was Higgs who had pointed out the presence of 'incomplete 
multiplets' i.e. the real part without the imaginary part of the scalar 
field.}.

The action in Eqn.~(\ref{eqn:ssbU1shifted}) still has a problem, since 
it contains the complex scalar field $\omega = {\rm Re}\;\omega + i{\rm 
Re}\;\omega$. but only the real field ${\rm Re}\;\omega$ has a finite 
mass term. The other real field ${\rm Im}\;\omega$ remains massless and 
indeed, it is the Goldstone boson of the theory. This is where the {\it 
third miracle} comes in, i.e.  the one devised by Anderson and the seven 
authors of 1964. The trick is go back and parametrise the complex scalar 
field $\varphi$ as
\begin{equation}
\varphi(x) \defeq \eta(x)\;e^{i\theta(x)}
\label{eqn:polar}
\end{equation} 
At this stage, the gauge symmetry is not broken, and hence, with 
hindsight, we can perform a $U(1)$ gauge transformation as in 
Eqn.~(\ref{eqn:U1gauge}) to obtain
\begin{equation}
\varphi(x) \to \varphi'(x) = \eta(x)\;e^{ie[\theta(x)+\alpha(x)]}
\label{eqn:polarU1gt}
\end{equation}
Since we are free to choose the gauge function $\alpha(x)$, we simply 
choose it such that
\begin{equation}
\alpha(x) \defeq - \theta(x)
\label{eqn:unitary} 
\end{equation}
in which case, 
\begin{equation}
\varphi(x) \to \varphi'(x) = \eta(x)
\label{eqn:unitarygauge}
\end{equation}
The physical system remains the same at this stage, because of the gauge 
invariance. The action in this so-called {\it unitary} gauge now becomes
\begin{eqnarray}
S & = & \int d^4x \left[ (D^\mu \eta)^*(x) \ D_\mu \eta(x) 
-\frac{1}{4}\;F^{\mu\nu}(x) \, F_{\mu\nu}(x) - V(\eta) \right] \nonumber 
\\ V(\eta) & = & -\frac{\mu^2}{2} \; \eta^2(x) + \frac{\lambda}{4!} 
\eta^4(x)
\label{eqn:ssbU1unitary}
\end{eqnarray}
where only the real scalar field $\eta(x)$ appears. This is still 
tachyonic, but we now note that the minimum of the potential lies at 
$\eta$ given by $\pm v = \pm \sqrt{12\mu^2/\lambda}$. Choosing the 
positive value, and making the shift $\eta(x) = v/\sqrt{2} + h(x)$, now 
produces both the desirable features of the sigma model, viz., a mass 
term for the photon $A_\mu(x)$ and a mass term of the proper sign for 
the Higgs boson $h(x)$. Obviously, the degree of freedom denoted by the 
field $\theta$ has vanished from the action in 
Eqn.~(\ref{eqn:ssbU1unitary}). It is still there, however --- as 
Anderson had realised --- disguised as the longitudinal polarisation 
mode of the photon.

For the Higgs mechanism to work, then, we require a tachyonic scalar 
field theory with a Mexican hat potential, where we make two successive 
reparametrisations of the complex scalar field --- the first is a gauge 
choice in the unbroken theory, and then, when the symmetry breaks, a 
constant shift in the remaining real scalar theory. This produces the 
three desirable features ('miracles') mentioned above. viz.,
\vspace*{-0.2in}
\begin{itemize}
\item The photon acquires a mass. 
\item The real scalar mode acquires a mass. 
\item The massless scalar mode disappears from the theory. 
\end{itemize}   
\vspace*{-0.2in}
Even today, more than a quarter of a century later, the deep insight 
into Nature provided by spontaneous symmetry-breaking and the Higgs 
mechanism is quite awe-inspiring. It was soon to provide the key to 
developing a theory of the electroweak interactions. We must recognise, 
however, that the choice of parameters in the potential is quite 
arbitrary, and is in no way demanded by the gauge symmetry. The 
spontaneously-broken gauge theory, therefore, is a mixed marriage of a 
pure gauge theory and a specific potential choice, which is driven 
purely by phenomenological demands.

\section{High Renaissance: the Glashow-Salam-Weinberg Model} 

Given the breakthrough of 1964, it took surprisingly long for someone to 
come up with a generalisation of the Abelian $U(1)$ Higgs model to the 
non-Abelian case of an $SU(2)\times U(1)$ local gauge theory {\it \`a 
la} Glashow or Sakurai. However, it was elegantly done in a 1967 paper 
by T.W.B.~Kibble, one, indeed, of the 'magnificent seven' who had 
proposed the Abelian Higgs mechanism \cite{Kibble}. The essence of this 
work is very close to the sigma model\footnote{Actually, Kibble's paper 
first discusses a general non-Abelian gauge group before taking up the 
special case of the Glashow-Sakurai group $SU(2)\times U(1)$.}. Instead 
of one complex scalar $\varphi(x)$, the theory has four complex scalars 
$\varphi_0(x)$ and $\vec{\varphi}(x) = \left\{\varphi_i(x)|i=1,2,3 
\right\}$ forming a doublet of $SU(2)$
\begin{equation}
\Phi(x) \defeq \mathbb{I} \; \varphi_0(x) + i \; \vec{\mathbb{T}} \cdot \vec{\varphi}(x)
\label{eqn:Kibble} 
\end{equation}    
and the potential energy terms $V(\Phi)$ will be identical to those in 
Eqn.~(\ref{eqn:Mexicanhat}).  We start, as before, with the action
\begin{equation}
S \defeq \int d^4x \ \left[ \left(\mathbb{D}^\mu \Phi\right)^\dagger 
\mathbb{D}_\mu \Phi - \frac{1}{8} \; {\rm Tr}\left[ \mathbb{F}^{\mu\nu} 
\, \mathbb{F}_{\mu\nu}\right] - \frac{1}{4}\,B^{\mu\nu} B_{\mu\nu} - 
V(\Phi) \right]
\label{eqn:SU2xU1action} 
\end{equation}
where $\mathbb{W}_\mu \defeq \vec{\mathbb{T}}\cdot \vec{W}_\mu = 
\frac{1}{2}\vec{\sigma}\cdot \vec{W}_\mu$, and
\vspace*{-0.1in} 
\begin{eqnarray}
\mathbb{D}_\mu & \defeq & \mathbb{I}\,\partial_\mu -ig\,\mathbb{W}_\mu - 
\frac{ig'}{2}\,\mathbb{I}\,B_\mu
\nonumber \\
\mathbb{F}_{\mu\nu} & \defeq & \partial_\mu \mathbb{W}_\nu - 
\partial_\nu \mathbb{W}_\mu -ig \left[ \mathbb{W}_\mu, \mathbb{W}_\nu 
\right] \nonumber \\
B_{\mu\nu} & = & \partial_\mu B_\nu - \partial_\nu B_\mu
\label{eqn:nonabelian} 
\end{eqnarray}
Following the Abelian case, we can make a polar parametrisation, writing
\begin{equation}
\Phi(x) \defeq \left(\!\begin{array}{c} 0 \\ \eta(x) \end{array} \right) 
\; e^{i\,\vec{\mathbb{T}}\,\cdot\,\vec{G(x)}}
\label{eqn:polardoublet} 
\end{equation}
and then apply on it a $SU(2)$ local gauge transformation to get rid of 
the exponentiated (Goldstone) fields, leaving
\begin{equation}
\Phi(x) = \left(\!\begin{array}{c} 0 \\ \eta(x) \end{array} \right) 
\label{eqn:polarunitary} 
\end{equation}
The potential term now becomes identical to that in 
Eqn.~(\ref{eqn:ssbU1unitary}) and a shift $\eta(x) = v/\sqrt{2} + H(x)$ 
will produce gauge boson mass terms of the form
\begin{eqnarray}
S_m & = & \int d^4x \
\left(\! \begin{array}{cc} 0 & v/\sqrt{2} \end{array} \!\right) 
\left(\!g\,\mathbb{W}^\mu + \frac{g'}{2} \mathbb{I}\,B^\mu \!\right)
\left(\!g\,\mathbb{W}_\mu - \frac{g'}{2} \mathbb{I}\,B_\mu \!\right) 
\left(\!\! \begin{array}{c} 0 \\ v/\sqrt{2} \end{array} \!\!\right)  \\
& = & \int d^4x \   \left[ \ \frac{g^2v^2}{8}
\left( W_1^\mu - iW_2^\mu \right) \left( W_{1\mu} + iW_{2\mu} \right)  
 \right. \nonumber \\  
 && \left. \hspace*{0.6in} + \frac{v^2}{8} \left(\! \begin{array}{cc} 
 W_3^\mu & B^\mu \end{array} \! \right)
\left(\! \begin{array}{cc} g^2 & -gg' \\ 
                           -gg' & g'^2 \end{array} \!\right)    
\left(\! \begin{array}{c} W_{3\mu} \\ B_\mu \end{array} \!\right)                  
\right]   
\label{eqn:WBmasses}
\end{eqnarray}
Kibble's paper stopped at this point, simply stating that the gauge 
bosons would have masses\footnote{In fact, Kibble stated that {\it all} 
the gauge bosons have mass.}. However, the thread was brilliantly taken 
up by Weinberg \cite{Weinberg} later the same year. It was Weinberg who 
pointed out that Eqn.~(\ref{eqn:WBmasses}) means that the $\vec{W}_\mu$ 
and $B_\mu$ fields are unphysical because of the presence of mixing 
terms. However, this can be dealt with quite easily. From the first two 
$W$'s we can define conjugate complex fields
\begin{equation}
W_\mu^\pm \defeq \frac{1}{\sqrt{2}} (W_{1\mu} \mp i W_{2\mu})
\label{eqn:Wbosons}
\end{equation}
and for the $W_{3\mu}$ and the $B_\mu$, the mass matrix is diagonalised 
by the orthonormal physical states
\begin{equation}
Z_\mu \defeq \frac{gW_{3\mu} + g'B_\mu}{\sqrt{g^2 + g'^2}} 
\qquad\qquad\qquad
A_\mu \defeq \frac{g'W_{3\mu} - gB_\mu}{\sqrt{g^2 + g'^2}}
\label{eqn:Weinbergmixing}
\end{equation} 
corresponding to eigenvalues $g^2 + g'^2$ and $0$ respectively. 
Eqn.~(\ref{eqn:WBmasses}) then reduces to
\begin{equation}
S = \int d^4x \ \left[ \frac{g^2v^2}{4} W^{+\mu} W^-_\mu + \frac{v^2}{8} 
(g^2 + g'^2) Z^\mu Z_\mu \right]
\label{eqn:WZmasses}
\end{equation}
We thus have charged vector bosons $W^\pm_\mu$ with mass $M_W = 
\frac{1}{2}gv$, a neutral vector boson $Z_\mu$ with mass $M_Z = 
\frac{1}{2}v\sqrt{g^2 + g'^2}$ and a massless vector boson $A_\mu$.  
This certainly fits the pattern of massive weak bosons and a photon -- 
if the $A_\mu$ can be identified with the photon. For this, it must 
couple to fermion exactly like the photon. It was here that Weinberg 
pointed out that we can recast Eqn.~(\ref{eqn:Weinbergmixing}) in the 
exact form of Eqn.~(\ref{eqn:WBmixing}) in Glashow's model, provided we 
set
\begin{equation}
\tan\theta \defeq \frac{g'}{g}
\label{eqn:WeinbergAngle}
\end{equation} 
Now this, if we write $g = g_2$ and $g' = g_1$, is precisely what 
Glashow had obtained --- see Eqn.~(\ref{eqn:angletuning}) --- by fine 
tuning the mixing so that the photon coupling to neutrinos vanishes. We 
thus have a beautiful understanding of where this particular relation 
comes from -- it arises from the pattern of spontaneous 
symmetry-breaking and is not a fine tuning at all!  Naturally, this 
relation also means that we can identify the $A_\mu$ field as Glashow's 
photon.  It is for this insight that the mixing angle $\theta$ in 
Eqn.~(\ref{eqn:angletuning}) has, ever since, been known as the {\it 
Weinberg angle} and denoted $\theta_W$. In terms of this we have the 
useful relation
\begin{equation}
M_Z = \frac{M_W}{\cos\theta_W}
\label{eqn:rhoSM}
\end{equation}
Profound as it may be, this $SU(2)\times U(1)$ electroweak model is 
predicated on the existence of a bunch of scalar fields, of which --- 
before 2012 --- there was no evidence. What made Weinberg's model a 
runaway success was its ability to explain fermion masses and couplings. 
Here again, the spadework had been done by Salam and Ward \cite{SalamB}, 
as early as 1964. They did not, however, incorporate any solution for 
the gauge boson masses, contenting themselves with the comment that the 
weak bosons must be 'outrageously' heavy.

Weak interaction theory, in fact, has a serious difficulty, not just 
with gauge boson masses, but also with fermion masses. The reason lies 
in their parity-violating nature. This had been proposed as early as 
1956 by Yang and Lee\cite{YangLee}, and was soon proved in beautiful 
experiments by Wu {\it et al}\cite{MadameWu} and by Goldhaber {\it et 
al}\cite{Goldhaber}. This was followed by the discovery \cite{VminusA} 
that the weak interaction currents are of the $V$-$A$ form, which, in 
the IVB theory would mean that the couplings of the $W$ boson are of the 
form
\begin{eqnarray}
S_W & \sim & \int d^4x \ \overline{\psi}(x) \gamma^\mu \left(\! 1 - 
\gamma_5 \!\right) \psi(x) \ W_\mu(x) + {\rm H.c.} \nonumber \\ & \sim & 
\int d^4x \ \overline{\psi_L}(x) \gamma^\mu \psi_L(x) \ W_\mu(x) + {\rm 
H.c.}
\label{eqn:VmAform}
\end{eqnarray}     
where we define
\begin{equation}
\psi_L(x) \defeq \frac{1}{2} \left(\! 1 - \gamma_5 \!\right)\, \psi(x) 
\qquad\qquad \psi_R(x) \defeq \frac{1}{2} \left(\! 1 + \gamma_5 
\!\right)\, \psi(x)
\label{eqn:chirality}  
\end{equation}
However, the $W$ boson is charged and it couples to a $SU(2)$ doublet of 
scalars. Accordingly, we must create an $SU(2)$ doublet of fermions. For 
the moment, we just choose the electron and its neutrino, {\it \`a la} 
Weinberg, and create a doublet with the left-handed components
\begin{equation}
L_L(x) \defeq \left(\!\!\!\begin{array}{c} \nu_{eL}(x) \\ e_L(x) \end{array}\!\!\!\right)
\qquad\qquad
\overline{L_L}(x) = \left(\!\!\!\begin{array}{cc} \overline{\nu_{eL}}(x) & \overline{e_L}(x) 
\end{array}\!\!\!\right)
\label{eqn:leptondoublet}
\end{equation}   
with a gauge-kinetic term
\begin{equation}
S_L = \int d^4x \ i \,\overline{L_L}(x) \, \gamma^\mu \, 
\mathbb{D}^{(L)}_\mu \, L_L(x)
\label{eqn:Wleptondoublet}
\end{equation}
where 
\begin{equation}
\mathbb{D}^{(L)}_\mu \defeq \mathbb{I}\,\partial_\mu + ig\,\mathbb{W}_\mu 
- \frac{ig'}{2}Y_L\,\mathbb{I}\,B_\mu
\label{eqn:covariantleftlepton} 
\end{equation}
to be added to the rest of the electroweak action. Since the 
right-handed components do not interact with the $W$ boson (but the 
$e_R$ most certainly interacts with the photon), we make them $SU(2)$ 
{\it singlets}, with gauge-kinetic terms
\begin{equation}
S_R = \int d^4x \  i \, \overline{e_{R}}(x) \, \gamma^\mu \, D^{(e)}_\mu e_R(x) 
\label{eqn:leptonright}
\end{equation}
where
\begin{equation}
D^{(e)}_\mu \defeq \partial_\mu - \frac{ig'}{2}Y_e\,B_\mu
\label{eqn:covarantrightlepton}
\end{equation}
also to be added to the rest of the electroweak action. In view of the 
experiments of Goldhaber {\it et al} \cite{Goldhaber}, no right-handed 
neutrino field is included in Eqn.~(\ref{eqn:covarantrightlepton}).

It may be noted that the gauge-covariant derivatives of 
Eqns.~(\ref{eqn:covariantleftlepton}) and 
(\ref{eqn:covarantrightlepton}) differ from those in 
Eqns.~(\ref{eqn:nonabelian}) and (\ref{eqn:scalarcovariant}) 
respectively because of the presence of the {\it weak hypercharges} 
$Y_L$ and $Y_e$. To find these quantum numbers of the chiral fermions, 
we need to expand the interaction terms in 
Eqns.~(\ref{eqn:Wleptondoublet}) and (\ref{eqn:leptonright}), thereby 
obtaining
\begin{eqnarray}
S_{\rm int}^{\rm cc} = \int d^4x \ \frac{g}{2\sqrt{2}} \; \left[ \overline{\nu}_e(x) \gamma^\mu 
\left(\!1 - \gamma_5 \!\right) e(x) W_\mu^+(x) + {\rm H.c.} \right]
\label{eqn:CCinteractions}
\end{eqnarray}
for the {\it charged current interactions} i.e. terms with currents\footnote{These currents will themselves transform like charged objects under the $U(1)$ of QED.} that couple to the charged $W^\pm$. We can now write the Fermi interaction of Eqn.~(\ref{eqn:currentcurrent}) (with a $V$-$A$ modification) as a second order effective interaction of this theory in the low-energy limit to get the identification
\begin{equation}
\frac{G_F}{\sqrt{2}} \defeq \frac{g^2}{8M_W^2}
\label{eqn:Fermicoupling}
\end{equation}
where $G_F$ is the Fermi coupling constant. The presence of $\gamma^\mu$ 
as well as $\gamma^\mu \gamma_5$ interactions ensures that both 
Fermi-type and Gamow-Teller type interactions are taken care of.

The corresponding {\it electromagnetic interactions} are now obtained by 
substituting Eqn.~(\ref{eqn:WBmixing}) in 
Eqns.~(\ref{eqn:Wleptondoublet}) and (\ref{eqn:leptonright}) to get
\begin{eqnarray}
S_{\rm int}^{\rm em} & = & \int d^4x \ \left[ -\frac{1}{2}g\sin\theta_W 
\left(1 - Y_L\right) \ \overline{\nu_{eL}} \gamma^\mu \nu_{eL} A_\mu 
\right. \nonumber \\
&& \hspace*{0.55in} \left. 
+ \frac{1}{2}g\sin\theta_W \left(1 + Y_L\right) \ \overline{e_L} 
\gamma^\mu e_L A_\mu + \frac{1}{2}g\sin\theta_W Y_e \ \overline{e_R} 
\gamma^\mu e_R A_\mu \right]
\label{eqn:emleptons}
\end{eqnarray}  
where we also substitute $g'= g \tan\theta_W$ from 
Eqn.~(\ref{eqn:WeinbergAngle}). This will yield the QED interactions if 
we put
\begin{equation}
Y_L = 1 \qquad\qquad Y_e = 2 \qquad\qquad g\sin\theta_W = e
\label{eqn:leptonhypercharge}
\end{equation} 
Once the quantum numbers $Y_L$ and $Y_e$ are known, we can determine the 
{\it neutral current interactions}, i.e. those involving the vector 
boson $Z_\mu$, as
\begin{equation}
S_{\rm int}^{\rm nc} = \int d^4x \ \left[ \frac{g}{4\cos\theta_W} 
\overline{\nu_e} \gamma^\mu {\left(1 - \gamma_5\right) \nu_e Z_\mu } - 
\frac{g}{4\cos\theta_W} \overline{e} \gamma^\mu \left( 1 - 
4\sin^2\theta_W - \gamma_5\right) e Z_\mu
\right] 
\label{eqn:NCinteractions}
\end{equation}
All of this falls into place very nicely, and today it has all been 
verified experimentally in a myriad ways.

\section{Late Renaissance: quarks, mixing and a third generation} 

Despite the elegance of the Weinberg-Salam construction, in 1967, there 
still remained three serious issues with this version of the electroweak 
model of leptons.
\vspace*{-0.1in}
\begin{itemize}
\item The first of these was the same problem which had plagued 
Glashow's 1961 model, viz., we cannot write this interaction for a 
nucleon doublet, for there would then be a non-vanishing neutron-photon 
coupling. In fact, the neutron does have a magnetic moment, and hence it 
does couple to the magnetic field components in the photon, but the 
above form of interaction would couple it to the electric field. \\ 
[-3mm]
\item The second problem was that the separation of chiral fermions into 
left-handed doublets and right handed singlets does not permit the 
fermions to have a mass. For example, an electron mass term
\begin{equation}
{\cal L}_m^{(e)} = m_e \, \overline{e} \, e = m_e \left( \overline{e_L} \, e_R + \overline{e_R} \, e_L 
\right) 
\label{eqn:fermionmass}
\end{equation}
will not be gauge invariant, since (i) $e_L$ has weak isospin $T_3 = 
-1/2$ and $e_R$ has $T_3 = 0$, and (ii) $e_L$ has $Y = Y_L = -1$ and 
$e_R$ has $Y - Y_e = -2$. The only way to retain the gauge invariance 
would be to set $m_e = 0$, whereas the electron is known to have a 
finite mass\footnote{One might argue that the electron mass is small and 
may be treated as an explicit breaking of the gauge symmetry. However, 
with hindsight, we know that the same argument applies to the top quark, 
and its huge mass çan by no means be regarded as an explicit 
symmetry-breaking.}.  \\ [-3mm]
\item A more subtle issue attaches itself to the axial vector coupling 
of the $Z$. The axial vector current should be conserved in the limit of 
unbroken gauge symmetry, but sny axial current is well known to have a 
chiral anomaly. This arises from the $B_\mu$ part of the interaction and 
is, therefore, proportional to the hypercharge $Y$. To have a 
renormalisable quantum field theory, we must arrange for the total 
anomaly to cancel\cite{BenLee, GrossJackiw}, i.e. we demand $\sum_i Y_i 
= 0$ for the particles coupling to the $Z$ boson. Clearly, if the 
particles in the theory are $\nu_{eL}$, $e_L$ and $e_R$, then $\sum_i Y 
= Y_L + Y_L + Y_e = -4$. This non-cancellation of the anomaly was the 
third problem.
\end{itemize}
\vspace*{-0.1in}
The first and third problem were solved only with the introduction of 
quarks, but the second problem -- that of how chiral fermions acquire 
masses -- was solved by Weinberg \cite{Weinberg}. The idea is that the 
symmetry-breaking mechanism would generate the fermion masses. This is 
because the symmetries of the leptonic theory permit the existence of 
Yukawa couplings
\begin{eqnarray}
S_{\rm Yuk} & \defeq & \int d^4x \left[ y_e \overline{L_L}(x) \, \Phi(x) 
\, e_R(x) + {\rm H.c.} \right] \nonumber \\
& = &  \int d^4x \left[ y_e \overline{\nu_{eL}}(x) \varphi^+(x) e_R(x)  
+ y_e \overline{e_L}(x) \varphi^0(x) e_R(x)  + {\rm H.c.}  \right]
\label{eqn:leptonYukawa}
\end{eqnarray}
After symmetry-breaking, we replace $\varphi^0 \to \frac{v}{\sqrt{2}} + 
\eta^0(x)$, and obtain
\begin{eqnarray}
S_{\rm Yuk} & = & \int d^4x \left[ y_e \overline{\nu_{eL}}(x) 
\varphi^+(x) e_R(x) + y_e \overline{e_L}(x) \eta^0(x) e_R(x) \right. 
\nonumber \\ && \left. \hspace*{1.7in} + \frac{y_e v}{\sqrt{2}} 
\overline{e_L}(x) \, e_R(x) + {\rm H.c.}  \right]
\label{eqn:leptonYukawabroken}
\end{eqnarray}
The last of these terms is clearly a mass term. Writing it separately, 
we obtain
\begin{equation}
S_{\rm m}^{(e)} = \int d^4x \ \frac{y_e v}{\sqrt{2}} \left[ 
\overline{e_L}(x) \, e_R(x) + \overline{e_R}(x) \, e_L(x) \right] = \int 
d^4x \ m_e \overline{e} e
\label{eqn:electronmass}
\end{equation}
where
\begin{equation}
m_e \defeq \frac{y_e v}{\sqrt{2}}
\label{eqn:massYukawa}
\end{equation}
The electron mass is thus a product of the vacuum expectation value $v$ 
with the Yukawa coupling\footnote{Conversely, we can write the Yukawa 
coupling $y_e = \sqrt{2}m_e/v$.} $y_e$, and obviously will vanish as $v 
\to 0$. This may be regarded as a {\it fourth miracle} of electroweak 
theory. There will, however, be no neutrino mass, since there is no 
$\nu_R$, as pointed out by Goldhaber {\it et al}\cite{Goldhaber} and 
hence, there will be no neutrino Yukawa coupling.

The 1969 discovery of quarks \cite{DIS} provided a new impetus to 
attempts to include strongly-interacting states in the electroweak model 
\cite{Schechter}. For the $u$ and $d$ quarks, it was a simple matter.  
They can be combined into a $SU(2)$ doublet for the left-handed 
components
\begin{equation}
Q_L(x) \defeq \left(\!\!\!\begin{array}{c} u_L(x) \\ d_L(x) \end{array}\!\!\!\right)
\qquad\qquad
\overline{Q_L}(x) = \left(\!\!\!\begin{array}{cc} \overline{u_L}(x) & \overline{d_L}(x) 
\end{array}\!\!\!\right)
\label{eqn:quarkdoublet}
\end{equation}
while the right-handed components $u_R$ and $d_R$ are singlets under 
$SU(2)$. We now write the gauge kinetic term for these quarks as
\begin{equation}
S_q = \int d^4x \ \left[ i \,\overline{Q_L}(x) \, \gamma^\mu \, 
\mathbb{D}^{(Q)}_\mu \, Q_L(x) + i \, \overline{u_{R}}(x) \, \gamma^\mu 
\, D^{(u)}_\mu u_R(x) + i \, \overline{u_{R}}(x) \, \gamma^\mu \, 
D^{(u)}_\mu u_R(x) \right]
\label{eqn:gaugequark}
\end{equation}   
where, as in the leptonic case, we have covariant derivatives 
\begin{equation}
\mathbb{D}^{(Q)}_\mu \defeq \mathbb{I}\,\partial_\mu + ig\,\mathbb{W}_\mu 
- \frac{ig'}{2}Y_Q\,\mathbb{I}\,B_\mu \qquad
D^{(u)}_\mu \defeq \partial_\mu - \frac{ig'}{2}Y_u\,B_\mu \qquad
D^{(d)}_\mu \defeq \partial_\mu - \frac{ig'}{2}Y_d\,B_\mu
\label{eqn:covariantquark} 
\end{equation}
Expanding this, we get charged current interactions
\begin{eqnarray}
S_{\rm int}^{\rm cc} = \int d^4x \ \frac{g}{2\sqrt{2}} \left[ 
\overline{u}(x) \gamma^\mu \left(\!1 - \gamma_5 \!\right) d(x) \, 
W_\mu^+(x) + {\rm H.c.} \right]
\label{eqn:CCquarks}
\end{eqnarray}
and electromagnetic interactions
\begin{eqnarray}
S_{\rm int}^{\rm em} & = & \int d^4x \  \left[ 
\frac{1}{2}e \left(1 + Y_Q\right) \ \overline{u_L} \gamma^\mu u_L A_\mu 
+ \frac{1}{2}e Y_u \ \overline{u_R} \gamma^\mu u_R A_\mu 
\right. \nonumber \\
&& \hspace*{0.45in} \left. 
- \frac{1}{2}e \left(1 - Y_Q\right) \ \overline{d_L} \gamma^\mu d_L A_\mu 
+ \frac{1}{2}e Y_d \ \overline{d_R} \gamma^\mu d_R A_\mu 
\right]
\label{eqn:emquarks}
\end{eqnarray} 
using Eqns.~(\ref{eqn:WeinbergAngle}) and (\ref{eqn:quarkhypercharge}) 
to replace $g'$ and $g$ by $e$. It is now easy to obtain the proper 
quark charges by setting
\begin{equation}
Y_Q = \frac{1}{3} \qquad\qquad Y_u = \frac{4}{3} \qquad\qquad Y_d = -\frac{2}{3}
\label{eqn:quarkhypercharge}
\end{equation}  
leading to the usual electromagnetic interaction
\begin{equation}
S_{\rm int}^{\rm em} = \int d^4x \  \left[ \frac{2e}{3} \,\overline{u}(x) \gamma^\mu u(x) A_\mu(x)
- \frac{e}{3} \,\overline{d}(x) \gamma^\mu d(x) A_\mu(x) \right]   
\label{eqn:QEDquark}
\end{equation}
Both these quarks have nonzero charge, but when combined into $n = udd$, 
of course, the net charge is zero. This solves the first problem, and 
also explains why the neutron has a magnetic (dipole) moment.

Inclusion of these quarks also solves the second problem, but with a 
twist. If we naively sum the hypercharges of the neutrino-electron and 
up-down pairs (constituting a {\it generation}), we will get
\begin{equation}
2Y_L + Y_e + 2Y_Q + Y_u + Y_d = - 2 - 2 + \frac{2}{3} + \frac{4}{3} - \frac{2}{3} = - \frac{8}{3}
\label{eqn:anomalynocolor}  
\end{equation}
indicating that the chiral anomaly does not cancel. However, if we take into account that every quark comes in 3 colours, then the addition must be
\begin{equation}
2Y_L + Y_e + 3\times (2Y_Q + Y_u + Y_d) = - 2 - 2 + 3\left(\frac{2}{3} + \frac{4}{3} - \frac{2}{3} 
\right) = 0   
\label{eqn:anomalycancellation}
\end{equation} 
We now have almost everything fitting in the Glashow-Salam-Weinberg 
model, except for the strange quark $s$ \footnote{It is an interesting 
social phenomenon that the nomenclature of hadronic $SU(2)$ doublets has 
proceeded from the classical (proton, neutron), to the colloquial (up, 
down), to the fanciful (charm, strange) and finally the downright 
irreverent (top, bottom). It is, perhaps, for the best that there are no 
more.}. That it cannot be a singlet of $SU(2)$ is clear from the fact 
that we have weak decays like $K^+ (u \bar{s}) \to \mu^+ \nu_\mu$ which 
must involve a $\bar{s}d W$ vertex. This puzzle led to the creation of 
different models \cite{Prentki}, of which Lee \cite{BenLee} pointed out 
that the simplest, and, as it later turned out, correct one was that due 
to Glashow, Iliopoulos and Maiani (or GIM for short) \cite{GIM}. This 
was simply to postulate the existence of a new quark $c$ --- called the 
{\it charm quark} --- which would be the $SU(2)$ partner of the $d$ 
quark, just as the $u$ is the $SU(2)$ partner of the $d$ quark. Thus, 
the GIM model envisages {\it two} fermion generations, viz.,
\begin{eqnarray}
L_L^{(1)} & = & \left(\!\!\!\begin{array}{c} \nu_{eL} \\ e_L \end{array}\!\!\!\right) \quad
\ell_R^{(1)} = e_R \qquad
Q_L^{(1)} = \left(\!\!\!\begin{array}{c} u_L \\ d_L \end{array}\!\!\!\right) \quad
u_R^{(1)} = u_R \qquad d_R^{(1)} = d_R \nonumber \\
L_L^{(2)} & = & \left(\!\!\!\begin{array}{c} \nu_{\mu L} \\ \mu_L \end{array}\!\!\!\right) \quad
\ell_R^{(2)} = \mu_R \qquad
Q_L^{(2)} = \left(\!\!\!\begin{array}{c} c_L \\ s_L \end{array}\!\!\!\right) \quad
u_R^{(2)} = c_R \qquad d_R^{(2)} = s_R
\label{eqn:twogenerations}
\end{eqnarray}    
with the chiral anomaly cancelling separately over each generation. This 
was brilliantly confirmed by the 1974 discovery \cite{charmonium} of the 
$c$-quark --- or rather of a $c\bar{c}$ bound state, called the 
$J/\psi$.

The existence of two generations also provided an explanation for the 
mystery of flavour-mixing, which had been postulated by Cabibbo 
\cite{Cabibbo} as early as 1963. This was elaborated by GIM, using the 
Higgs doublet, as follows. The existence of two generations permits the 
existence of mixed Yukawa couplings in the quark sector of the form
\begin{equation}
{\cal S}_{\rm Yuk}^{(q)} = \int d^4x \ \sum_{a=1}^2 \sum_{b=1}^2 \left[ y_{ab}^{(u)} \; 
\overline{Q_L^{(a)}}(x) \, \widetilde{\Phi}(x) \, u_R^{(b)}(x) + y_{ab}^{(d)} \; 
\overline{Q_L^{(a)}}(x) \, \Phi(x) \, d_R^{(b)}(x) + {\rm H.c.} \right] 
\label{eqn:mixedYukawa} 
\end{equation}  
using the notation of Eqn.~(\ref{eqn:twogenerations}), where, in order 
to keep the $U(1)_Y$ invariance, the first term uses the charge 
conjugate doublet
\begin{equation}
\widetilde{\Phi}(x) \defeq i\sigma_2 \Phi^*(x) = \left(\!\!\!\begin{array}{c} \varphi^0 \\ - \varphi^- \end{array}\!\!\!\right) 
\end{equation}
Expanding and shifting $\varphi^0 = v/\sqrt{2} + \eta^0(x)$, we obtain 
mass terms of the form
\begin{equation}
S_m = \int d^4x \ \sum_{a=1}^2 \sum_{b=1}^2 \left[ 
\frac{y_{ab}^{(u)}v}{\sqrt{2}} \overline{u_L} u_R + 
\frac{y_{ab}^{(d)}v}{\sqrt{2}} \overline{d_L} d_R + {\rm H.c.} 
\right]
\end{equation}
which can be written in matrix form as
\begin{equation}
S_m = \int d^4x \ \left[ \; \overline{U_L} \, \mathbb{M}^{(u)} \, U_R + 
\overline{D_L} \, \mathbb{M}^{(d)} \, D_R + {\rm H.c.} \right]
\label{eqn:matrixform}
\end{equation}
where $\overline{U_{L,R}} = \left(\!\!\!\begin{array}{cc} 
\overline{u_{L,R}} & \overline{c_{L,R}}
\end{array}\!\!\!\right)$
and $\overline{D_{L,R}} = \left(\!\!\!\begin{array}{cc} 
\overline{d_{L,R}} & \overline{s_{L,R}}
\end{array}\!\!\!\right)$, while the mass matrices are
\begin{equation}
\mathbb{M}^{(u)}_{ab} \defeq \frac{y_{ab}^{(u)}v}{\sqrt{2}} \qquad\qquad
\mathbb{M}^{(d)}_{ab} \defeq \frac{y_{ab}^{(d)}v}{\sqrt{2}}
\label{eqn:massYukawa3d}
\end{equation}
Each of these mass matrices $\mathbb{M}^{(u)}$ and $\mathbb{M}^{(d)}$ 
can be diagonalised by bi-unitary transformations \cite{ChengLi}
\begin{equation}
U_L = \mathbb{V}_L^{(u)} U_L^0 \qquad\qquad U_R = 
\mathbb{V}_R^{(u)}U_R^0 \qquad\qquad \mathbb{V}_L^{(u)} \mathbb{M}^{(u)} 
\mathbb{V}_R^{u\dagger} = \left(\!\!\!\begin{array}{cc} m_u & 0 \\ 0 & 
m_c \end{array} \!\!\!\right)
\end{equation} 
and
\begin{equation}
D_L = \mathbb{V}_L^{(d)} D_L^0 \qquad\qquad D_R = \mathbb{V}_R^{(d)} 
D_R^0 \qquad\qquad \mathbb{V}_L^{(d)} \mathbb{M}^{(d)} 
\mathbb{V}_R^{d\dagger} = \left(\!\!\!\begin{array}{cc} m_d & 0 \\ 0 & 
m_s \end{array} \!\!\!\right)
\label{eqn:biunitary}
\end{equation} 
from which
\begin{equation}
S_m = \int d^4x \ \left[ m_u \, \overline{u_L^0} u_R^0 + m_c \, 
\overline{c_L^0} c_R^0 + m_d \, \overline{d_L^0} d_R^0 + m_s \, 
\overline{s_L^0} s_R^0 + {\rm H.c.} \right]
\end{equation}
where the superscript $0$ denotes the physical quark state (often 
referred to as the {\it mass basis}.)
 
When we come to the charged current interactions of 
Eqn.~(\ref{eqn:CCquarks}), however, we can rewrite them as
\begin{eqnarray}
S_{\rm int}^{\rm cc} & = & \int d^4x \ \left[ \frac{g}{2\sqrt{2}} \; 
\overline{U_L} \gamma^\mu (1 - \gamma_5) D_L W_\mu^+ + {\rm H.c.} 
\right] \nonumber \\ & = & \int d^4x \ \left[ \frac{g}{2\sqrt{2}} \; 
\overline{U_L^0} \mathbb{V}_L^{(u)\dagger} \gamma^\mu (1 - \gamma_5) 
\mathbb{V}_L^{(d)} D_L^0 W_\mu^+ + {\rm H.c.} \right] \nonumber \\ & = & 
\int d^4x \ \left[ \frac{g}{2\sqrt{2}} \; \overline{U_L^0} \gamma^\mu (1 
- \gamma_5) \, \mathbb{K}_2 D_L^0 W_\mu^+ + {\rm H.c.} \right]
\end{eqnarray}
where $\mathbb{K}_2 \defeq \mathbb{V}_L^{(u)\dagger} \mathbb{V}_L^{(d)}$ 
is known as the {\it mixing matrix}. The subscript $2$ indicates that it 
is a $2\times2$ matrix, and, of course, since all the $\mathbb{V}$'s are 
unitary, $\mathbb{K}$ is also unitary. Such matrix can be parametrised 
in terms of one angle parameter and three phase parameters; however, it 
can be shown \cite{ChengLi} that these three phase parameters can be 
absorbed in the phases of the four quark fields. This makes 
$\mathbb{K}_2$ real, i.e. it has the form
\begin{equation}
\mathbb{K}_2 \defeq \left(\!\!\!\begin{array}{cc} 
\cos\theta_C & \sin\theta_C \\
-\sin\theta_C &  \cos\theta_C \end{array}\!\!\!\right) 
\end{equation}    
where $\theta_C$ is the Cabibbo angle --- postulated by Cabibbo in 1963. 
The matrix $\mathbb{K}_2$ may, therefore, be referred to as the {\it 
Cabibbo matrix}\footnote{Strictly speaking, in 1963 Cabibbo 
\cite{Cabibbo} himself wrote down only the mixing between the $d$ and 
the $s$ quarks, or rather, the corresponding charged currents with a 
$\bar{u}$. The full matrix, which requires the $c$ quark, was written 
down only in 1970 by Glashow, Iliopoulos and Maiani \cite{GIM}.}.

The electroweak model developed above was almost complete, but one 
important feature was left unexplained. When parity ${\cal P}$ was shown 
not to be conserved \cite{YangLee,MadameWu}, Landau had speculated 
\cite{CP} that what is conserved is the combination\footnote{This is 
really to say that while elementary particles may have a 
parity-violating excess, the number of antiparticles in the Universe 
will have an exactly complementary deficit, or vice versa, keeping the 
total number unchanged.} of parity ${\cal P}$ and charge conjugation 
${\cal C}$, i.e., the discrete quantum number ${\cal CP}$. This is 
precisely what happens in the GIM model of the electroweak interactions, 
even if the Yukawa couplings $y_{ab}^{(u,d)}$ are taken to be complex, 
since the phases of the mixing matrix $\mathbb{K}_2$ will not appear. 
${\cal CP}$, however, is not conserved, as was proved in 1964 by the 
iconic Cronin-Fitch experiment \cite{CPV}. In fact, unless ${\cal CP}$ 
is violated, there would be no baryon asymmetry in the Universe, as 
pointed out by Sakharov \cite{Sakharov}, and hence the Universe of 
matter which we live in and carry out experiments and observations could 
not exist.  It was in a perhaps desperate attempt\footnote{Reminiscent 
of Pauli's 'desperate attempt' to save energy conservation by 
postulating the neutrino \cite{neutrino}.} to save the electroweak model 
that Kobayashi and Maskawa, in 1973, suggested \cite{KM} that there may 
be a {\it third} generation of fermions, viz.
\begin{equation}
L_L^{(3)} = \left(\!\!\!\begin{array}{c} \nu_L^{(3)} \\ \ell_L^{(3)} \end{array}\!\!\!\right) \qquad
\ell_R^{(3)} \qquad\qquad 
Q_L^{(3)} = \left(\!\!\!\begin{array}{c} u_L^{(3)} \\ d_L^{(3)} \end{array}\!\!\!\right) \qquad
u_R^{(3)} \qquad d_R^{(3)}
\end{equation}  
such that the Yukawa couplings and mass matrices would become $3\times 
3$ matrices, and we would have to replace the $2\times 2$ mixing matrix 
$\mathbb{K}_2$ by a $3\times 3$ version $\mathbb{K}_3$, which is now 
called the Cabibbo-Kobayashi-Maskawa matrix, or {\it CKM matrix} for 
short. Now, a $3\times 3$ unitary matrix has 3 angles and 6 phases, of 
which 5 phases can be absorbed into the phases of the six quark fields. 
One phase, however, remains, and this, according to Kobayashi and 
Maskawa \cite{KM}, is the source of ${\cal CP}$ violation.
 
Like Pauli's bold prediction of the neutrino and GIM's bold prediction 
of the charm quark, the equally bold speculation of Kobayashi and 
Maskawa was proved absolutely correct, when the fermions of the third 
generation began to be discovered one by one. First came the tau lepton 
$\ell^{(3)} = \tau^-$ in 1975 \cite{Perl}, closely followed by the 
bottom quark $d^{(3)} = b$ in 1977 \cite{bdiscovery}.  There followed a 
17-year hiatus till the 1994 discovery of the top quark $u^{(3)} = t$ 
\cite{top}, and another 6 years wait till the existence of the tau 
neutrino $\nu^{(3)} = \nu_\tau$ was confirmed \cite{nutau} in 2000.
            
By the turn of the century, it was clear that the electroweak model --- 
dubbed the 'Standard Model', or SM for short, by Pais and Treiman 
\cite{Pais} as early as 1975 --- provides a logical and coherent 
structure for the weak and electromagnetic interactions, and also 
intimately ties in with QCD through the importance of three colours for 
anomaly cancellation. All that remained was the cornerstone of the whole 
theory, viz., the Higgs boson, which had proved elusive to experimental 
searches for several decades We may quickly recall that it is by 
undergoing a phase transition in the early Universe, as it was cooling 
down, that the Higgs field gave mass to all the other particles in the 
SM (except photons and gluons with which it does not interact). And it 
is the large masses of $W,Z$ bosons that makes the weak interaction 
weak, since they are responsible for the smallness of the Fermi coupling 
constant. Therefore, the Higgs boson lies at the very heart of weak 
interactions theory. However, the electroweak action does not tell us 
anything about the mass of the Higgs boson, which is a free parameter in 
the Glashow-Salam-Weinberg model. Fortunately, there are elegant 
arguments why the mass of the Higs boson cannot be very high. These ars 
outlined below.

It one considers the elastic scattering of $W$ bosons --- a permitted 
process in the GSW model --- then it is easy to see that the 
longitudinal mode, i.e.
\begin{equation}
W_L + W_L \longrightarrow W_L + W_L 
\end{equation}
will have bad high energy behaviour because in that limit 
$\varepsilon_L^\mu (p) \simeq p^\mu/M_W$. In fact, in the limit when the 
$W$ and $Z$ masses can be neglected (but not the Higgs boson mass), the 
amplitude for this process assumes the form
\begin{equation}
{\cal A}_L(s,t) \approx -\sqrt{2} M_H^2 \left( \frac{s}{s - M_H^2}
+ \frac{t}{t-M_H^2} \right)
\label{eqn:WW scattering}
\end{equation} 
If we allow $M_H$ to grow without limit, we find that as $M_H \to 
\infty$ the amplitude reduces to
\begin{equation}
{\cal A}_L(s,\theta^*) \approx \frac{G_F}{\sqrt{2}} \, s\left( 1 + 
\cos\theta^*\right)
\end{equation}  
where $\theta^*$ is the scattering angle. Obviously, as $s$ increases, 
this will grow without limit, violating unitarity. It follows that we 
cannot let $M_H$ grow without limit.

This argument can be sharpened \cite{Quigg} by invoking partial-wave 
unitarity of the above amplitude, which can be expanded as
\begin{equation}
{\cal A}_L(\theta^*) = \sum_{\ell=0}^\infty (2\ell+1) A_\ell 
P_\ell(\cos\theta^*)
\end{equation}
with the requirement $|A_j|^2 \leq |{\rm Im} A_j|$, which can be 
rewritten as $\mid\!A_j\!\!\mid\leq \frac{1}{2}$. Applying this to the 
$j = 0$ partial wave (which happens to give the best bound among all the 
values of $j$)
\begin{equation}
A_0 = \frac{1}{16\pi s} \int_{-s}^\infty dt \ {\cal A}_L(s,t)
\end{equation} 
one obtains an upper bound
\begin{equation}
M_H^2 \leq \frac{4\sqrt{2}\pi}{3G_F} \approx \left( 700~{\rm GeV} \right)^2
\end{equation}       
Thus it was clear that the Higgs boson could not be very heavy, for in 
1977, thoughts of a 20~TeV LHC and a 40~TeV SSC were already in the air. 
In fact, these machines (of which the SSC never materialised) were 
designed as machines where the Higgs boson would certainly be 
discovered. However, successive experiments came and went, but there was 
no sign\footnote{It is well-known that when Leon Lederman wrote a 
popular book with Dick Teresi on the hard-to-find Higgs boson, he wanted 
to name his book {\it The Goddamn Particle} to express the frustration 
of the high energy physics community. It is equally famous that his 
scandalised publisher changed the title to {\it The God Particle} 
\cite{God}, thereby creating an endless source of confusion in the minds 
of the public when the elusive particle finally showed up.} of the 
elusive particle. This explains the jubilation when it was finally 
discovered at CERN \cite{Hdiscovery} in 2012, and of course, the prompt 
award of a much-belated Nobel Prize to François Englert and Peter Higgs 
in the following year\footnote{Robert Brout, Englert's co-author, had 
died in 2011.}. The long road to this discovery makes, of itself, a 
fascinating story, and it has been covered in several books published 
after 2012 \cite{Higgsbooks}.

With the discovery of the Higgs boson, the long development of the 
Standard Model was complete. Till date, we do not have any better model 
for explaining the experimental measurements made in all terrestrial 
experiments.


\begin{center}
\Large{\textbf{{\textsc{Part 2}} : After the Higgs, What?}}
\end{center}
\vspace*{0.1in}

During the US Senate hearings on the SSC project, one Senator is 
reported to have commented "You have quite a few quarks already. What 
will you do with another one?" While this story may be apocryphal, a 
similar question can be asked in the context of extended Higgs sectors 
and, more generally, of all physics beyond the SM. We need to briefly 
justify this before moving on to more Higgs-specific discussions.

\section{Impressionism: Beyond the Standard Model} 

The initial impetus to go beyond the SM (bSM) came from the fact that 
one requires three colours --- the basis of QCD --- to get an 
anomaly-free theory of electroweak interactions. The fact that QCD is 
also a gauge theory led to speculations that a further level of 
unification, viz. strong and electroweak, may occur if we embed both in 
a common gauge theory with a larger gauge group.  The earliest ideas in 
this direction came from Pati and Salam \cite{PatiSalam}, who, in 1974, 
created a gauge theory based on the group $SO(4)$, where lepton number 
is a fourth colour. A more comprehensive theory, based on the group 
$SU(5)$ was written down by Georgi and Glashow soon after 
\cite{GeorgiGlashow}. This set off a vast industry of {\it Grand Unified 
Models} or GUTs, which generically provided strong-electroweak 
unification in addition to explaining one or more of the arbitrary 
features of the SM.

A common feature of GUTs, however, is the prediction of proton decay 
through the process
\begin{equation}
p \longrightarrow \pi^0 + e^+ 
\end{equation}
which would be mediated by a (massive) gauge boson $X$ of the higher 
(broken) symmetry. The decay width will be suppressed by the mass of 
this heavy gauge boson $X$, and the lifetime can be estimated as 
\cite{Chanowitz}
\begin{equation}
\tau_p \sim \frac{M_X^4}{m_p^5}
\end{equation}
Though originally conceived by Sakharov \cite{Sakharov}, this prediction 
of proton decay in the context of GUTs was first made by Pati and Salam 
\cite{PatiSalam} in 1974. Since then, there have been intensive searches 
for proton decay\footnote{Including pioneering searches at the Kolar 
Gold Fields in India \cite{KGF}.}, but none have been successful in 
finding any evidence whatsoever. In the 1970s, the lower bound on proton 
decay\footnote{Since the age of the Universe is estimated to be only 
about $1.38 \times 10^{10}$ years \cite{age}, there is no imminent 
danger of the Universe disappearing into a puff of pions and leptons.} 
stood at around $10^{29}$ years, which means that $M_X$ should be at 
least $10^{15}$~GeV. This was known as the GUT scale. However, over the 
years, the lower bound on the proton lifetime has crept up and up to 
around $1.67 \times 10^{34}$ years today \cite{HyperK}, requiring $M_X$ 
to be greater than about $3 \times 10^{16}$ GeV. This is still about 400 
times smaller then the Planck scale $M_P = \left(\hbar c/G_N 
\right)^{1/2}$, where $G_N$ is the gravitational constant, and 
therefore, it should be just about possible to construct a GUT without 
having to seriously take into account the threshold effects of strong 
gravity.

\begin{figure}[h!]
\begin{center}
\includegraphics[width=0.5\textwidth]{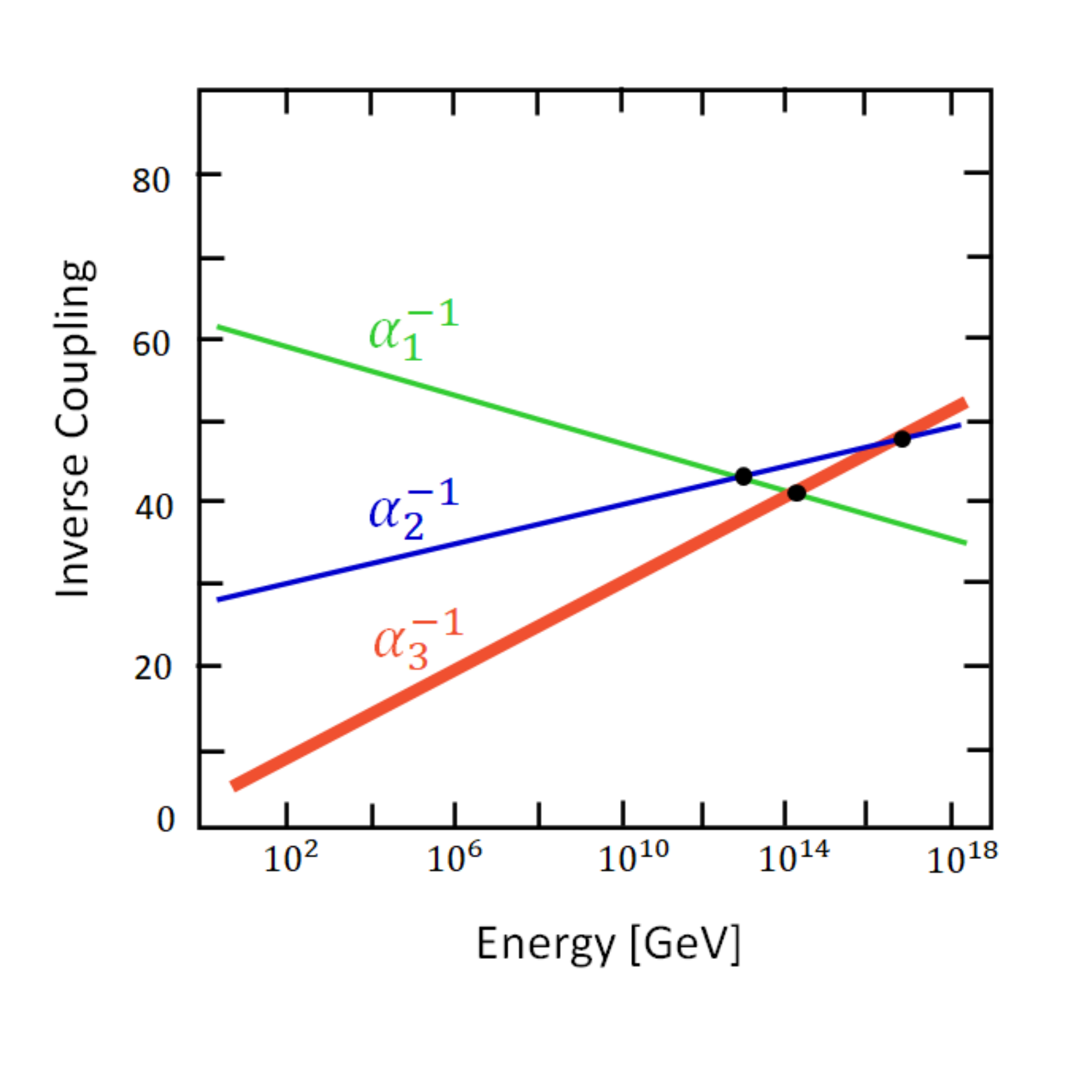}  
\vspace*{-0.3in}
\caption{\small Running coupling constants in the SM.}
\label{fig:GUT}
\end{center}
\end{figure}
\vspace*{-0.2in}

A more serious problem with GUTs arose after 1991 when the LEP-1 
collider at CERN produced its precision measurements of all the three 
gauge couplings $\alpha_1$, $\alpha_2$ and $\alpha_3$ corresponding to 
$U(1)$, $SU(2)$ and $SU(3)$ respectively, at the mass scale $M_Z$. One 
could now determine the beta functions of these couplings, and thereby 
predict their running till the Planck scale. This exercise was done by 
several groups \cite{Amaldi} with a startling result. Instead of 
evolving logarithmically so that all three met at a common scale where 
the expected GUT unification would occur, the renormalisation group 
evolution of these couplings seemed to lead to pairwise crossing at 
three different scales (the black dots in Figure~4) --- just as if there 
is no GUT unification. This was to prove a blow from which GUTs never 
really recovered. Of course, it was almost immediately shown that the 
single-scale unification could be restored by embedding the SM in a 
supersymmetric model, and later it was established that one could also 
achieve it with a bigger GUT group, with symmetry-breaking in multiple 
stages, where the beta functions would change at each threshold 
\cite{GUTrescue}. But the pristine beauty of $SU(5)$ grand unification 
was gone, and thenceforth, slowly but surely, GUT theories moved from 
the mainstream to the fringes of high energy research.

\begin{minipage}{3.4in}
Before fading into the background, however, GUT theories brought to the 
fore a serious issue with the Higgs boson mass. The Higgs doublet, in 
order to transform under the $SU(2)\times U(1)$ gauge group, must also 
transform under the larger GUT gauge group $G \supset SU(2)\times U(1)$.  
The kinetic term for Higgs bosons will, then, induce seagull terms of 
the form
\end{minipage}
\hskip 12pt
\begin{minipage}{3.0in}
\centerline{\includegraphics[width=0.65\textwidth]{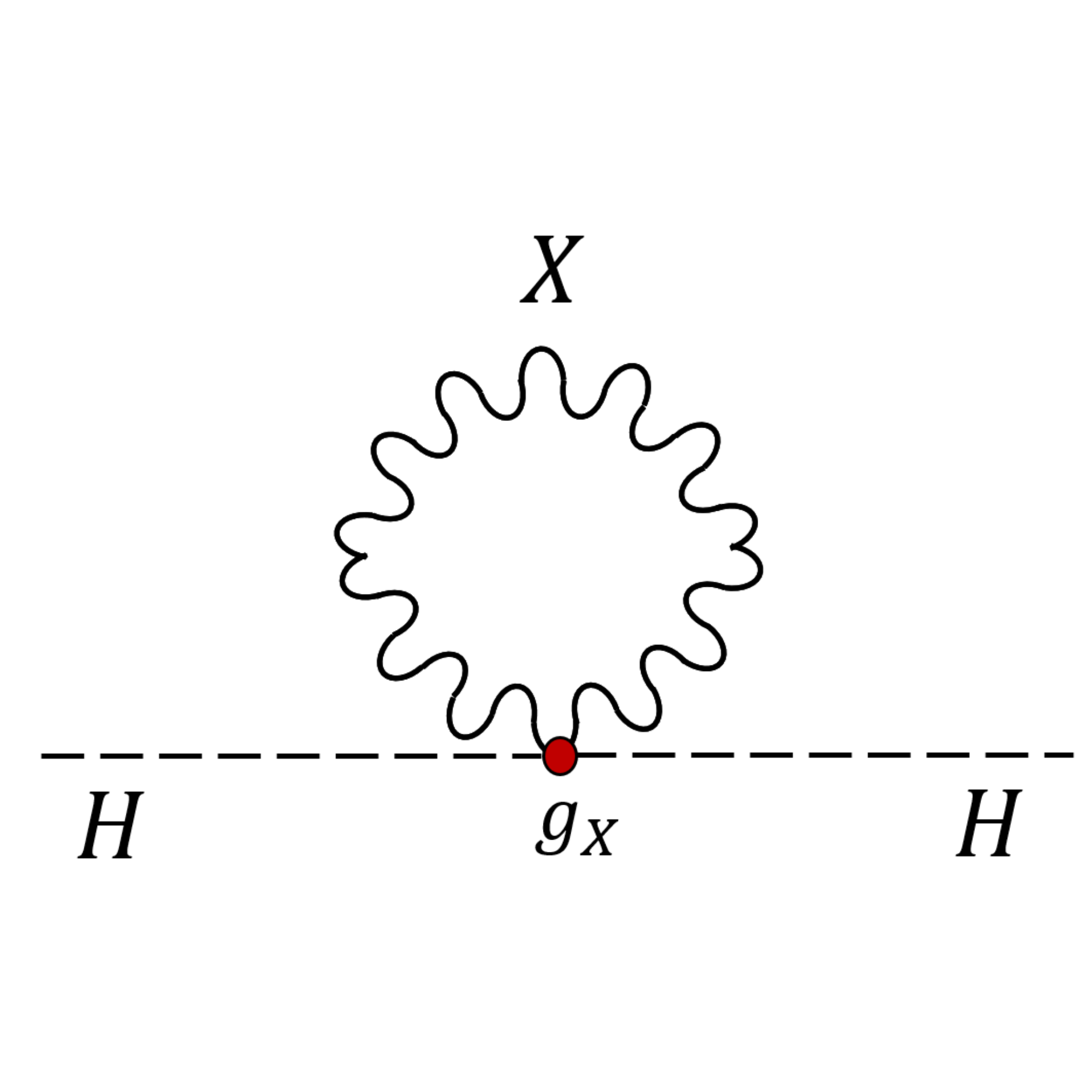} } 
\end{minipage}

\begin{equation}
S_{\rm sg} = \int d^4x \ g_X  X^\mu(x) X_\mu(x) \Phi^\dagger(x) \Phi(x0 
\end{equation}
leading to the so-called 'sunset' diagrams as shown in the figure, with 
a heavy $X$ boson contributing to the Higgs boson self-energy at the 
one-loop level. This contribution may be calculated, using the 
dimensional regularisation scheme in $4-\varepsilon$ dimensions, as
\begin{equation}
\delta M_H^2(M_X) = \frac{g_X}{32\pi^2} M_X^2 \left( 
\frac{2}{\varepsilon} - \gamma + \ln\pi - 1 + \ln M_X^2 \right)
\label{eqn:quadratic} 
\end{equation}
where $\gamma$ is the Euler-Mascheroni constant \cite{tHooft}. Even 
apart from the divergence as $\varepsilon \to 0$, this contribution is 
intolerably large, given that $M_X \sim 10^{16}$~GeV and $M_H$ is 
125.4~GeV. This is known as the {\it hierarchy problem} and it can 
appear whenever there are two widely-separated energy scales in a 
quantum field theory \cite{hierarchy}. In this case, the renormalisation 
technique cannot save the situation. One can, of course, introduce a 
counterterm in the action to cancel this divergence in the 
$\overline{\rm MS}$ scheme. However, if we recall that the $M_X$ is a 
physical mass and will surely run with energy, this means that the 
divergence will reappear at a different energy. One would then have to 
introduce a running counterterm, which is hard to justify. This is known 
as the {\it fine-tuning problem}.

It is important to point out at this stage that the hierarchy and 
fine-tuning problems arise only when there is a GUT or other energy 
scale $M_X$ lying between the electroweak scale $M_H$ and the Planck 
scale $M_P$. If there is no such scale, and the SM holds all the way up 
to the Planck scale $M_P$, there will be no hierarchy problem, since 
$\delta M_H^2$ will depend only on $M_P$ --- which does not run and can 
therefore be removed once and for all in an $\overline{\rm MS}$ 
cancellation. We therefore need to look for a stronger reason to have an 
intermediate scale $M_X$ -- other than a naive belief that there is a 
GUT somewhere in the background. This may be sought in a consideration 
of the stability of the electroweak vacuum against quantum corrections.

As is the case with all the coupling constants in the SM, the scalar 
quartic coupling $\lambda$ in Eqn.~(\ref{eqn:ssbU1unitary}) will also 
run. At the electroweak scale, we can solve the two mass relations
\begin{equation}
M_W = \frac{1}{2} gv  \qquad\qquad  M_H = \sqrt{2\lambda}v
\end{equation} 
to obtain
\begin{equation}
\lambda = \frac{\pi\alpha}{2 \sin^2\theta_W} \left(\!  \frac{M_H}{M_W} 
\!\right)^2 \simeq 0.129
\end{equation}
plugging in the measured values. This value of $\lambda$ is already 
quite small, and, as it happens, with increasing energy it decreases 
steadily, its evolution being driven mostly by the large top quark 
Yukawa coupling. At a scale somewhere between $10^{9}$~GeV and 
$10^{20}$~GeV, in fact, it will become zero, and thenceforth run to 
negative values. The state of the art in this analysis is reproduced un 
Figure~5 from Ref.~\cite{Giudice}, where the computation has been 
carried to the three-loop level in the SM. The grey shaded region shows 
the uncertainty in the extrapolated value of $\lambda$ as it runs with 
increasing energy ($\mu$ in the figure). The two main sources of this 
error are the values of the strong coupling constants $\alpha_3$ and the 
uncertainty in the Yukawa coupling of the top quark, as currently 
measured.
 
\vspace*{-0.2in}
\begin{figure}[h!]
\begin{center}
\includegraphics[width=0.6\textwidth]{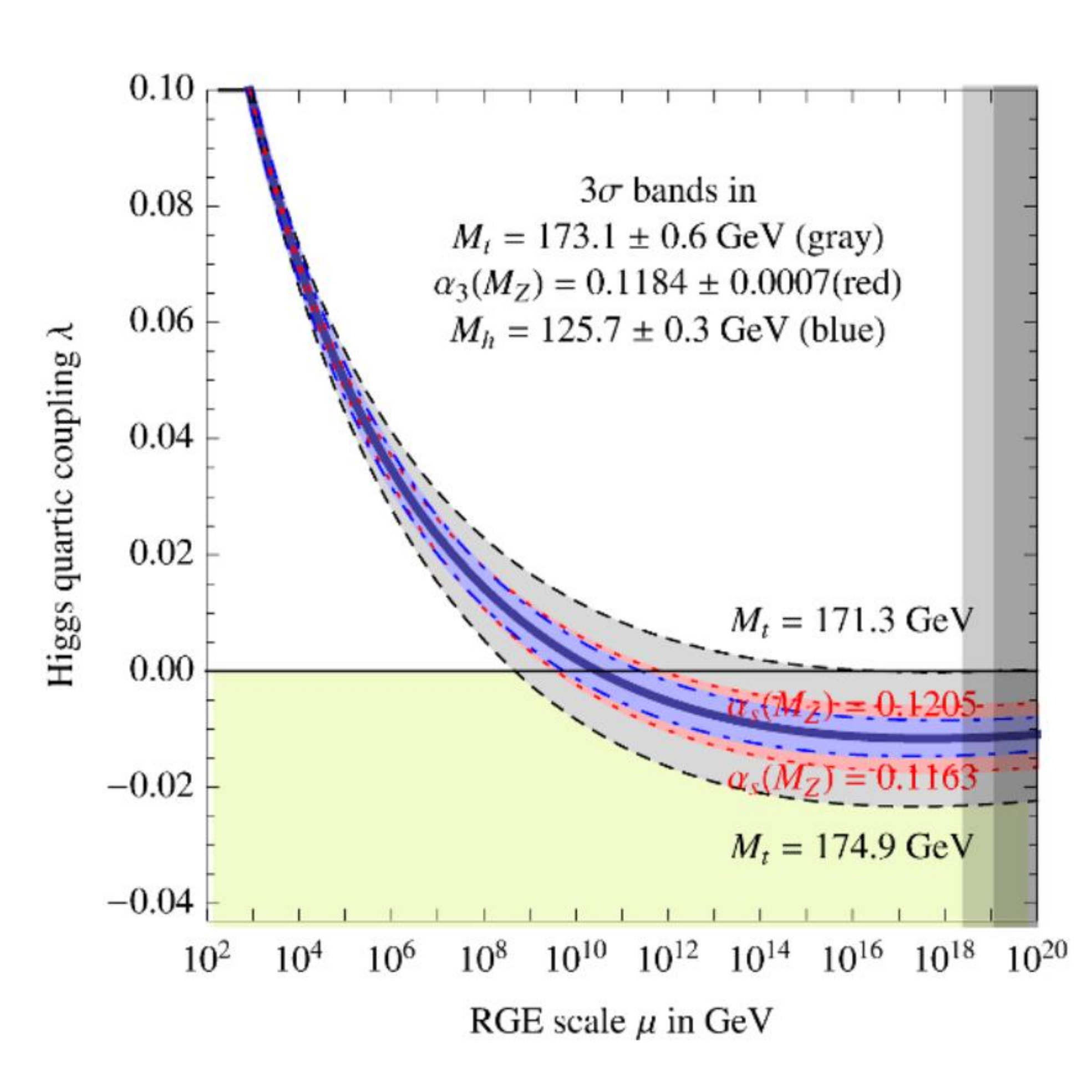}  
\caption{\small Running of the Higgs quartic coupling in the SM. Reproduced 
from Ref.~\cite{Giudice}). The parameters have values as measured in 
2013. Values of $\lambda$ is the light green-shaded region will 
correspond to an unstable vacuum.}
\label{fig:phi4run}
\end{center}
\end{figure}
\vspace*{-0.2in}
A glance at Figure~\ref{fig:phi4run} will make it obvious, that if 
$\lambda$ is non-positive, it will be an absolute disaster for the SM, 
for then the potential will become unbounded from below, and it will be 
possible for any particle to go down the potential, emitting an infinite 
amount of energy, i.e. there will be a catastrophic explosion of the 
Universe. Of course, a catastrophe is not precluded just because it is a 
catastrophe. The argument here is that just after the Big Bang, the 
newly-created Universe consisted of particles at energies at the Planck 
scale, or just below it. If the electroweak vacuum had been unstable at 
these energies, there would have been plenty of radiated energy to keep 
the Universe in that hot condition, or make it even hotter. This clearly 
did not happen, since the Universe has cooled down to as low as 2.73~K. 
Ergo, the electroweak vacuum has stayed stable throughout, and the 
$\lambda$ coupling has never gone negative.  This can only happen if, at 
some higher scale, there is new physics beyond the SM, which changes the 
running of the $\lambda$ coupling and keeps it from falling to zero.

Once can therefore make a strong claim that if $\lambda$ becomes 
negative at a high energy scale below the Planck scale, we are bound to 
have new physics at that scale, and this will immediately reinstate the 
hierarchy problem for the Higgs boson mass. But can we be sure that 
$\lambda$ does, indeed become negative at a value below the Planck 
scale? Figure~5 shows that because of the uncertainty in 
parameters\footnote{There are several uncertainties in the value of the 
top quark mass which goes into the RGE evolution of $\lambda$. The 
principal problem is that there is no measurable pole mass for the top 
quark, since quark states are not asymptotically free. The 
experimentally measured mass at the LHC is an invariant mass of final 
states where the interpretation of hadronisation effects is model 
dependent. The best inference of the top quark 'pole' mass comes from 
the measurement of the $t\bar{t}$ production cross section, but that is 
dependent on PDF uncertainties as well as the renormalisation scheme. 
Most of these uncertainties do not affect common processes, but even 
small changes affect the RGE scale, which has the top quark-dependent 
parameters in the argument of the exponential function.} we cannot yet 
be sure about this, since the range of $\lambda $ values has one edge 
that lies just around the limiting value, i.e. there is still a 
possibility that the vacuum will remain stable all the way to the Planck 
scale without invoking new physics.

There have been attempts to save the electroweak vacuum by postulating a 
more complicated structure, equivalent to putting in a 
(non-renormalisable) $\epsilon \varphi^6(x)$ term in the potential. In 
these models, there is another, deeper vacuum, which is stable, and the 
current vacuum is a 'false vacuum', i.e., a metastable state 
\cite{Coleman1977}. The tunnelling time for this has been calculated as 
$10^{10}$ years \cite{Giudice}, which is reassuring, since it makes the 
probability that it will happen in the next 10,000 years around 
$10^{-129}$. However, it has been pointed out \cite{Spencer2013} that if 
the Universe underwent an inflationary stage, as is commonly believed, 
then the tunnelling time would have been much shorter --- in fact short 
enough for the transition to the true vacuum to have occurred. 
However,it was then pointed out \cite{Espinosa2015} that this is true 
only in a flat space calculation; if there is Higgs-curvature mixing 
with a mixing parameter $\xi$, then it can be tuned to the so-called 
'conformal point' $\xi = -1/6$ to keep the electroweak vacuum stable 
throughout the inflation period. This was reinforced by a calculation of 
reheating effects \cite{Gross2015} which established that the fine 
tuning $\xi = -1/6$ is the only way to save the electroweak vacuum. 
There the case rests.

However, long before the value of $\lambda$ was measured at any scale, 
the influence of GUTs had inspired several bSM solutions for the 
hierarchy problem. These are of two classes, as follows.
\newpage
\begin{enumerate}
\item Models where the cutoff scale is brought down to just above the 
electroweak scale.
\begin{itemize}
\item {\it Technicolor}: It was Wilson in 1971 who first pointed out 
\cite{hierarchy} that gauge boson and fermion self-energy corrections 
are protected from having quadratic divergences by the gauge symmetry 
and the chiral symmetry, respectively. For a scalar, however, there is 
no such symmetry, and hence, quadratic divergences, such as 
Eqn.~(\ref{eqn:quadratic}), cannot be avoided. Wilson went on to point 
out that this was not a problem for scalars like the pions, since they 
were composites, and the scale set by the binding energy would be a 
natural cutoff for self-energy corrections. For pions, this is the QCD 
scale, which, at around 280~MeV, is not so much larger than the pion 
mass. \\ [-3mm]

By the later half of the decade, as the SM went from triumph to triumph, 
it was clear that the hierarchy problem in the Higgs boson mass needed 
to be addressed. The first solution to this was proposed by Weinberg 
\cite{techniWeinberg} in 1976 and backed up by Susskind \cite{Susskind} 
in 1978.  This was more-or-less a repetition of Wilson's argument for 
pions. It was assumed that the Higgs boson is actually a composite of a 
new kind of confined fermions called {\it techni-fermions}, analogous to 
qiarks, which were bound into a scalar by a $SU(N)$ gauge interaction 
called {\it technicolor}. The cutoff scale could then be around a few 
TeV, the analogy of the QCD scale, and there would be no hierarchy 
problem. \\ [-3mm]
  
An obvious experimental check of the technicolour model would be the 
presence of a heavy techni-rho $\rho_T^0$, being the spin-1 bound state 
partnering the spin-0 Higgs boson, in the same way as the spin-1 
$\rho^0$ partners the $\pi^0$. This could be searched for through decays 
$\rho_T^0 \to W_L^+W_L^-$, where the $W_L^\pm$ are 
longitudinally-polarised $W$ bosons. Needless to say, no such signals 
have been found at the LHC or other experiments. An even stronger 
argument against technicolor ideas came in 1990, when Peskin and 
Takeuchi \cite{Takeuchi} showed that the prediction of the so-called 
oblique parameter $S$ in technicolor models was much too large to be 
compatible with the value obtained from experimental measurements at the 
CERN LEP-1 collider. Since then, technicolor models have had few takers, 
except for a few persistent souls who have crafted modifications of the 
minimal technicolor model \cite{technimodern} to evade all experimental 
constraints. However, these models have been acquiring an increasingly 
baroque nature, just as modern GUT models have\footnote{The 
partly-pejorative term 'epicycles' is often applied to these efforts, 
which is not without justification. However, a sobering thought is that 
epicycles are just a two-dimensional version of the Fourier series, 
discovered many centuries after Ptolemy.}. \\ [-3mm]

\item {\it Extra Dimensions}: Extra dimensions had been introduced as a 
vehicle for unification of gravity with electromagnetism in the early 
twentieth century \cite{Kaluza}, but largely abandoned when it became 
clear that such ideas could not be verified experimentally. In the 
1970s, as gauge theories stated to become popular, extra dimensions 
became a mainstay of string theories \cite{string}, which visualised 
extra compact dimensions with a size around $M_P^{-1}$ and were not 
unduly worried about experimental verification. In 1998, a sensation was 
created when the trio of Arkani-Hamed, Dimopoulos and Dvali, or ADD for 
short, proposed a model \cite{ADD} where there would be extra dimensions 
of the size of a millimetre, with the SM fields (which don't seem to 
sense any extra dimensions) confined to a wafer-thin slice of 
four-dimensional space inside the higher-dimensional bulk\footnote{The 
thickness of this slice must be smaller that the smallest length scale 
to which the SM has curently been probed, i.e. about $10^{-17}$~cm}. 
This scenario was, in fact, nicely fulfilled by a topological defect, 
called a $D_3$-brane, predicted in Type-II string theories 
\cite{Antoniadis}, and this led to all such subspaces being dubbed {\it 
branes}, whether the model was embedded in a string theory or not. The 
ADD construction would immediately bring the effective Planck scale down 
to around a TeV, obviously removing the hierarchy problem. The model 
predicted novel spin-2 interactions from towers of Kaluza-Klein 
excitations of the graviton, which would lead to observable consequences 
in high energy experiments.  \\ [-3mm]

Soon, however, it became apparent that the ADD model suffers from its 
own hierarchy problem --- not in the mass of the Higgs boson, but in the 
size of the extra dimensions. Here too, quantum corrections were the 
problem, for they would tend to shrink these dimensions down to the 
smallest length scale in the theory, i.e. the Planck scale --- thereby 
reinstating the hierarchy problem {\it in toto}. To solve this problem, 
an ingenious construction was effected by Randall and Sundrum in 1999 
\cite{RS1}.  The Randall- Sundrum model, or RS model for short, 
envisages two branes, separated by a small distance not much larger than 
the Planck length, which communicate through gravitational interactions. 
By fine-tuning cosmological constants on the two branes and in the 
intervening bulk (which has to have the topology of a circle folded 
about one of its diameters \cite{Horava}, i.e. a 
$\mathbb{S}^1/\mathbb{Z}_2$ orbifold) RS obtained a solution for the 
gravitational field which falls off exponentially from one brane to the 
other. It is thus possible for gravity to have electroweak strength on 
one brane (dubbed the {\it Planck brane}) and its usual ultra-weak 
Newtonian strength on the other (dubbed the {\it TeV brane}).  The 
electroweak scale is the only scale in this model and all large numbers 
are generated from this by exponentials with small arguments. This 
theory predicts graviton Kaluza-Klein modes with masses and coupling 
constants of electroweak strength, which can be easily looked for as 
resonances in high energy collisions \cite{Davoudiasl}. This model also 
has an issue with the stability of the inter- brane distance, which was 
brilliantly solved by Goldberger and Wise \cite{Wise} through the 
introduction of a bulk scalar, which appears on the TeV brane as a 
Higgs-like particle, called the {\it radion}.  \\ [-3mm]

It is worth mentioning, in this context, a third model with an 
$\mathbb{S}^1/\mathbb{Z}_2$ orbifolded extra dimension --- not because 
it solves the hierarchy problem (it does not!), but because it is 
relevant for Higgs boson physics. This model places all the SM fields in 
the bulk, with, presumably some topological defects at the two ends of 
the extra dimension. In this model, called the Universal Extra 
Dimensions, or UED, scheme \cite{Appelquist}, all the SM fields have 
Kaluza-Klein excitations and the only new parameter in the theory is the 
radius of the bulk. \\ [-3mm]

All three extra-dimensional models are of {\it decoupling} nature, i.e. 
the observable effects can be made arbitrarily small by tuning some 
parameter -- generally the size of the extra dimension. Indeed, this has 
been forced upon us by the non-observation of any trace of extra 
dimensions. The negative aspect of this is that the effective Planck 
mass keeps rising, and thereby we have a {\it little} hierarchy problem, 
i.e. a ratio of order $10^4$ between the Higgs boson mass and the 
minimum (bulk) Planck scale. This is a slightly uncomfortable situation, 
but by no means as intolerable as having a cutoff at the GUT scale of 
$10^{16}$~GeV.  \\
\end{itemize}   
\vspace*{-0.2in} 
\item Models where the quadratic divergences in the Higgs self-energy 
cancel.
\begin{itemize}
\item {\it Supersymmetry}: In Wilson's pioneering work \cite{hierarchy}, 
it had been pointed out that there are symmetries protecting the fermion 
and gauge boson masses from quadratic divergences, but none protecting 
the scalar Higgs boson mass. If one could find such a symmetry, clearly 
there would be no quadratic divergence in $M_H^2$ and hence no hierarchy 
problem. As it happens, by the late 1970s, such a symmetry lay ready at 
hand. This was {\it supersymmetry}. As is well known, quantum mechanics 
assigns different roles to particles of spin-$\frac{1}{2}$ which are 
fermions and particles of integer spin which are bosons --- this being 
known as the {\it spin-statistics theorem}.  Supersymmetry, however, 
transcends this by allowing fermionic degrees of freedom to mix with 
bosonic degrees of freedom and vice versa. This revolutionary idea was 
first used in string theory, where the basic string excitations are all 
bosonic, and therefore one needs supersymmetry to generate fermionic 
degrees of freedom \cite{Ramond}. It was discovered by Wess and Zumino 
in 1974 \cite{Wess} that in fact the effect of this extra symmetry is to 
protect the Higgs boson mass from developing quadratic divergences, 
thereby removing the hierarchy problem. The Wess-Zumino model is the 
simplest supersymmetic model, being the supersymmetric version of a 
$\lambda\varphi^4$ theory. In 1976, Pierre Fayet was the first to 
develop a supersymmetric SM \cite{Fayet}, but a phenomenologically 
viable model came only with the work of Dimopoulos and Georgi in 
1981\cite{Dimo}. In all of these models, the Higgs boson mass remains 
free of quadratic divergences.  \\ [-3mm]

In the minimal supersymmetric SM, called the MSSM, and all of its 
different variations, every boson in the SM has a fermionic partner and 
every fermion in the SM has a bosonic partners. These additional fields 
are called {\it superpartners}. In the limit of exact supersymmetry, the 
SM particles and their respective superpartners have the same masses and 
coupling strengths. However, since no such superpartners have been 
observed, it must be assumed that they are too heavy to have been 
produced in experiments till date, i.e. supersymmetry is explicitly 
broken by large masses of the superpartners\footnote{and by some other 
operators besides.} Different patterns of this symmetry-breaking create 
a landscape of supersymmetric models, almost all of which have been 
diligently searched for at high energy experiments, including the LHC. 
Unfortunately, nothing has been seen as of date, as a result of which 
the lower bounds on superpartner masses are getting steadily pushed up. 
Once again, there is nothing wrong in principle with such heavy 
superpartners; however their failure to show up has led to a certain 
degree of despondency in the high energy community.  \\ [-3mm]

One important feature of the MSSM and its variants is that there 
necessarily has to be a second Higgs doublet. This has inspired a whole 
new field of investigation, which is described in the next section.  \\ 
[-3mm]

\item {\it Little Higgs Models}: An ingenious construction, whose 
proximate inspiration was no doubt extra-dimensional models, was 
proposed in 2008. However, this idea really goes back to the times when 
the pion was thought to be an elementary particle and the pion-nucleon 
Lagrangian had a spontaneously-broken global symmetry, of which the pion 
was a Goldstone boson --- or rather (since the pion gets a small mass 
through explicit symmetry-breaking), a pseudo-Goldstone boson. This was 
an elegant argument to keep the pion light, and the idea of little Higgs 
models was to explain the lightness of the Higgs boson using similar 
ideas. There are different versions of the model, but the essential 
features are captured by a version where the scalar sector has a local 
$\left[SU(2)\times U(1)\right]^2$ at high energies. This breaks 
spontaneously around 8--10~TeV to the SM group $SU(2)_L \times U(1)_Y$, 
and the Higgs boson is the Goldstone boson of this theory. This symmetry 
again breaks spontaneously to $U(1)_{\rm em}$ at the electroweak scale, 
and it is at that scale that the Higgs boson acquires a mass, becoming a 
pseudo-Goldstone boson. However, quantum corrections to this cannot be 
too large, since masslessness is guaranteed at the not-too-high scale 
8--10~TeV. In practice, the way this happens is that the one-loop 
corrections from the pair of gauge bosons corresponding to the two 
$SU(2)\times U(1)$ symmetries, and these will have opposite signs due to 
group theoretic factors which are induced by the unbroken symmetry. 
Therefore the Higgs boson mass is generated at two loops and is 
perturbatively small. In the so-called 'littlest Higgs model', the 
$\left[SU(2)\times U(1)\right]^2$ symmetry is obtained from the breaking 
chain $SU(5) \to SO(5) \to SU(3)_c\times\left[SU(2)\times 
U(1)\right]^2$. \\ [-3mm]
 
An immediate problem in such a model is that these are extra 
contributions to the $S$ and $T$ precision oblique parameters at LEP, 
and these have a non-decoupling component which makes them incompatible 
with the data, which are very close to the Standard Model predictions. 
If there are {\it two} sets of gauge bosons, however, we can assign a 
$\mathbb{Z}_2$ symmetry called $T$-parity, such that the SM particles 
have $T = 1$ and the new particles have $T = -1$. This would ensure a 
cancellation between the corresponding contributions to the oblique 
parameters, and indeed to any loop corrections, with any residual 
effects being due to the mass differences. An immediate consequence is 
that the lightest particle with $T = -1$ will not decay and can be a 
candidate for dark matter, just as the LSP is.
\end{itemize}
\end{enumerate} 
\vspace*{-0.2in}
A great deal of the research with bSM Higgs bosons has been in the 
context of one or other of these models. In fact, there was a time when 
supersymmetry was widely regarded as the 'standard' bSM theory, but that 
alas! is no longer the case.

We now come to the greatest failing of the SM, viz., its complete 
inability to explain most of the composition of the Universe. These 
mysterious components have been dubbed {\it dark matter} and {\it dark 
energy}, and the SM does not have any candidate for either.

Dark matter was postulated by Zwicky in 1937, when he could not 
reconcile the proper motion of galaxies in the Virgo super cluster with 
the mass inferred from observation of the luminous matter \cite{Zwicky}.  
Essentially the galaxies were moving at speeds higher than the escape 
velocity corresponding to this mass, and yet the system remains in a 
bound state, presumably under Newtonian gravity. The solution, according 
to Zwicky, was that the supercluster contains a great deal of matter 
which is not luminous, i.e. it does not radiate any energy in the 
electromagnetic spectrum, and is, therefore,'dark'. Nearly four decades 
later, Rubin \cite{Rubin} measured the transverse (rotational) velocity 
of stars in the neighbouring Andromeda galaxy and a few others. The 
expectation from Newtonian mechanics was that if the mass $M$ of the 
galaxy is concentrated in the central region, the transverse velocity of 
stars in the outer arms would fall off as $v(r) = \sqrt{G_NM/r}$.  Rubin 
found, however, that the measured velocities remain essentially constant 
as $r$ increases.  She correctly interpreted thus to mean that the 
galaxy is immersed in a ball of invisible --- dark --- matter, such that 
the mass inside a sphere of radius $r$ is $M(r) \propto r$. This 
immediately means that the dark matter distribution inside a galaxy is 
$\rho(r) \propto r^{-2}$, with a cusp at the centre, compatible with the 
existence of a black hole there. Since the 1970s, it has been shown that 
practically every galaxy, and cluster of galaxies shows similar 
behaviour. The existence of dark matter is also proved by a 
consideration of gravitational lensing phenomena. Moreover, there is an 
increasing body of evidence that the dark matter is particulate in 
nature and does not consist of compact objects like mini black holes 
\cite{Schramm}.

The problem with dark matter is that there is no particle in the SM 
which can be described as dark matter. Since dark matter is believed to 
be non-baryonic \cite{bullet} and neutral, the only possible candidates 
are neutrinos and the Higgs boson. The latter is ruled out because it is 
unstable, having a very short lifetime of around $1.6\times 10^{-22}$~s. 
Neutrinos, on the other hand, may be stable, but they are too light to 
satisfy the requirement that the dark matter has cooled to the observed 
temperature (2.73 K) of the Universe. In fact, the ideal dark matter 
candidate is a weakly-interacting massive particle (WIMP) which is 
stable, or at least has a lifetime greater than the age of the Universe 
\cite{Schramm}. This is simply not available in the SM.

The mystery that is dark matter pales in significance compared to what 
is called 'dark energy', which essentially means a positive cosmological 
constant. The cosmological constant $\Lambda$ was originally introduced 
by Einstein in 1917 to get a static Universe \cite{EinCosmo}. The best 
way to understand it \cite{WeinGR} is to write down the Robertson-Walker 
metric for a spherically-symmetric Universe
\begin{equation}
ds^2 \defeq dt^2 - R^2(t) \left( d\theta^2 + \sin^2\theta \ d\varphi^2 
\right)
\end{equation}    
and require the radial factor $R(t)$ to satisfy the Friedmann equation
\begin{equation}
\frac{\ddot{R}}{R} = - \frac{4\pi G_N}{3} \left( \rho + 3p \right) + 
\frac{\Lambda}{3}
\end{equation}
where $p$ and $\rho$ stand for pressure and density respectively 
(assumed uniform over cosmic scales) and $\Lambda$ is the cosmological 
constant. Assuming $p$ and $\rho$ are positive, as would be the case for 
both matter and energy, then $\Lambda = 0$ would lead to $\ddot{R} < 0$, 
i.e. a Universe which is expanding but at a decelerated rate. The truth 
is exactly the opposite -- it has been shown by a study of very distant 
supernovae that the Universe is, in fact expanding at a small {\it 
accelerated} rate \cite{Perlmutter}. This makes it imperative to make 
the right side of the Friedmann equation positive, and the obvious way 
to do this is to postulate a value $\Lambda > 0$ which will lead to a 
small positive residue on the right side.

Since the cosmological constant acts as a uniform matter density in the 
Universe, it is commonly referred to as {\it dark energy}. Studies of 
the cosmic microwave background indicate that the Universe is composed 
of $68\%$ of this mysterious material, about $27\%$ of dark matter, and 
the remaining $5\%$ of baryonic, i.e. SM matter. It would be easiest to 
identify the cosmological constant or dark energy density with the 
vacuum energy associated with the electroweak transition -- 
unfortunately, that leads to an overestimate by a factor of the order of 
$10^{60}$, leading to the so-called {\it cosmological constant problem}. 
Many solutions of this have been proposed, from modification of 
Newtonian dynamics (MOND) \cite{MOND} to the use of effective field 
theories to cancel the vacuum energy. There has also been a suggestion 
of a new kind of particle called {\it quintessence}, which on cosmic 
length scales, has a repulsive effect \cite{quintessence}, thereby 
changing the sign of the pressure term and making a nonzero cosmological 
constant unnecessary. These ideas are only peripheral to this article, 
becoming relevant only when the Higgs sector acts as a portal to these 
exotic particles through some kind of feeble coupling.

The signal failure of the first three runs of the LHC to find any 
evidence for physics beyond the Standard Model has brought to the fore 
an alternative paradigm. This is the development of bottom-up theories, 
i.e. those which retain the gauge and global symmetries of the SM and 
make minimal (or just beyond minimal) extensions to either the field 
content, or the specific interactions. Among the first type are 
extensions of the gauge boson sector (often associated with extra gauge 
symmetries of the simplest kind), extensions of the Higgs sector, 
extensions of the fermion sector, or addition of exotica such as 
leptoquarks and dileptons -- almost all leading to new interactions with 
the SM Higgs boson. There are also suggestions for modification of the 
SM interaction vertices by introducing anomalous terms, which can mostly 
be generated in effective field theories where there are new heavy 
fields whose interactions at a high scale can be integrated out leaving 
effective operators which then modify the SM vertices. In a combination 
of these, models of dark matter often assume the existence of dark 
particles, fermions or bosons, which interact with the SM fields through 
mediator fields, and this can also modify the Higgs boson interactions.

The USP of these bottom-up approaches is claimed to be their 
model-independent nature. However, while these models generally do not 
attempt to solve the basic problems with the SM itself, they often make 
assumptions about new fields and couplings which amount to a degree of 
model-building. A comprehensive description of these ideas would require 
a different article altogether. Hence, in the present work, we shall 
concentrate only on that part of these extensions which is relevant 
specifically to the Higgs sector.

\section{Expressionism: Extended Higgs Sectors} 

If we do not consider models which are derivable from some symmetry at a 
high energy scale, then there is just one guideline to the kind of extra 
scalars which can be added to the Standard Model. This comes from the 
so-called $\rho$ parameter (also called $T$ parameter), given by
\begin{equation}
\rho \defeq \frac{M_W^2}{M_Z^2 \cos^2\theta_W}
\end{equation}  
In the Standard Model, $\rho = 1$, and indeed its experimental value is 
\cite{PDG-rho}
\begin{equation}
\rho_{\rm exp} = 1.00038 \pm 0.00020
\label{eqn:rho-expt}
\end{equation}
In a model with several extra Higgs multiplets $\Phi_1, \Phi_2, \dots$ 
with different weak isospin and weak hypercharge assignments $(T_i,Y_i)$ 
where $i = 1,2,\dots$, the rho parameter (at tree-level) can be shown to 
have the form
\begin{equation}
\rho = \frac{2v^2 + \sum_i \left\{4T_i(T_i+1) - Y_i^2\right\} v_i^2} 
{2v^2 + \sum_i 2Y_i^2 v_i^2}
\end{equation}
where $v_i$ is the vev of the neutral component in the $i$-th multiplet. 
It is trivial to check that this remains unity if we add any number of 
scalar doublets ($T=\frac{1}{2},Y=\pm 1$) or singlets ($T = 0, Y = 0$), 
irrespective of the vacuum expectation values $v_i$. For other 
multiplets, the $v_i$ must be fine-tuned suitably. It is not surprising, 
therefore, that the most popular extensions of the Higgs sector are 
singlets and doublets. Some of the specific ideas are now discussed in 
more detail.

\subsection{\sl Extra Real Singlet Scalar} 

The simplest extension of the Higgs sector is to add on a real scalar 
which is a gauge singlet of $SU(2)_L \times U(1)$. An early exploration 
of this \cite{AKSR} in the wake of the top quark discovery tried to 
satisfy the Veltman condition of cancellation of quadratic divergences 
in the Higgs boson self-energy, but this led to the prediction of a 
Higgs boson mass above 300~GeV, which is now known to be unphysical. 
However, more recent analyses \cite{Dawson} take a less ambitious 
approach, confining themselves to studying the parameter space and 
discovery limits of such models. A singlet scalar does not, by 
definition, have gauge interactions and hence its decays are solely 
governed by the dynamics of the (extended) Higgs sector. This makes it, 
in many models, practically stable, making it an attractive candidate 
for dark matter \cite{singletDM}.

Denoting the real singlet by $\sigma$, the scalar potential of the 
theory in such models becomes
\begin{equation}
V(\Phi,\sigma) = -\mu^2 \Phi^\dagger \Phi + \lambda \left( \Phi^\dagger 
\Phi \right)^2 + \lambda_1 \sigma + \frac{1}{2} \lambda_2 \sigma^2 + 
\lambda_3 \sigma^3 + \lambda_4 \sigma^4 + \lambda'_1 \Phi^\dagger \Phi 
\sigma + \lambda'_2 \Phi^\dagger \Phi \sigma^2
\end{equation}
where $\Phi^\dagger \defeq \left( \begin{array}{cc} \varphi^- & 
\varphi^{0\ast} + \frac{v}{\sqrt{2}} \end{array} \right)$. Taking 
advantage of the singlet nature of the scalar $\sigma$, it is always 
possible to make a redefinition such that its vacuum expectation value 
vanishes, i.e. $\langle\sigma\rangle = 0$. Minimising the potential now 
leads to a mixing between the erstwhile Higgs boson and the singlet 
scalar, as follows
\begin{eqnarray}
h & = & \ \ {\rm Re}\, \varphi^0 \ \cos\alpha + \sigma \ \sin\alpha \nonumber \\
H & = & - {\rm Re}\, \varphi^0 \ \sin\alpha + \sigma \ \cos\alpha
\end{eqnarray}    
where $M_h = 125.4$~GeV is the mass of the elementary scalar discovered 
at the LHC and $v = 246.2$~GeV as in the SM. The free parameters of this 
theory can be taken to be \cite{Dawson}
\begin{displaymath}
M_H \qquad\qquad \sin\alpha \qquad\qquad \lambda_3 \qquad\qquad 
\lambda_4 \qquad\qquad \lambda'_2
\end{displaymath}
in terms of which
\begin{eqnarray}
\lambda & = & \frac{1}{2v^2} \left( M_h^2 \cos^2\alpha + M_H^2 
\sin^2\alpha \right) \nonumber \\
\lambda_2 & = & M_h^2 \sin^2\alpha + M_H^2 \cos^2\alpha - \lambda'_2 v^2 
\nonumber \\
\lambda'_1 & = & \frac{1}{2v} \left( M_h^2 - M_H^2 \right) \sin 2\alpha 
\label{eqn:singlet_parameters}
\end{eqnarray}
The mixing results will modify all the couplings of the $h$ and this 
will lead to testable consequences at experimental facilities. For 
example, if $M_h < 2M_H$, it is easy to see that the Higgs production 
cross-section will scale as $\cos^2\alpha$. Taking the experimental data 
from the LHC experiments, it is shown in Ref.~\cite{Dawson} that
\begin{equation}
\mid \!\sin\alpha \!\mid \, < \, 0.2
\label{eqn:singlet_angle}
\end{equation}
In case $M_h > 2M_H$, the decay channel $h \to HH$ will open up and the 
bound will then be dependent on the other parameters as well. There will 
also be a unitarity bound from the consideration of $h + h \to h + h$, 
which appears as
\begin{equation}
M_H^2 \leq \frac{16\pi v^2}{3} \csc^2\alpha - M_h^2 \cot^2\alpha  
\end{equation} 
Taken in conjunction with Eqn.~(\ref{eqn:singlet_angle}) this leads to 
an upper bound $M_H \lesssim 5$~TeV, which gets relaxed as $\sin\alpha 
\to 0$. Only about two-fifths of this range on $M_H$ would be accessible 
at the LHC.

In a variant of this idea, an extra $\mathbb{Z}_2$ symmetry $\sigma \to 
-\sigma$ is included in the theory. In this case, $\lambda_1 = \lambda_3 
= \lambda'_1$, but a glance at the third line of 
Eqn.~(\ref{eqn:singlet_parameters}) will show that this is too 
simplistic. In fact, in such models, we must set $\langle\sigma\rangle = 
v' \neq 0$ in order to get nontrivial results.
 
In a more exotic variant of this idea, a model with a $SU(2)_L \times 
U(1)_Y \times U(1)_{Y'}$ is considered, where there is an extra $Z'$ 
boson, which is massless even when the electroweak symmetry is broken. 
Since there is no evidence for such a massless gauge boson, it must 
acquire mass somehow. One way to do this is to break the extra 
$U(1)_{Y'}$ symmetry spontaneously by introducing some extra Higgs 
multiplet which develops a vacuum expectation value of its own. Another 
possibility, which has been investigated recently \cite{Radhika}, is to 
add a St\"uckelberg-type real scalar $\sigma$ (which transforms under 
the adjoint representation of the gauge group, and is therefore not a 
singlet {\it per se}) preserving the gauge symmetry, but giving a mass 
to the $Z'$. This model is highly predictive, since the $U(1)_{Y'}$ 
symmetry forbids many operators which would otherwise have been present.
  
\subsection{\sl Two Extra Real Singlets} 

If one can add on one real scalar singlet, why not two? This Two Real 
Singlet Model (TRSM) \cite{2singlet} is equivalent to adding a single 
complex scalar doublet (CxSM) \cite{zinglet}. The main motivation of 
such models is to obtain a viable dark matter candidate in one of the 
physical scalars which result when these extra scalars mix among 
themselves and with the SM scalar. The basic formalism is not too 
different from a single real scalar. If we have two extra real scalars 
$\sigma$ and $\sigma'$, the most general scalar potential becomes very 
messy. It is more usual, therefore, to assume an additional 
$\mathbb{Z}_2 \times \mathbb{Z}'_2$ symmetry under transformations of 
the form
\begin{eqnarray}
\mathbb{Z}_2 & : & \sigma \to - \sigma \qquad\qquad\qquad  \sigma' \to \sigma'
\nonumber \\
\mathbb{Z}'_2 & : & \sigma \to \sigma \qquad\qquad\qquad \ \ \;  \sigma' \to -\sigma'
\end{eqnarray}  
where the SM fields are left invariant by both the transformations. The 
scalar potential now becomes
\begin{eqnarray}
V(\Phi,\sigma,\sigma') & = & -\mu^2 \Phi^\dagger \Phi + \lambda \left( 
\Phi^\dagger \Phi \right)^2 \\ && + \frac{1}{2} \lambda_2 \sigma^2 + 
\lambda_4 \sigma^4 + \frac{1}{2} \lambda'_2 \sigma'^2 + \lambda'_4 
\sigma'^4 + \eta_1 \Phi^\dagger \Phi \sigma^2 + \eta_2 \Phi^\dagger \Phi 
\sigma'^2 + \eta_3 \sigma^2 \sigma'^2 \nonumber
\end{eqnarray} 
where
\begin{equation}
\Phi = \left( \begin{array}{c} \varphi^+ \\ \frac{\varphi_0 + 
v}{\sqrt{2}} \end{array} \right) \qquad\qquad \sigma = \frac{s + 
w}{\sqrt{2}} \qquad\qquad \sigma' = \frac{s' + w'}{\sqrt{2}}
\end{equation}
where the vacuum expectation values $\langle\sigma\rangle = w$ and 
$\langle\sigma'\rangle = w'$ result in the spontaneous breaking of the 
$\mathbb{Z}_2 \times \mathbb{Z}'_2$ symmetry.

It is convenient to take all the quartic couplings, viz., $\lambda, 
\lambda_4, \lambda'_4, \eta_1, \eta_2, \eta_3$ as independent 
parameters. The remaining parameters are then given by the minimisation 
conditions as
\begin{eqnarray}
\mu^2 & = & \lambda v^2 + \frac{1}{2} \eta_1 w^2 +  \frac{1}{2} \eta_2 w'^2
\nonumber \\
\lambda_2 & = & \frac{1}{2} \lambda_4 w^2 + \frac{1}{4} \eta_1 v^2 + \frac{1}{4} \eta_3 w'^2 \nonumber \\
\lambda'_2 & = & \frac{1}{2} \lambda'_4 w'^2 + \frac{1}{4} \eta_2 v^2 + \frac{1}{4} \eta_3 w^2 
\end{eqnarray}
where we can set $v = 246.2$~GeV and choose $w, w'$ appropriately as two 
more free parameters of the theory. With these conditions, the potential 
can be expanded to obtain mass terms resulting in a mixing between the 
neutral scalar $\varphi_0$ of the SM and the two new scalars, of the 
form
\begin{equation}
\left( \begin{array}{c} h_1 \\ h_2 \\ h_3 \end{array} \right)
= \mathbb{O} 
\left( \begin{array}{c} \varphi_0 \\ \sigma \\ \sigma' \end{array} \right)
\end{equation} 
where the $h_i$ masses $M_i$ satisfy $M_1 \leq M_2 \leq M_3$ and the 
orthogonal mixing matrix $\mathbb{O}$ satisfies the equations
\begin{eqnarray}
\sum_{i=1}^3 M_i^2 O_{i1}^2 & = & 2\lambda v^2
\quad
\sum_{i=1}^3 M_i^2 O_{i2}^2 = 2\lambda_4 w^2
\quad
\sum_{i=1}^3 M_i^2 O_{i3}^2 = 2\lambda'_4 w'^2
\\
\sum_{i=1}^3 M_i^2 O_{i1}O_{i2} & = & 2\eta_1 vw
\quad
\!\!\! \sum_{i=1}^3 M_i^2 O_{i1}O_{i3} = 2\eta_2 vw'
\quad
\sum_{i=1}^3 M_i^2 O_{i2}O_{i3} = 2\eta_3 ww'
\nonumber 
\label{eqn:twosinglets-mixing}
\end{eqnarray}
It is simple to now parametrise the mixing matrix $\mathbb{O}$ in terms 
of 3 mixing angles $\theta_1, \theta_2, \theta_3$, or rather their sines 
$s_i = \sin\theta_i \ (i = 12,3)$ and take these as well as as the 3 
physical masses $M_1, M_2, M3$ to replace the 6 free parameters in the 
potential using Eqns.~(\ref{eqn:twosinglets-mixing}). We can also set 
one of the $M_i = 125.4$~GeV and $v = 246.2$~GeV. This leaves 7 free 
parameters --- 3 mixing angkes, 2 scalar masses and 2 vev's of the 
singlets. The potential and all the couplings of the model are then 
determined in terms of these, and phenomenological studies can be made.

There are mild unitarity bounds on this model, but the phenomenological 
bounds from electroweak precision tests and from high-energy collider 
searches can be fairly complicated, depending on which of the different 
channels of $h_i \to h_j + h_k$, where $ijk$ is some combination of 
$123$ is kinematically permitted. Detailed studies may be found in 
Ref.~\cite{2singlet}. What emerges is that the current data permit a 
range of parameter space where one of the physical scalars $h_i$ is 
coupled very lightly with matter. It can be produced by co-annihilation 
of two of its sister scalars $h_j + h_k \to h_i \ (j,k \neq i)$, but 
cannot decay back because of kinematic constraints, i.e. $M_i < M_j + 
M_k$. This gives it the stability to become a candidate for dark matter.
  
\subsection{\sl Two Higgs Doublet Models} 

While one can, in principle, go on adding more singlets scalars to the 
SM, there is not much motivation for going beyond two singlets. Instead, 
the bulk of efforts in this field has been in adding an extra doublet of 
scalars, leading to the so-called two-Higgs doublet models (2HDM).

The basic formalism resembles the SM or its singlet extensions. If there 
are two Higgs doublets $\Phi_1$ and $\Phi_2$ with identical gauge 
quantum numbers, the most general potential has the form
\begin{eqnarray}
V(\Phi_1,\Phi_2) & = & 
- \mu_1^2 \Phi_1^\dagger \Phi_1 
- \mu_2^2 \Phi_2^\dagger \Phi_2 
- \mu_3^2 \left( \Phi_1^\dagger \Phi_2 + {\rm H.c.} \right) \nonumber \\
&&  + \lambda_1 (\Phi_1^\dagger \Phi_1)^2 
+ \lambda_2 (\Phi_2^\dagger \Phi_2)^2
+ \lambda_3 (\Phi_1^\dagger \Phi_1) (\Phi_2^\dagger \Phi_2)
\nonumber \\
&& + \lambda_4 \left\{ (\Phi_1^\dagger \Phi_2)^2 + (\Phi_2^\dagger 
\Phi_1)^2 \right\} + \lambda_5 (\Phi_1^\dagger \Phi_2) (\Phi_2^\dagger 
\Phi_1)
\label{eqn:2HDM_potential} 
\end{eqnarray}
As usual, we write
\begin{equation}
\Phi_1 \defeq \left( \begin{array}{c} \varphi_1^+ \\  
\frac{1}{\sqrt{2}} (\eta_1^0 + i g_1^0 + v_1)
\end{array} \right)
\qquad\quad
\Phi_2 \defeq \left( \begin{array}{c} \varphi_2^+ \\ 
\frac{1}{\sqrt{2}} (\eta_2^0 + i g_2^0 + v_2) 
\end{array} \right)
\end{equation}
The minimisation conditions for this potential are
\begin{equation}
\mu_1^2 = 2\lambda_1 v_1,  \qquad
\mu_2^2 = 2\lambda_2 v_2, \qquad
\mu_3 = 0, \qquad
\lambda_5 = -2\left(\lambda_3 + \lambda_4\right) 
\end{equation}
Substituting these in the potential and expanding, we obtain mass terms 
of the form
\begin{eqnarray}
{\cal L}_{\rm m} & = & 
- \frac{1}{2}\left(3\lambda_3 + 8\lambda_4\right) 
\left(\begin{array}{cc} g_1^+ & g_2^+ \end{array} \right) 
\left(\begin{array}{cc} v_2^2 & - v_1 v_2 \\ - v_1 v_2 & v_1^2 
\end{array} \right) \left(\begin{array}{c} g_1^- \\ g_2^- \end{array} 
\right) \nonumber \\
&& \hspace*{0.8in} - \left(\begin{array}{cc} h_1^+ & h_2^+ \end{array} \right) 
\left(\begin{array}{cc} - \lambda_1 v_1^2 & \lambda_3 v_1 v_2 \\ 
\lambda_3 v_1 v_2 & (\lambda_1 - 2\lambda_2) v_2^2 \end{array} \right)
\left(\begin{array}{c} h_1^- \\ h_2^- \end{array} \right)  \nonumber \\
&& \hspace*{0.8in} + \lambda_3 \left(\begin{array}{cc} \varphi_1^+ & 
\varphi_2^+ \end{array} \right) \left(\begin{array}{cc} v_2^2 & - v_1 
v_2 \\ - v_1 v_2 & v_1^2 \end{array} \right) \left(\begin{array}{c} 
\varphi_1^- \\ \varphi_2^- \end{array} \right)
\end{eqnarray}
Diagonalising these mass matrices, we obtain the following mixed states 
as the neutral mass eigenstates of the theory
\begin{eqnarray}
h^0 & = & \eta_1^0 \cos\alpha + \eta_2^0 \sin\alpha \qquad\qquad
\ G^0   =   g_1^0 \cos\beta + g_2^0 \sin\beta \nonumber \\
H^0 & = & \!\!\!-\eta_1^0 \sin\alpha + \eta_2^0 \cos\alpha \qquad\qquad
A^0   =  \!\!\!-g_1^0 \sin\beta + g_2^0 \cos\beta 
\label{eqn:2HDM-neutralmix}
\end{eqnarray}
as well as the charged mass eigenstates
\begin{eqnarray}
G^+ & = &  \varphi_1^+ \cos\beta + \varphi_2^+ \sin\beta  \nonumber \\
H^+ & = & \!\!\!-\varphi_1^+ \sin\beta + \varphi_2^+ \cos\beta 
\label{eqn:2HDM-chargedmix}
\end{eqnarray} 
of these, the $G^0, G^\pm$ are the massless Goldstone bosons of the 
theory, which can be removed by a redefinition of the fields (which we 
call choosing the unitary gauge) and the remaining 5 are physical 
fields, viz. the charged Higgs bosons $H^\pm$ with mass $M_+$, the 
${\cal CP}$-odd neutral Higgs boson $A^0$ with mass $M_A$ and the two 
${\cal CP}$-even neutral Higgs bosons $h^0$ and $H^0$ with masses $M_h$ 
and $M_H$ respectively, where it is usual to identify $M_h = 125.4$~GeV. 
In the above equations, we have
\begin{equation}
v_1^2 + v_2^2 = v^2 = (246.2~{\rm GeV})^2 
\qquad\qquad\qquad
\tan\beta = \frac{v_2}{v_1}
\end{equation}    
while the physical masses $M_+$ and $M_A$ are given by
\begin{equation}
M_+^2 = \lambda_3 v^2  \qquad\qquad\qquad  M_A^2 = -\frac{1}{2}\left(3\lambda_3 + 8\lambda_4\right) v^2
\end{equation}
and $M_h$ and $M_H$ can be obtained from the roots of the secular equation
\begin{equation}
x^2 + \left\{\lambda_1(v_1^2 - v_2^2) + 2\lambda_2v_2^2\right\}\,x
-\left(\lambda_1^2 - 2\lambda_1 \lambda_2 + \lambda_3^3\right)v_1^2 v_2^2 = 0
\end{equation} 
while the mixing angle $\alpha$ satisfies
\begin{equation}
\tan 2\alpha = \frac{\lambda_3 \sin 2\beta}{2\lambda_2 \cos^2\beta - \lambda_1 \cos 2\beta}
\end{equation}  
The free parameters of the theory are $\lambda_1$, $\lambda_2$, 
$\lambda_3$, $\lambda_4$ and $\tan\beta$. It is common to trade the 
$\lambda_{1,2,3,4}$ for the physical masses, and take the free parameter 
set as $M_\pm$, $M_A$, $M_h$, $M_H$ and $\tan\beta$. Since $M_h$ is a 
known value, we are left with a 4-parameter theory.

The above description is incomplete because we have assumed that the 
scalar potential is ${\cal CP}$-conserving, for which there is no a 
priori justification. If we include ${\cal CP}$ violating terms, we will 
have one more coupling $\lambda_6$ and a phase $\xi$. Interested readers 
are referred to Ref.~\cite{Hunters} for more details.

However, a new complication arises because a doublet of scalars can have 
Yukawa couplings with sequential fermions and this could affect the 
flavour structure of the model. To see this, we go back to 
Eqns.~(\ref{eqn:gaugequark}) and (\ref{eqn:covariantquark}), from the 
expansion of which we can obtain the {\it neutral current} (nc) 
interactions
\begin{eqnarray}
S_{\rm int}^{\rm nc} & = & \int d^4x \ \frac{g}{2\cos\theta_W} \left[ 
  g_L^u \overline{U_L} \gamma^\mu U_L + g_R^u \overline{U_R} \gamma^\mu 
  U_R \right.  \nonumber \\
&& \left. \hspace*{0.9in} + g_L^d \overline{D_L} \gamma^\mu D_L + g_R^d 
\overline{D_R} \gamma^\mu D_R \right] Z_\mu
\end{eqnarray}
where $(1 - g_L^u)/2 = -g_R^u/2 = 1 + g_R^d = g_L^d = 2\sin^2\theta_W/3$ and 
as in Eqn.~(\ref{eqn:matrixform}) we have
\begin{equation}
U_{L,R} = \left( \begin{array}{c} u_{L,R} \\ c_{L,R} \\ t_{L,R} \end{array}
\right)
\qquad\qquad\qquad
D_{L,R} = \left( \begin{array}{c} d_{L,R} \\ s_{L,R} \\ b_{L,R} \end{array}
\right) 
\end{equation}
In the SM we can replace the unphysical fields $U_{L,R}$ and $D_{l,R}$ 
with the physical fields $U_{L,R}^0$ and $D_{L,R}^0$ defined in 
Eqn.~(\ref{eqn:biunitary}) to get
\begin{eqnarray}
S_{\rm int}^{\rm nc} & = & \int d^4x \ \frac{g}{2\cos\theta_W} \left[
  g_L^u \overline{U_L^0} \mathbb{V}_L^{(u)\dagger} \gamma^\mu \mathbb{V}_L^{(u)} U_L^0 
+ g_R^u \overline{U_R^0} \mathbb{V}_R^{(u)\dagger} \gamma^\mu \mathbb{V}_R^{(u)} U_R^0
+ (u \leftrightarrow d) \right] Z_\mu \nonumber \\
& = & \int d^4x \ \frac{g}{2\cos\theta_W} \left[
  g_L^u \overline{U_L^0}  \gamma^\mu \left(\mathbb{V}_L^{(u)\dagger}\mathbb{V}_L^{(u)}\right) U_L^0 
+ g_R^u \overline{U_R^0} \gamma^\mu \left(\mathbb{V}_R^{(u)\dagger}\mathbb{V}_R^{(u)}\right) U_R^0
+ (u \leftrightarrow d) \right] Z_\mu \nonumber \\
& = & \int d^4x \ \frac{g}{2\cos\theta_W} \left[
  g_L^u \overline{U_L^0} \gamma^\mu U_L^0 
+ g_R^u \overline{U_R^0} \gamma^\mu U_R^0
+ (u \leftrightarrow d) \right] Z_\mu
\label{eqn:neutralcurrent}
\end{eqnarray} 
since $\mathbb{V}_L^{(u)\dagger}\mathbb{V}_L^{(u)} = 
\mathbb{V}_R^{(u)\dagger}\mathbb{V}_L^{(u)} = \mathbb{I}$. Expanding 
this, we obtain the explicit interactions
\begin{eqnarray}
S_{\rm int}^{\rm nc} & = & \int d^4x \ \frac{g}{2\cos\theta_W} \left[
  g_L^u \overline{u_L^0} \gamma^\mu u_L^0 + g_R^u \overline{u_R^0} \gamma^\mu u_R^0
\right. \nonumber \\
&& \hspace{1.02in} + g_L^u \overline{c_L^0} \gamma^\mu c_L^0 + g_R^u \overline{c_R^0} \gamma^\mu c_R^0
\nonumber \\
&& \left. \hspace{1.02in} + g_L^u \overline{t_L^0} \gamma^\mu t_L^0 + g_R^u \overline{t_R^0} \gamma^\mu c_t^0  + (u \leftrightarrow d) \right] Z_\mu  
\end{eqnarray}
Unlike the charged current interactions, the neutral currents connect 
only quarks and antiquarks of the same flavour, i.e. flavour-changing 
neutral currents (FCNC) are absent in the tree-level action of the gauge 
sector. This is a strong prediction of the SM which is backed up by a 
massive amount of experimental data.

When we consider the Yukawa couplings of the Higgs boson, we note that 
when the quarks are in the unphysical (gauge) basis, it has the form of 
Eqn.~(\ref{eqn:mixedYukawa}), which can be rewritten as
\begin{equation}
S_{\rm Yuk}^{(H)} = \int d^4x \ \left[ \; \overline{U_L} \, \mathbb{Y}^{(u)} \, U_R + \overline{D_L} \, \mathbb{Y}^{(d)} \, D_R + {\rm H.c.} \right] H^0
\label{eqn:matrixukawa}
\end{equation}
where the Yukawa couplings are given in Eqn.~(\ref{eqn:massYukawa3d}). 
Writing this in terms of the physical basis, we get
\begin{eqnarray}
S_{\rm Yuk}^{(H)} & = & \int d^4x \ \frac{\sqrt{2}}{v} \left[ \; \overline{U_L} \, \mathbb{M}^{(u)} \, U_R + \overline{D_L} \, \mathbb{M}^{(d)} \, D_R + {\rm H.c.} \right] H^0  \\
& = & \int d^4x \ \frac{\sqrt{2}}{v} \left[ \; \overline{U_L^0} \mathbb{V}_L^{(u)\dagger} \, \mathbb{M}^{(u)} \, \mathbb{V}_L^{(u)} U_R^0 + \overline{D_L^0} \mathbb{V}_R^{(d)\dagger} \, \mathbb{M}^{(d)} \, \mathbb{V}_R^{(d)} D_R^0 + {\rm H.c.} \right] H^0 \nonumber
\label{eqn:HYukawa_partial}
\end{eqnarray}
However, we have, following Eqn.~(\ref{eqn:biunitary}),
\begin{equation}
\mathbb{V}_L^{(u)\dagger} \, \mathbb{M}^{(u)} \, \mathbb{V}_L^{(u)} = 
\left( \begin{array}{ccc} m_u & 0 & 0 \\ 0 & m_c & 0 \\ 0 & 0 & m_t 
\end{array} \right) 
\quad
\mathbb{V}_L^{(d)\dagger} \, \mathbb{M}^{(d)} \, \mathbb{V}_L^{(d)} = 
\left( \begin{array}{ccc} m_d & 0 & 0 \\ 0 & m_s & 0 \\ 0 & 0 & m_b 
\end{array} \right) 
\end{equation}  
using which Eqn.~(\ref{eqn:HYukawa_partial}) reduces to the simple form
\begin{eqnarray}
S_{\rm Yuk}^{(H)} & = & \int d^4x \ \frac{\sqrt{2}}{v} \left[
 m_u \, \overline{u_L^0} u_R^0 + m_c \, \overline{c_L^0} c_R^0 
 + m_t \, \overline{t_L^0} t_R^0 \right. \nonumber \\ 
&& \hspace*{0.65in}  
\left. + m_d \, \overline{d_L^0} d_R^0 + m_s \, \overline{s_L^0} s_R^0 + m_b \, \overline{b_L^0} b_R^0 + {\rm H.c.} \right] H^0 
\end{eqnarray} 
which shows that there are no tree-level FCNCs in the scalar sector either. 

Stated in words, since the matrix of Yukawa coupkings in either sector 
is proportional to the mass matrix, the basis in which the mass matrices 
are diagonalised will also have diagonal Yukawa couplimngs, and hence no 
FCNCs.  This result --- that the SM has no tree-level FCNCs in either 
gauge or Yukawa sectors --- was discovered as early as 1977 and is 
generally known as the {\it Glashow-Weinberg theorem} \cite{GlasWein}.

Let us now consider a model with two similar Higgs doublets 
\begin{equation}
\Phi_1 \defeq \left( \begin{array}{c} \varphi_1^+ \\ \varphi_1^0 + \frac{v_1}{\sqrt{2}}\end{array} \right)
\qquad\qquad\qquad
\Phi_2 \defeq \left( \begin{array}{c} \varphi_2^+ \\ \varphi_2^0 + \frac{v_2}{\sqrt{2}}\end{array} \right)
\end{equation}
where the vacuum expectation values $v_1$ and $v_2$ are, in general, 
different. In this case, the Yukawa part of the SM will have to be 
changed from that in Eqn.~(\ref{eqn:mixedYukawa}) to
\begin{eqnarray}
{\cal S}_{\rm Yuk}^{(q)} & = & \int d^4x \ \sum_{a=1}^2 \sum_{b=1}^2 \left[ 
y_{ab}^{(u,1)} \; \overline{Q_L^{(a)}}(x) \, \widetilde{\Phi_1}(x) \, u_R^{(b)}(x) + y_{ab}^{(d,1)} \; \overline{Q_L^{(a)}}(x) \, \Phi_1(x) \, d_R^{(b)}(x)  
\right.  \\
&& \left. \hspace{0.85in} 
+ y_{ab}^{(u,2)} \; \overline{Q_L^{(a)}}(x) \, \widetilde{\Phi_2}(x) \, u_R^{(b)}(x) + y_{ab}^{(d,2)} \; \overline{Q_L^{(a)}}(x) \, \Phi_2(x) \, d_R^{(b)}(x)
+ {\rm H.c.} \right] \nonumber
\label{eqn:mixedYukawa1} 
\end{eqnarray}
Expanding this leads to the mass terms
\begin{eqnarray}
S_m & = & \int d^4x \ \sum_{a=1}^2 \sum_{b=1}^2 \left[ 
\left(\frac{y_{ab}^{(u,1)}v_1}{\sqrt{2}} + \frac{y_{ab}^{(u,2)}v_2}{\sqrt{2}}\right) \overline{u_L} u_R \right. \nonumber \\
&& \hspace*{0.85in} + \left. \left(\frac{y_{ab}^{(d,1)}v_1}{\sqrt{2}} + \frac{y_{ab}^{(d,2)}v_2}{\sqrt{2}}\right) \overline{d_L} d_R + {\rm H.c.} \right]
\end{eqnarray}
which can be written in the matrix form of Eqn.~(\ref{eqn:matrixform}) 
where the mass matrices are now
\begin{equation}
\mathbb{M}^{(u)}_{ab} \defeq \frac{y_{ab}^{(u,1)}v_1}{\sqrt{2}} 
+ \frac{y_{ab}^{(u,2)}v_2}{\sqrt{2}}
\qquad\quad
\mathbb{M}^{(d)}_{ab} \defeq \frac{y_{ab}^{(d,1)}v_2}{\sqrt{2}}
+ \frac{y_{ab}^{(d,2)}v_2}{\sqrt{2}}
\label{eqn:massYukawa3d2HDM}
\end{equation}
We can still diagonalise these mass matrices by bi-unitary 
transformations $\grave{\rm a}$ la Eqn.~(\ref{eqn:biunitary}) and this 
will lead to a charge current interaction with a CKM matrix and no FCNC 
in the gauge neutral current. However, the mass matrices are no longer 
proportional to any of the Yukawa couplings individually, but to a 
linear combination of them. Therefore if $\mathbb{V}_L^{(u)\dagger} \, 
\mathbb{M}^{(u)} \, \mathbb{V}_L^{(u)}$ is diagonal, there is no 
guarantee that $\mathbb{V}_L^{(u)\dagger} \, \mathbb{Y}_1^{(u)} \, 
\mathbb{V}_L^{(u)}$ or $\mathbb{V}_L^{(u)\dagger} \, \mathbb{Y}_2^{(u)} 
\, \mathbb{V}_L^{(u)}$ will individually be diagonal, unless indeed 
$\mathbb{Y}_1^{(u)}$ and $\mathbb{Y}_2^{(u)}$ are proportional to each 
other. Similar considerations apply to the case of $\mathbb{M}^{(d)}$ 
and its components $\mathbb{Y}_1^{(d)}$ and $\mathbb{Y}_2^{(d)}$. As a 
result, the Yukawa couplings of both $\varphi_1^0$ and $\varphi_2^0$ 
will, in general, have off-diagonal elements, leading to FCNC in the 
tree-level action, i.e. a violation of the Glashow-Weinberg Theorem.

What are the different ways to have a 2HDM without violating the 
Glashow-Weiberg theorem? The simplest way, as mentioned above, is to 
make the $\mathbb{Y}_1^{(u,d)}$ and $\mathbb{Y}_2^{(u,d)}$ proportional 
to each other. However, it has been shown by Pich and Tuzon \cite{Pich} 
that this proportionality breaks down under radiative corrections. On 
the other hand, we have an ingenious work due to Botella, Cornet-Gomez 
and Nebot \cite{Batela} where the $\mathbb{Y}_1^{(u,d)}$ and 
$\mathbb{Y}_2^{(u,d)}$ are constructed in such a way that they are not 
proportional and yet {\it simultaneously} diagonalisable. However, 
though the mathematical sleight-of-hand involved in this construction is 
to be admired, the constructed mass matrices have many textures (i.e. 
zero elements) and it would require a rash of ad hoc assumptions to 
justify them theoretically, apart from the difficulty of maintaining the 
textures as zero at all energy scales.

If we discount the above ideas, then we must end up with a model in 
which the mass matrices have only one Yukawa matrix term, for then it 
will become similar to the SM case with simultaneous diagonalisation of 
the mass and Yukawa matrices. There are 4 ways to achieve this and these 
are enlisted below.
\vspace*{-0.2in}
\begin{itemize}
\item {\sf Type I models} \medskip

Here the $\Phi_1$ is {\it fermiophobic} and does not couple to the 
quarks or leptons; all the Yukawa couplings are those of the $\Phi_2$, 
and hence the $\mathbb{Y}_2^{(u,d)}$ get diagonalised with the masses 
just as in the SM. The Yukawa sector would be of the form
\begin{eqnarray}
{\cal S}_{\rm Yuk}^{(q)} & = & \int d^4x \ \sum_{a=1}^2 \sum_{b=1}^2 \left[ 
 y_{ab}^{(u)} \; \overline{Q_L^{(a)}}(x) \, \widetilde{\Phi_2}(x) \, u_R^{(b)}(x)
 \right. \nonumber \\
&& \hspace*{0.85in} \left. + y_{ab}^{(d)} \; \overline{Q_L^{(a)}}(x) \, \Phi_2(x) \, d_R^{(b)}(x)
+ {\rm H.c.} \right] 
\label{eqn:2HDM-Type1} 
\end{eqnarray}
To make the $\Phi_1$ fermiophobic, all that we need to do is to assume a 
$\mathbb{Z}_2$ symmetry $\Phi_1 \to - \Phi_1$ which leaves all the other 
fields invariant. All terms with an odd number of $\Phi_1$'s will then 
be forbidden, but the $\Phi_1$ will still couple to the gauge bosons and 
the other scalar doublet $\Phi_2$ of course, which would lead to 
phenomenological signals.

Type I models have been studied phenomenologically but not in the same 
extensive way as the Type II models (see below). However, an interesting 
variant, the so-called {\it inert doublet} model, where the $\Phi_1$ 
does not acquire a vacuum expectation value, has an excellent dark 
matter candidate in the heavier ${\cal CP}$-even scalar $H^0$, which has 
been extensively studied in the literature \cite{inert}.

\item{\sf Type II models} \medskip

Here the two scalars $\Phi_1$ and $\Phi_2$ carry opposite hypercharges 
$Y = -1$ and $Y = +1$ respectively. Gauge invariance then demands that 
the $\Phi_1$ couple only to $u$-type quarks and the $\Phi_2$ couple only 
to $d$-type quarks, i.e the Yukawa interaction would have the form
\begin{eqnarray}
{\cal S}_{\rm Yuk}^{(q)} & = & \int d^4x \ \sum_{a=1}^2 \sum_{b=1}^2 \left[ 
 y_{ab}^{(u)} \; \overline{Q_L^{(a)}}(x) \, \Phi_1(x) \, u_R^{(b)}(x) \right. \nonumber \\ 
&& \hspace*{0.85in} \left. + y_{ab}^{(d)} \; \overline{Q_L^{(a)}}(x) \, \Phi_2(x) \, d_R^{(b)}(x) + {\rm H.c.} \right] 
\label{eqn:2HDM-Type2} 
\end{eqnarray}
where it may be noted that the doublets $\Phi_1$ and $\Phi_2$ will have the explicit forms
\begin{equation}
\Phi_1 \defeq \left( \begin{array}{c} \frac{1}{\sqrt{2}} (\eta_1^0 + i g_1^0 + v_1) \\  
\varphi_1^-
\end{array} \right)
\qquad\quad
\Phi_2 \defeq \left( \begin{array}{c} \varphi_2^+ \\ 
\frac{1}{\sqrt{2}} (\eta_2^0 + i g_2^0 + v_2) 
\end{array} \right)
\end{equation}  
and it is not necessary to introduce the charge-conjugated doublets 
$\widetilde{\Phi}_{1,2}(x)$ at all. Such a condition arises naturally in 
the MSSM (see below), and hence this models is said to be {\it 
MMSM-like}.

It is possible to obtain a model of this form by imposing {\it two} 
$\mathbb{Z}_2$ symmetries,, viz.
\begin{eqnarray}
\Phi_1 \to - \Phi_1  \qquad\qquad\qquad  \Phi_2 \to - \Phi_2 \nonumber \\
u_R \to - u_R \qquad\qquad\qquad  d_R \to - d_R 
\end{eqnarray} 
where each keeps all the other fields invariant. These will then prevent 
any operators which have $\Phi_1$ with $d_R$ and $\Phi_2$ with $u_R$.

\item{\sf Type X models} \medskip

Till now, we have not brought the leptons into the discussion, since 
flavour-violation in the lepton sector (if at all) is a very small 
effect depending on neutrino masses. However, the charged leptons 
($e,\mu,\tau$) do get masses from their Yukawa couplings. If the leptons 
are brought into play, we get the so-called Type X (or Type III) or {\it 
leptophilic} model, where the $\Phi_2$ couples only to quarks (as in 
Type 1 models), and the $\Phi_1$, instead of decoupling from fermions 
altogether, couples to the leptons. The Yukawa sector of the model now 
looks like
\begin{eqnarray}
{\cal S}_{\rm Yuk}^{(q)} & = & \int d^4x \ \sum_{a=1}^2 \sum_{b=1}^2 \left[ 
 y_{ab}^{(u)} \; \overline{Q_L^{(a)}}(x) \, \widetilde{\Phi_2}(x) \, u_R^{(b)}(x)  \right.
\\ 
&& \left. + y_{ab}^{(d)} \; \overline{Q_L^{(a)}}(x) \, \Phi_2(x) \, d_R^{(b)}(x) + y_{ab}^{(\ell)} \; \overline{L_L^{(a)}}(x) \, \Phi_1(x) \, e_R^{(b)}(x) + {\rm H.c.} \right] \nonumber 
\label{eqn:2HDM-TypeX} 
\end{eqnarray}
and it is possible to achieve this by the single $\mathbb{Z}_2$ symmetry
\begin{eqnarray}
\Phi_1 & \to & - \Phi_1 \nonumber \\
e_R & \to & - e_R 
\end{eqnarray}
Clearly in this case, each of the Yukawa matrices will be proportional 
to a separate mass matrix, and these will all get diagonalised in 
tandem.

\item{\sf Type Y models} \medskip

Of course, the roles of the $\Phi_1$ and the $\Phi_2$ can always be 
flipped; this makes little or no difference in the Type 1 and Type 2 
cases. However, the leptonic $\mathbb{Z}_2$ induces different behaviour, 
and hence we have a fourth model, which is called Type Y (or Type IV) or 
{\it flipped leptophilic}, with the discrete symmetry
\begin{eqnarray}
\Phi_2 & \to & - \Phi_2 \nonumber \\
e_R & \to & - e_R 
\end{eqnarray} 
and the Yukawa interactions have the form 
\begin{eqnarray}
{\cal S}_{\rm Yuk}^{(q)} & = & \int d^4x \ \sum_{a=1}^2 \sum_{b=1}^2 \left[ 
 y_{ab}^{(u)} \; \overline{Q_L^{(a)}}(x) \, \widetilde{\Phi_1}(x) \, u_R^{(b)}(x)  \right.
 \\ 
&& \left.  + y_{ab}^{(d)} \; \overline{Q_L^{(a)}}(x) \, \Phi_1(x) \, d_R^{(b)}(x) + y_{ab}^{(\ell)} \; \overline{L_L^{(a)}}(x) \, \Phi_2(x) \, e_R^{(b)}(x) + {\rm H.c.} \right] \nonumber
\label{eqn:2HDM-TypeY} 
\end{eqnarray}
\end{itemize}  
As we have just seen, the fact that the two doublets transform 
differently under these discrete symmetries profoundly affects the 
Yukawa sector, but it hardly changes the scalar potential in 
Eqn.~(\ref{eqn:2HDM_potential}), except to require $\mu_3 = 0$, which is 
anyway set by the minimisation condition. Therefore, the Goldstone 
bosons and physical scalar states will remain the same in all these 
versions of the 2HDM. The actual phenomenology will, of course, be 
different, and there exists an extensive literature on the subject 
\cite{2HDM-12XY}.

\subsection{\sl MSSM Higgs Sector} 

Supersymmetry, the much sought-after symmetry which mixes boson and 
fermion states, has been a mainstay of theories and experimental 
searches beyond the Standard Model, and it still commands the largest 
proportion of searches at the LHC. Though the literature on this is 
vast, a short summary of the phenomenological motivations of the {\it 
minimal} supersymmetric Standard Model (MSSM) and its so-called 
'constrained' version (cMSSM), together with the theoretical and 
experimental constraints on it may be found in Ref.~\cite{How}.

One of the basic features of the MSSM -- and indeed of any 
supersymmetric model -- is the fact that it always calls for two Higgs 
doublets. There are two paths to understanding this. One, which may be 
called the high road, notes that if we have a single Higgs doublet 
$\Phi$, then we require its charge conjugate $\widetilde{\Phi}$ (which 
includes a complex conjugation) to give mass to both $u$ and $d$ types 
of quarks. In a supersymmetric model, these would have to be embedded in 
superfields $\hat{\Phi}$ and $\hat{\Phi}^\dagger$ and then incorporated 
into the superpotential to get the interaction Lagrangian. However, it 
is known that the superpotential will not give a Lagrangian invariant 
under supersymmetry unless it is {\it holomorphic} in the superfields, 
i.e. once cannot use the $\hat{\Phi}^\dagger$. Its place must be taken 
by a different superfield, with opposite hypercharge, and thus there 
will be two scalar doublets, just as in the Type II models described 
above. A more pedestrian argument, which may be called the low road, 
notes that every Higgs boson will have a fermionic partner, called the 
{\it higgsino}. Since the Higgs boson couples to $Z$ bosons, so will the 
higgsino, and being a fermion, we will get an anomaly proportional to 
the hypercharge $Y$. Since all the other anomalies cancel out (see 
Eqn.~(\ref{eqn:anomalycancellation})), this will leave a residual 
anomaly. There is no help for it, then, other than to introduce another 
Higgs doublet, with opposite hypercharge $-Y$, so that these 
higgsino-induced anomalies mutually cancel out.

Moreover, the requirement that the action be invariant under 
supersymmetric transformations imposes stringent restrictions on the 
scalar potential of the theory. In particular, the quartic couplings get 
related to the gauge couplimgs and are no longer free parameters. The 
potential now has the form \cite{Hunters}
\begin{eqnarray}
V(\Phi_1,\Phi_2) & = & 
-\mu_1^2 \Phi_1^\dagger \Phi_1 -\mu_2^2 \Phi_2^\dagger \Phi_2
- \mu_3^2 \left( \Phi_1 \times \Phi_2 + {\rm H.c.} \right)  \nonumber \\
&& + \frac{1}{8} (g^2 + g'^2) \left( \Phi_1^\dagger \Phi_1 - \Phi_2^\dagger \Phi_2 \right)^2 + \frac{1}{2} g^2 \left(\Phi_1^\dagger \Phi_2 \right)^2 
\label{eqn:MSSMpotential} 
\end{eqnarray}   
with
\begin{equation}
\Phi_1 = \left( \begin{array}{l} \Phi_1^1 = \frac{1}{\sqrt{2}} (\eta_1^0 + i g_1^0 + v_1) \\  
\Phi_1^2 = \varphi_1^-
\end{array} \right)
\quad
\Phi_2 = \left( \begin{array}{l} \Phi_2^1 = \varphi_2^+ \\ 
\Phi_2^2 = \frac{1}{\sqrt{2}} (\eta_2^0 + i g_2^0 + v_2) 
\end{array} \right)
\end{equation}    
and $\Phi_1 \times \Phi_2 = \Phi_1^1 \Phi_2^2 - \Phi_1^2 \Phi_2^1$, 
which is a $SU(2)$-invariant construction. There are only three free 
parameters in this potential, viz., $\mu_1$, $\mu_2$ and $\mu_3$. 
Imposition of the minimisation conditions leads to
\begin{equation}
\mu_3^2 = -\mu_1^2 \frac{v_1}{v_2} = - \mu_2^2 \frac{v_2}{v_1}
\end{equation}  
There is thus, just one free parameter in the potential apart from the 
ratio $\tan\beta = v_2/v_1$, i.e., just 2 free parameters in the MSSM 
Higgs sector. This makes it highly predictive and leads to a very 
important constraint on the mass of the lighter ${\cal CP}$-even scalar.

Expanding the potential and obtaining the mass matrices leads to the 
same mixing pattern as in the general 2HDM, i.e. 
Eqns.~(\ref{eqn:2HDM-neutralmix}) and (\ref{eqn:2HDM-chargedmix}) where, 
it is now common to choose $M_A$ and $\tan\beta$ as the free parameters. 
The remaining parameters are now given by \cite{Hunters}
\begin{eqnarray}
M_+^2 & = & M_A^2 + M_W^2 \nonumber \\
M_h^2 & = & \frac{1}{2} \left[ M_A^2 + M_Z^2 - \sqrt{(M_A^2 + M_Z^2)^2 - 4M_Z^2 M_Z^2 \cos^2 2\beta} \right] \nonumber \\
M_H^2 & = & \frac{1}{2} \left[ M_A^2 + M_Z^2 + \sqrt{(M_A^2 + M_Z^2)^2 - 4M_Z^2 M_Z^2 \cos^2 2\beta} \right] \nonumber \\
\tan 2\alpha & = & \tan 2\beta \ \frac{M_A^2 + M_Z^2}{M_A^2 - M_Z^2}
\end{eqnarray}
Allowing $M_A$ and $\tan\beta$ to range freely, this leads to some 
interesting inequalities, viz.,
\begin{equation}
M_+ \geq M_W \qquad\qquad M_h \leq M_Z \qquad\qquad M_H \geq M_Z 
\qquad\qquad M_A > M_h
\end{equation}
In particular, the constraint $M_h < M_Z$ would immediately rule out the 
MSSM (since, after all, $M_h \simeq 125.4$~GeV), had it not been for the 
fact that radiative corrections drive $M_h$ to higher values. However, 
there is a limit beyond which perturbative processes cannot drive it and 
hence, we have an upper limit around 135~GeV \cite{Higgsmass}, and this 
can be extended by maximum 8-10~GeV even when further extensions of the 
MSSM are taken into account \cite{Kola}. If, indeed, the Higgs boson had 
been found with a mass above this limit, e.g., at 150~GeV, then the MSSM 
and its more restrictive variants would have been definitively ruled 
out. As it happens, however, the mass of the discovered particle lies 
squarely within the permitted region. This is not, of course, a proof of 
the existence of supersymmetry (specifically the MSSM), but just a 
tantalising hint that it may exist. Not until one discovers one or more 
of the supersymmetric partners can anything definite be said on this 
account.

The MSSM story is not yet complete, however, for these Higgs scalars 
will have fermionic partners called the {\it higgsinos'}. When the 
electroweak symmetry breaks sponntaneously, these will mix with the 
fermionic partners of the gauge bosons -- the {\it gauginos} -- to 
produce physical states. Thus there will be mixing with the 'Wino' 
states $\widetilde{W}^\pm, \widetilde{H}^\pm$ to produce a pair of 
charged fermions $\widetilde{\chi}_1^\pm$ and $\widetilde{\chi}_2^\pm$ 
called the {\it charginos}. There will also be mixing of the 'photino' 
$\widetilde{\gamma}$ and 'Zino' $\widetilde{Z}$ states with the neutral 
Higgsinos $\widetilde{h}_1^0$ and $\widetilde{h}_2^0$ to produce four 
neutral fermions $\widetilde{\chi}_i^0 \ {i=1,2,3,4}$ called the {\it 
neutralinos}. These physical states are generally labelled in order of 
increasing mass, and the lightest neutralino $\widetilde{\chi}_1^0$ is 
generally the lightest supersymmetric particle (LSP) and, in a whole 
class of supersymmetric models, a prime candidate for dark matter. Since 
the parameters of the Higgs sector, especially $\tan\beta$, enter into 
the couplings of these new fermions with SM fields, clearly the 
phenomenological constraints arising from these new particles will 
affect the Higgs potential as well. This, one cannot treat the Higgs 
sector of the MSSM --- or any supersymmetric model --- independently, 
even as a first approximation. It is for this reason, that most modern 
analyses of supersymmetry involve making a global fit to a slew of 
physical measurables, and constraints on the Higgs sector arise as a 
by-product.
          
\subsection{\sl NMSSM Higgs Sector} 

Since singlet scalars can be added to the SM, they can also be added to 
the MSSM, since the $\rho$ parameter remains the same. Thus, adding a 
complex singlet $\sigma(x) = s(x) + ia(x)$ in a way which preserves 
supersymmetry leads \cite{NMSSM} to extra terms added to the potential 
in Eqn.~(\ref{eqn:MSSMpotential})
\begin{eqnarray}
V_\sigma & = & \frac{1}{2}\mu_\sigma^2 \sigma^2 
+ \lambda_1^2 \sigma^*\sigma \left(\Phi^\dagger \Phi_1 + \Phi_2^\dagger \Phi_2 \right)  \\  
&& + \left(\lambda_1 \Phi_1 \times \Phi_2 - \nu_1 \sigma^2 \right)
  \left(\lambda_1 \Phi_2 \times \Phi_1 - \nu_1 \sigma^{*2} \right)
  - \left(\lambda_2 \sigma \Phi_1 \times \Phi_2 + \nu_2 \sigma^3 + {\rm H.c.}\right]  \nonumber
\end{eqnarray} 
where we note that making $\sigma$ real would violate supersymmetry. 
Once again, the physical states may be found by the usual process of 
minimisation of the potential, isolation of the mass term and 
diagonalisation of the mass matrices. These are omitted here in the 
interests of brevity, but may be found in all details in 
Ref.~\cite{NMSSM}. It suffices to say that there will be mixing of 
triplets of scalars
\begin{equation}
\left(h_1^0, h_2^0, s\right) \to \left(h^0,H^0,H'^0\right) 
\qquad\qquad
\left(g_1^0, g_2^0, a\right) \to \left(G^0,A^0,A'^0\right) 
\end{equation} 
with $3\times3$ mass matrices, while the mixing of charged scalars is 
the same as that in the MSSM. The mass spectrum of these states is as 
follows.
\begin{eqnarray}
M_+^2 & = & M^2 + M_W^2 - \frac{1}{2}\lambda_1 v^2  \nonumber \\
M_A^2 & \approx & g \left( \frac{1}{2} \nu_1 v^2 \sin 2\beta + 3 \nu_2 v_s \right)  
\nonumber \\
M_{A'}^2 & \approx & M^2 \left( 1 + \frac{v^2}{8\lambda_1 v_s^2} \sin^2 2\beta \right)  \nonumber \\
M_h^2 & \approx & v_s \left(4\nu_1 v_s^2  -\nu_2\right) \nonumber \\
M_H^2 & \approx & M_Z^2 \nonumber \\
M_{H'}^2 & \approx & M^2 \left( 1 + \frac{v^2}{32\lambda_1 v_s^2} \sin^2 4\beta \right) 
\end{eqnarray} 
where $v_s \defeq \langle\sigma\rangle$ and
\begin{equation}
M^2 = 2v_s \csc 2\beta \left( \lambda_1 \nu_1 v_s + \lambda_2 \right) \ .
\end{equation}
Like the MSSM, this model must rely on radiative corrections to get a 
neutral ${\cal CP}$-even scalar up to $125.4$~GeV. There are 6 free 
parameters, viz., $\mu_s$, $\lambda_1$, $\lambda_2$, $\nu_1$, $\nu_2$ 
and $\tan\beta$, which can be traded for the 5 masses $M_A$, $M_{A'}$, 
$M_H$, $M_{H'}$ and $\tan\beta$.

The fermionic partner of the $\sigma$, i.e. the $\widetilde{\sigma}$, 
which is known as the {\it singlino}, will naturally not have any gauge 
couplings. Its only coupling will be to the Higgs sector through the 
mixing matrices and this can be arranged to be very feeble. Hence, this 
singlino is an excellent candidate for dark matter, even for regions of 
parameter space where the neutralino is ruled out as a dark matter 
candidate. All other phenomenology of the model will be rather similar 
to that of the MSSM, and once again, constraints on the parameter space 
from chargino and neutralino searches will impact the Higgs sector as 
well.
     
\subsection{\sl Higgs Portal Models for Dark Matter} 

Since the Higgs sector is the least known sector of the Standard Model, 
it offers a fertile ground for speculation that what has been seen, 
viz., a single, seemingly elementary, scalar, is merely the tip of the 
iceberg and that it may offer the first glimpse into a whole new world 
with new fields and new interactions. More specifically, we can create 
models of dark matter and dark energy where the dark fields do not 
couple to any of the sectors of the SM except the Higgs sector. These 
are known as {\it Higgs portal} models, for then it is only by studying 
the Higgs boson that we can get any empirical insights into the dark 
sector.

Since there is plenty of room for speculation about dark matter (less so 
for dark energy), an extensive literature has grown up around Higgs 
portal models. A proper discussion would merit a small review in its own 
right. Here we merely list a few of the more exciting ideas.
\vspace*{-0.2in}
\begin{enumerate}
\item The idea of the Higgs sector as a portal to 'hidden' sectors was 
first proposed by Patt and Wilczek in 2006 \cite{Patt}. In this article 
they mooted the idea of a 'hidden' copy of the Standard Model with the 
gauge symmetry $SU(3)' \times SU(2)' \times Y(1)'$ and a full complement 
of SM-like fields.
\item The idea was quickly taken up by Bertolami and Rosenfeld 
\cite{Berto}, who speculated on the Higgs sector being coupled to a 
singlet scalar which is a quintessence field (solving the dark energy 
problem) and showing that the symmetry-breaking also generates a dark 
matter candidate.
\item In the wake of the 2012 Higgs boson discovery, it was pointed out 
by Djouadi, Falkowski, Mambrini and Quevillon \cite{Yann} that the 
fairly stringent LHC limits on invisible decays of the observed $H^0$ 
would lead to severe constraints on Higgs portal models.
\item It was perhaps inevitable that speculation about a scalar dark 
matter candidate would be followed by speculation about a vector dark 
matter candidate. Higgs portal models with a dark vector field, with a 
$Z_2$ symmetry forbidding it to couple to all SM fields except the Higgs 
field where it can have seagull-type interactions were introduced by 
Lebedev, Lee and Mambrini \cite{Lebedev} in 2012 and have been followed 
up by several studies of a similar nature \cite{Vportal}.
\item In 2013, Weinberg \cite{Steve} speculated that an extra complex 
scalar carrying a conserved $U(1)$ quantum number, which was 
spontaneously broken through its mixing with the SM Higgs scalar, could 
be feebly interacting and might 'masquerade' as a cosmic ray neutrino. 
It was soon realised (2014) that if the singlet can be thought of as 
dark matter rather than a fake neutrino, then this is also a suitable 
model for Higgs portal dark matter \cite{Anchor}.

\item Taking into account the existence of neutrino masses, a 
right-chiral neutrino becomes a perfect candidate for Higgs portal dark 
matter, sicne it has no gauge interactions and couples to the Higgs 
sector only very feebly through its Yukawa couplings, which are 
proportional to the tiny neutrino mass \cite{Haba}.
\end{enumerate}  
\vspace*{-0.2in} 
This above list is illustrative but not exhaustive. It may be mentioned 
at this point that Higgs portal models belong to the genre of {\it 
simplified models} where a few extra fields are added to the SM. mostly 
to explain dark matter, without worrying about the other issues in the 
SM. Obviously, these theories have modest aims, and hence, they 
represent necessary but modest progress in the vast area of dark matter 
research.

\subsection{\sl Higgs Triplets} 

If there is an additional Higgs triplet, obviously it would cause 
deviations in the $\rho$ parameter, which is highly constrained, and 
this would require fine-tuning of the vacuum expectation value of the 
triplet state \cite{Mann}. An example of such a model is that due to 
Gelmini and Roncadelli \cite{Gelmini} which considers an extra triplet 
with $T = 1$ and $Y = 2$ and a small vacuum expectation value $v_t$ of 
the neutral component of the triplet. This leads to
\begin{equation}
\rho = \frac{2v^2 + 4v_t^2}{2v^2 + 8v_t^2} \simeq 1 - 2\left(\frac{v_t}{v}\right)^2
\end{equation}  
and the existing bound in Eqn.~(\ref{eqn:rho-expt}) would require $v_t < 
2.6$~GeV at $3\sigma$. Accepting this fine tuning, one can build a model 
with a Higgs triplet $\Phi_t$ (also parametrisable as a bidoublet 
$\Delta_t$ which are given by
\begin{equation}
\Phi_t = \left( \begin{array}{c} \varphi_t^{++} \\ \varphi_t^+ \\ \varphi_t^0 + \frac{v_t}{\sqrt{2}} \end{array} \right)
\qquad\qquad
\Delta_t = \frac{1}{\sqrt{2}} \left( \begin{array}{cc} \varphi_t^+ & \sqrt{2} \varphi_t^{++} \\ \sqrt{2} \varphi_t^0 + v_t & -\varphi_t^+ \end{array} \right)
\end{equation} 
The attractive feature of this model was that the low scale $v_t$ could 
be used to generate small Majorana neutrino masses through the lepton 
number-violating operator
\begin{equation}
{\cal L}_{\rm Maj} = \sum_{a = 1}^3 \sum_{b=1}^3 y_{ab}^{\rm Maj} (L_L^{a})^c i\sigma_2 \Delta_t L_L^b 
\end{equation}
leading to a neutrino Majorana mass matrix $\mathbb{M}_ab^{\nu} = 
\sqrt{2}v_t y_{ab}^{\rm Maj}$. This can be incorporated in a seesaw 
mechanism yielding correct neutrino masses, without making the 
neutrino-scalar couplings intolerably small. This is essentially the 
same, in a specific context, as the so-called Majoron mechanism 
\cite{Peccei}, and therefore it is common to call this the {\it triplet 
Majoron} model.
 
A more ingenious variant of this idea, due to Georgi and Machacek 
\cite{Machacek} is to introduce {\it two} extra triplets, one of them 
$\Phi_t$ with $Y = 2$, i.e., similar to the previous case, and one 
$\Phi'_t$ with $Y=0$. The contributions to the $\rho$ parameter then 
become
\begin{equation}
\rho = \frac{2v^2 + 4v_t^2 + 8{v'}_t^2}{2v^2 + 8v_t^2}
\end{equation}  
which can easily be made unity by choosing ${v'}_t^2 = 
\frac{1}{2}v_t^2$. This is a fine tuning, of course, but it can be 
induced by a global symmetry and obviously allows the value of $v_t$ to 
range freely, unlike in the previous case.  As a result, the 
Georgi-Machacek model, as it is called, has been the popular choice in 
which Higgs triplet studies have taken place.

However, the existence of a Higgs triplet permits a tree-level $H^\pm 
W^\mp Z$ vertex, which would have distinct signatures which so far have 
not been seen \cite{Kanemura}. There has also been no signal for the 
double-charged scalar $\varphi_t^{++}$, leading to a fairly robust lower 
bond around 200~GeV \cite{DCH}. Last, but not least, a triplet Majoron 
model would lead to an invisible decay mode for the $Z$ boson, on which 
there are stringent constraints, essentially ruling out SM extensions by 
a single multiplet with $Y > 1$ \cite{Nir}, which includes the 
Gelmini-Roncadelli model, but is evaded when there are more triplets, as 
in the Georgi-Machacek model.

Apart from collider signatures, in an interesting twist to the triplet 
story, it has been recently proposed \cite{Taiwan} that the discrepancy 
between the $W$ mass measurement from CDF and from the LHC 
collaborations can be explained by introducing triplets $\grave{\rm a}$ 
la Georgi-Machacek.

In the above discussion, we have discussed Higgs multiplets with $T = 
1,2$ and $3$. The literature on extended Higgs sectors is rich in ideas 
and it would require a much longer work to do justice to more exotic 
ideas such as Higgs-radion mixing \cite{Amit}, composite Higgses 
\cite{composite} and $S_3$-based models \cite{GG}.

\section{Post-Modernism: Higgs Effective Theories} 

As the first two runs of the LHC have failed to turn up any signals for 
physics beyond the SM, the euphoria caused by the Higgs discovery a 
decade ago has died down, to be replaced by a more agnostic attitude. 
This has led to the rise of {\it effective theories} as a tool to 
understand the data from the LHC and a host of other sources, including 
astrophysical ones. The basic idea is very simple. If there is new 
physics at a higher scale, it will induce new operators of dimension 
more than four in the action, in the same way as Fermi's six-dimensional 
current-current operator was induced by the four-dimensional operators 
of the IVB theory in the low-energy limit.The Standard Model, then, is 
the low-energy effective theory of the unknown high scale theory if we 
add on a full set of these higher-dimensional operators. These effective 
operators will contain only the SM fields and will respect all the 
symmetries of the SM.

Just as the Fermi coupling was the free parameter of Fermi's theory, 
each higher-dimensional operator will have an unknown coupling constant 
--- generically called a {\it Wilson coefficient} --- and these are the 
'free' parameters of the theory. Obviously, the same operator, or set of 
operators, can give rise to multiple observables, and thus it is 
economical to consider all these processes to obtain the values of the 
Wilson coefficients experimentally. This is in the same spirit as beta 
decay and muon decay could be compared to find the same Fermi constant 
and thereby discover charged current universality. This method is 
already the standard tool in flavour physics. It must be noted that the 
Wilson coefficients will have some dimension, just as the Fermi constant 
has dimension, and this will reflect the scale of the new physics, just 
as the Fermi constant reflects the scale of the electroweak 
symmetry-breaking. In fact, if we were to write the Fermi constant as
\begin{equation}
G_F \defeq \frac{c_F}{v^2}
\end{equation}
then $c_F = \frac{1}{\sqrt{2}}$ is the Wilson coefficient of the beta 
decay operator. It is, therefore usual to write the unknown couplings in 
terms of a common high scale $\Lambda$ as $c_5/\Lambda, c_6/\Lambda^2, 
c_7/\Lambda^3, \dots$ and use the label 'Wilson coefficients' for the 
$c_i$'s specifically and the subscript $i$ refers to the dimensionality 
of the operator.

Once we include higher-dimensional operators with couplings which are 
not dimensionless, the theory ceases to be renormalisable. This is not a 
problem, since we expect $\Lambda$ to be a cutoff scale, above which the 
new theory becomes valid, and that will, presumably, be renormalisable. 
This is analogous to the fact that the Fermi theory is not 
renormalisable, but when we go to higher energies close to the 
electroweak scale, it is replaced by the Glashow-Salam-Weinberg theory, 
which is renormalisable. However, just as the Fermi coupling $G_F$ runs 
with energy, so will the Wilson coefficients in an effective theory of 
the SM (SMEFT) run with energy. In doing so, they will also mix with 
other operators. Thus, we require to define clumps of operators which 
mix with each other, but not with others because of one conservation law 
or the other. These sets are then said to be {\it closed} under 
renormalisation group evolution. Early study in the context of flavour 
physics were carried out by Leung, Love and Rao \cite{Rao} and by Bigi, 
Kopp and Zerwas (1986) \cite{Bigi}, but a complete enumeration of all 
the SMEFT operators was done --- a real {\it tour de force} --- by 
Buchm\"uller and Wyler in 1986 \cite{BuWy}. Alas! there are 2499 
operators in the list when all flavours are taken into account, each 
with an unknown Wilson coefficient \cite{Manohar}. All is not lost 
however, for it has been pointed out (2010) by a group from Warsaw 
University \cite{Warsaw} that of the dimension-6 operators only a set of 
59 will affect electroweak measurables, as opposed to the total number 
80 in the Buchm\"uller-Wyler enumeration. This set of 59 operators has 
come to be known as the {\it Warsaw basis} and is a popular choice for 
SMEFT studies. However, there are many more basis choices, and it quite 
a task to translate between the different notations and conventions of 
these different choices \cite{Wcxf}.
    
Obviously many of the SMEFT operators will involve the Higgs doublet 
$\Phi(x)$. Isolating this set leads to a so-called Higgs EFT, or HEFT. 
In the following the HEFT outlined by Contino {\it et al} \cite{Contino} 
is briefly described. In this framework, the HEFT action is divided into 
three parts, having the form
\begin{equation}
S_{\rm HEFT} = S_{\rm HEFT}^{(1)} + S_{\rm HEFT}^{(2)} + S_{\rm HEFT}^{(3)}
\end{equation}   
where each of them is written as an operator sum
\begin{equation}
S_{\rm HEFT}^{(X)} \defeq \int d^4x \ \frac{1}{\Lambda^2} \sum_{i=1}^{n_X} c_i^{(X)} \ {\cal O}_i^{(X)} 
\end{equation}
where $n_X = 12, 8, 8$ for $X = 1,2,3$ respectively, $\Lambda$ is the 
scale of the new physics and the $c_i^{(X)}$ are the Wilson coefficients 
of the dimension-6 operators ${\cal O}_i^{(X)}$. The list of operators 
in $S_{\rm HEFT}^{(1)}$, which matches with the strongly-interacting 
light Higgs (SILH) scenario of Ref.~\cite{SILH}, is given below.
\begin{eqnarray}
{\cal O}_1^{(1)} & = & 
\mathbb{D}^\mu (\Phi^\dagger\Phi) \mathbb{D}_\mu (\Phi^\dagger\Phi)
\hspace*{1.0in}
{\cal O}_2^{(1)} =   
(\Phi^\dagger \overleftrightarrow{\mathbb{D}^\mu} \Phi) (\Phi^\dagger \overleftrightarrow{\mathbb{D}_\mu} \Phi) 
\nonumber \\
{\cal O}_3^{(1)} & = & 
- (\Phi^\dagger\Phi)^3
\hspace*{1.6in}
{\cal O}_4^{(1)} =
(\Phi^\dagger\Phi) \overline{Q}_L \widetilde{\Phi} u_R + {\rm H.c.}
\nonumber \\
{\cal O}_5^{(1)} & = & 
(\Phi^\dagger\Phi) \overline{Q}_L \Phi d_R + {\rm H.c.}
\hspace*{0.85in}
{\cal O}_6^{(1)} =
(\Phi^\dagger\Phi) \overline{L}_L \widetilde{\Phi} e_R + {\rm H.c.}
\nonumber \\
{\cal O}_7^{(1)} & = &
\Phi^\dagger (\mathbb{D}^\nu \mathbb{W}_{\mu\nu}) \overleftrightarrow{\mathbb{D}^\mu} \Phi
\hspace*{1.13in}
{\cal O}_8^{(1)} =
(\Phi^\dagger \overleftrightarrow{\mathbb{D}^\mu} \Phi) \partial^\nu B_{\mu\nu} 
\nonumber \\
{\cal O}_9^{(1)} & = &
(\mathbb{D}^\mu \Phi)^\dagger \mathbb{W}_{\mu\nu} (\mathbb{D}^\nu \Phi))
\hspace*{0.95in}
{\cal O}_{10}^{(1)} =
(\mathbb{D}^\mu \Phi)^\dagger (\mathbb{D}^\nu \Phi)) B_{\mu\nu}
\nonumber \\
{\cal O}_{11}^{(1)} & = &
(\Phi^\dagger\Phi) B_{\mu\nu} B^{\mu\nu}
\hspace*{1.27in}
{\cal O}_{12}^{(1)} =
(\Phi^\dagger\Phi) W_{\mu\nu}^a W^{a\mu\nu}
\end{eqnarray}  
where the covariant derivatives are defined in 
Eqn.~(\ref{eqn:nonabelian}). Similarly the operators for $S_{\rm 
HEFT}^{(2)}$ are
\begin{eqnarray}
{\cal O}_1^{(2)} & = &
(\overline{Q_L} \gamma^\mu Q_L) 
(\Phi^\dagger \overleftrightarrow{\mathbb{D}_\mu} \Phi)
\hspace*{0.9in}
{\cal O}_2^{(2)} =
(\overline{Q_L} \gamma^\mu \vec{\mathbb{T}} Q_L)\cdot 
(\Phi^\dagger \vec{\mathbb{T}} \overleftrightarrow{\mathbb{D}_\mu} \Phi)
\nonumber \\
{\cal O}_3^{(2)} & = &
(\overline{u_R} \gamma^\mu u_R) 
(\Phi^\dagger \overleftrightarrow{\mathbb{D}_\mu} \Phi)
\hspace*{0.97in}
{\cal O}_4^{(2)} =
(\overline{d_R} \gamma^\mu d_R) 
(\Phi^\dagger \overleftrightarrow{\mathbb{D}_\mu} \Phi)
\nonumber \\
{\cal O}_5^{(2)} & = &
(\overline{u_R} \gamma^\mu d_R)
(\Phi^\dagger \overleftrightarrow{\mathbb{D}_\mu} \Phi) + {\rm H.c.}
\hspace*{0.52in}
{\cal O}_6^{(2)} =
(\overline{L_L} \gamma^\mu L_L)
(\Phi^\dagger \overleftrightarrow{\mathbb{D}_\mu} \Phi)  
\nonumber \\
{\cal O}_7^{(2)} & = &
(\overline{L_L} \gamma^\mu \vec{\mathbb{T}} L_L) \cdot 
(\Phi^\dagger \vec{\mathbb{T}} \overleftrightarrow{\mathbb{D}_\mu} \Phi)
\hspace*{0.65in}
{\cal O}_8^{(2)} =
(\overline{e_R} \gamma^\mu e_R) 
(\Phi^\dagger \overleftrightarrow{\mathbb{D}_\mu} \Phi)
\end{eqnarray} 
and the operators for $S_{\rm HEFT}^{(3)}$ are
\begin{eqnarray}
{\cal O}_1^{(3)} & = & 
\overline{Q_L} \widetilde{\Phi} \sigma^{\mu\nu} u_R B_{\mu\nu} + {\rm H.c.} 
\hspace*{0.8in}
{\cal O}_2^{(3)} =
\overline{Q_L} \mathbb{W}_{\mu\nu} \widetilde{\Phi} \sigma^{\mu\nu} u_R + {\rm H.c.} 
\nonumber \\
{\cal O}_3^{(3)} & = &
\overline{Q_L} \widetilde{\Phi} \sigma^{\mu\nu} \mathbb{G}_{\mu\nu} d_R + {\rm H.c.} 
\hspace*{0.8in}
{\cal O}_4^{(3)} =
\overline{Q_L} \Phi \sigma^{\mu\nu} d_R B_{\mu\nu} + {\rm H.c.}
\nonumber \\
{\cal O}_5^{(3)} & = &
\overline{Q_L} \mathbb{W}_{\mu\nu} \Phi  \sigma^{\mu\nu} d_R + {\rm H.c.}
\hspace*{0.8in}
{\cal O}_6^{(3)} =
\overline{Q_L} \sigma^{\mu\nu} \mathbb{G}_{\mu\nu} d_R + {\rm H.c.} 
\nonumber \\
{\cal O}_7^{(3)} & = &
\overline{L_L} \Phi  \sigma^{\mu\nu} e_R B_{\mu\nu} + {\rm H.c.}
\hspace*{0.85in}
{\cal O}_8^{(3)} =
\overline{L_L} \mathbb{W}_{\mu\nu} \Phi \sigma^{\mu\nu} e_R + {\rm H.c.}
\end{eqnarray}

In the above formulae, we must take note of the iso-vector currents 
generated by the $\mathbb{T}$ as well as the fact that when the 
$3\times3$ matrix $\mathbb{G}_{\mu\nu}$ is introduced between quark 
fields, it is implicit that each quark field is a colour triplet. 
Moreover, these 28 operators have been written for a single flavour. If 
we assume that there is no flavour-violation from the new physics, we 
could simply write down three copies of these operators for the three 
fermion generations. Alternatively, if we allow there to be 
flavour-changing effects in the EFT, then we can write mixed flavour 
operators, e.g., \begin{equation} \left[{\cal O}_{4}^{(1)}\right]_{ij} 
\defeq (\Phi^\dagger\Phi) \overline{Q}_{Li} \widetilde{\Phi} u_{Rj} + 
{\rm H.c.} \end{equation} where $i,j = 1,2,3$ for the different 
generations. Thus, there will be a $3\times 3$ matrix of Wilson 
coefficients $[c_{4}^{(1)}]_{ij}$, which will complement the Yukawa 
couplings in generating flavour violation. In any case, even if we do 
not make this assumption, the above operators are written in terms of 
the gauge bosons and fermions in the (unbroken) gauge basis and some 
mixing will be induced when the electroweak symmetry is spontaneously 
broken.

Calculations in the HEFT are long and messy, but fortunately we live in 
the age of computer algebra, and hence several software packages have 
been created to carry out these calculations. A summary of the available 
tools may be found in Ref.~\cite{EFT-tools}. It would lead us too far 
afield in this article to discuss the details of EFT calculations and 
the different constraints obtainable on the Wilson coefficients. 
However, Higgs EFTs have become the framework of choice for the LHC 
Collaborations to present their (so far negative) results. A recent 
summary of these findings may be found in Ref.~\cite{Nikita}.

\section{Concluding Remarks} 

The Higgs boson represents both an end and a beginning. It most 
definitely marks the end of the long road to building up a viable 
framework which explains and correctly predicts the value of different 
measurables in the area of elementary particle interactions at high 
energies, a framework which we call the Standard Model. However (and 
this has been explained at some length), that that is precisely what it 
is --- a model and not a theory. There are too many {\it ad hoc} 
inclusions and experimentally-fitted parameters in the SM for it to 
really qualify as a theory. Nevertheless, that the Standard Model has 
been supremely successful in explaining a few thousand experimental 
results cannot be denied. Essentially, the SM is like a quack doctor 
whose motley treatments are suspect, but who has enormous success in 
curing people. After all (to flog a clich\'e), nothing succeeds like 
success.

For the last four decades, scientists have been beating at the stone 
wall of the SM, hoping to find a way to new physics, which is desirable 
for the many reasons explained in the text. In this wall, perhaps the 
weakest point is the Higgs sector, about which very little is really 
known, and hence provides a possible way to glimpse the new physics 
which lies at a higher scale. In this sense, the Higgs boson is also a 
beginning. Only the future will tell if this beginning is an auspicious 
one.

Before concluding this long story, it is worth noting the fact that the 
Higgs discovery required the linkage of elementary particle physics with 
several other branches of science. The idea of gauge theory itself came 
from general relativity and that of spontaneous symmetry-breaking came 
from condensed matter physics. The issues of dark matter and dark energy 
arose from astrophysical measurements. Among the tools of particle 
physicists, computers have always played a major role in the number 
crunching, but over the past couple of decades, they have begun to take 
over the burden of analytical calculations as well. This has permitted 
theorists to take up issues which would earlier have been totally 
intractable. Likewise, the enormous amounts of data which come pouring 
out of the LHC and similar machines, requires the handling of large data 
structures, and inevitably, the use of machine learning and artificial 
intelligence to interpret these masses of data. And yet, unless and 
until we find something which establishes the existence of physics 
beyond the SM in a terrestrial environment, much of all this effort and 
technical sophistication will be, to quote Lord Rutherford, 'mere 
stamp-collecting'. It may well be that a single small idea of a 
revolutionary nature is needed to break the impasse. One can only hope 
that this will not be a long time in coming.

{\small The author acknowledges support of the Department of Atomic 
Energy, Government of India, under Project Identification No. RTI 4002.}

\small
\setstretch{1.05}

\end{document}